\documentclass[a4paper,11pt]{article}
\usepackage{jheppub} 
\usepackage{lineno}
\usepackage{enumerate}
\usepackage{float}
\usepackage{changes}
\usepackage{subfig}
\usepackage{color}
\usepackage{slashed}
\usepackage{multirow}

\usepackage{chngcntr}
\usepackage{amssymb, amsmath,mathrsfs}
\usepackage{graphics}
\usepackage{graphicx}
\usepackage{epsf}
\usepackage{epsfig}
\usepackage{float}
\usepackage{makecell}
\usepackage{multirow}
\usepackage{color}
\usepackage{xcolor}
\usepackage[utf8]{inputenc}
\usepackage{subfig}
\usepackage[normalem]{ulem}
\usepackage{diagbox}
\usepackage{natbib}
\usepackage{tikz}
\usepackage[vcentermath]{youngtab}
\usepackage{slashed} 
\usepackage[font=small]{caption} 
\usepackage{times} 
\usepackage{bm}       
\usepackage{bbm} 
\usepackage{relsize}   
\usepackage{autobreak}  
\usepackage{makeidx} 
\usepackage{soul}
\usepackage{bbding} 
\usepackage{listings} 
\usepackage{comment} 
\usepackage{diagbox}
\usepackage[normalem]{ulem}
\usepackage{soul}
\def\<{\langle}
\def\>{\rangle}
\usepackage{array}
\newcommand{\ab}[1]{\langle #1 \rangle}
\newcommand{\sqb}[1]{[#1]}
\newcommand{\asb}[1]{\langle #1]}
\newcommand{\bs}[1]{\boldsymbol{#1}}
\newcommand{\capi}[1]{\tiny\mbox{$#1$}}
\newcommand{\capmi}[1]{\tiny\mbox{$\mathcal{#1}$}}

\newcommand{\lam}{\lambda}
\newcommand{\Lam}{\Lambda}
\newcommand{\eps}{\epsilon}
\newcommand{\veps}{\varepsilon}

\title{\boldmath Explicit Conditions for Diagnosing Tree-Level Unitarity}

\author[1]{Jaehoon Jeong,}
\author[1]{Pyungwon Ko,}
\author[2]{and Yu-Hui Zheng}
\affiliation{Quantum Universe Center (QUC), KIAS 85 Hoegi-ro, Seoul 02455, Korea}
\affiliation{School of Physics, KIAS 85 Hoegi-ro, Seoul 02455, Korea}

\abstract{We explicitly present all coupling conditions required for tree-level unitarity (tree unitarity) in theories with a finite number of massive and massless particles of spin up to 1. They allow us to diagnose tree unitarity of a system using only its particle content in the mass basis, without reconstructing the full Lagrangian. We show that all four-point amplitudes whose high-energy growth is canceled by tree unitarity conditions are on-shell constructible, thereby motivating the recursive construction of four-point amplitudes. By examining their high-energy growth, we derive tree unitarity conditions for four-point amplitudes. Imposing these conditions to simplify the Lagrangian structure, we use the St\"uckelberg formulation to derive the tree unitarity conditions arising from all higher-point amplitudes. We show that all tree unitarity conditions are fully captured up to five-point amplitudes, ensuring no necessity of examining higher-point ones. 
We apply our results to systematically examine tree unitarity conditions in the 
dark sector with a massive dark photon and dark matter particles 
of spin up to 1, and extract the essential features for mass generation in the massive
dark photon case. 
In addition, we show that our results allow us to conclude that theories without 
scalars require an infinite tower of vectors and fermions for tree unitarity.
Finally unitarity and related issues in the Higgs portal VDM are discussed in brief. 
}

\begin{document}

\begin{flushright}
KIAS-Q26005
\end{flushright}

\maketitle
\flushbottom

\section{Introduction}

Unitarity is a fundamental and indispensable principle in physics, ensuring the conservation of probability and the consistency of quantum theories. In practice, however, model building is often carried out by specifying a set of particle contents and interactions from an effective-field-theory (EFT) perspective. While this provides a useful parametrization of possible low-energy effects, such descriptions inevitably violate unitarity above their cutoff, and one is therefore led to examine how sensitively unitarity is violated in a given setup. In particular, if new physics admits a UV-complete description, where the cutoff scale does not explicitly appear in the Lagrangian, as in the Standard Model (SM)~\cite{Glashow:1961tr,Weinberg:1967tq,Salam:1968rm}, one is eventually led to assess the UV safety of the theory under consideration, rather than focusing only on dominant contributions in specific processes.

This motivates us to explicitly derive all coupling conditions for tree unitarity, such that they can be used to diagnose the tree unitarity of a system using only its particle content in the mass basis. In other words, these conditions allow us to assess tree unitarity without reconstructing the full Lagrangian. For this purpose, it is essential to examine tree unitarity of a system with a general set of particles. About five decades ago, the authors of Ref.~\cite{Cornwall:1974km} showed that tree-unitary theories correspond to spontaneously broken gauge theories (SBGT)~\cite{Higgs:1964pj,Englert:1964et,Guralnik:1964eu} supplemented by additional Abelian vector mass terms. To reach this conclusion, they derived part of the tree unitarity conditions on  couplings by examining several four-point (4-pt) amplitudes in theories including a finite number of massive particles with spin up to 1. However, the remaining conditions were left implicit, thereby limiting the direct applicability of these results in practical analyses.

Recently, in Ref.~\cite{Liu:2022alx}, the authors employed modern on-shell methods~\cite{Elvang:2015rqa,Arkani-Hamed:2017jhn,Conde:2016izb,Durieux:2019eor,Durieux:2020gip,Christensen:2018zcq,Li:2022tec}, but without using recursive shift constructions, to derive all tree unitarity conditions arising from 4-pt on-shell amplitudes, in the same theory setup as in Ref.~\cite{Cornwall:1974km}. However, these conditions have not been fully explored, particularly for higher-point (higher-pt) amplitudes or for theories including massless particles. For this reason, in this work, we construct tree-unitary theories including both massive and massless particles of spin up to 1. To achieve this goal, we first examine tree unitarity of 4-pt on-shell amplitudes in the extended set of particles. In this procedure, we adopt a fully on-shell approach using the recently developed All-Line Transverse (ALT) shift~\cite{Elvang:2015rqa,Ema:2024vww,Ema:2024rss}.

To validate the recursive construction, we first show that smooth massless limits can always be defined for amplitudes in tree-unitary theories, and that all 4-pt on-shell amplitudes whose high-energy growth is canceled by tree unitarity conditions are on-shell constructible. These two statements lead us to construct only the 3-pt on-shell amplitudes admitting smooth massless limits, and then to recursively construct the 4-pt on-shell amplitudes using the derived 3-pt ones. Subsequently, by examining their high-energy scaling, we derive all coupling conditions required for tree unitarity of 4-pt amplitudes.

Collecting all tree unitarity conditions for 4-pt amplitudes, we present them in a unified form as the Lie algebra of a group~\cite{Liu:2022alx}. To complete the tree-unitarity examination, one can in principle examine the high-energy scaling of higher-pt amplitudes after constructing them recursively, but this is not practically feasible. Thus, we adopt an alternative approach presented in Ref.~\cite{Cornwall:1974km}. After simplifying the Lagrangian to satisfy tree unitarity for 4-pt amplitudes, we take the St\"uckelberg formulation~\cite{Stueckelberg:1938hvi,Ruegg:2003ps}, which changes the field contents. After establishing the equivalence between the tree-level amplitudes from the mass-basis Lagrangian and those obtained from the St\"uckelberg Lagrangian, we show that in the St\"uckelberg formulation, all tree-unitarity-violating contributions arise from higher-pt contact terms in the scalar potential.

This leads us to conclude that, in the St\"uckelberg formulation, the tree unitarity conditions for all higher-pt amplitudes are translated into gauge-invariance conditions on the couplings in the scalar potential. In this procedure, we reach the same conclusion as in Ref.~\cite{Cornwall:1974km}, namely that tree-unitary theories correspond to SBGTs with additional Abelian vector mass terms. In our derivation, we keep the mass-basis couplings until the end. This clarifies that, unlike in previous approaches where the conditions are either implicit or limited to low-point amplitudes, all tree unitarity conditions can be formulated explicitly in the mass basis. Furthermore, keeping the coupling basis leads us to find that all tree unitarity conditions are fully captured up to 5-pt amplitudes.

As concrete applications, we use the derived conditions to diagnose the tree unitarity of dark photon scenarios with two dark matter (DM) particles of spin up to 1, which provide a minimal extension of the Standard Model through an additional Abelian gauge sector~\cite{Holdom:1985ag,Pospelov:2008zw,Fabbrichesi:2020wbt}. 
In this procedure, we find that for scalar and fermion DM particles, inelastic scenarios can be realized only by introducing an additional scalar responsible for spontaneous symmetry breaking (SSB). However, in the vector DM case, we show that even generating a mass splitting between the dark photon and vector DM particles is not straightforward in this way, because the three-vector coupling is strongly constrained by the gauge structure of the theory. Thus, our analysis implies that tree unitarity provides nontrivial and highly restrictive constraints on inelastic dark photon scenarios with vector DM particles.

Finally, we show that the coupling conditions can also be applied to diagnose tree unitarity in theories without scalars. In this case, a finite number of vectors and fermions cannot satisfy the tree unitarity conditions. However, an infinite tower of such particles can preserve tree unitarity, implying a connection to extra-dimensional Higgsless models~\cite{Csaki:2003dt,Csaki:2003zu,Csaki:2003sh,Csaki:2004sz}. 

The remainder of this work is organized as follows. In Section~\ref{sec:tree_unitarity}, we examine the definition of tree unitarity and derive a bound on the mass dimension of couplings. After reviewing the improved ALT shift, we compare the bounds on the coupling mass dimension in tree-unitary and on-shell constructible theories, ensuring the on-shell constructibility of tree-unitary 4-pt on-shell amplitudes. In Section~\ref{sec:3-pt_amps_operators}, we construct all tree-unitary on-shell 3-pt amplitudes and the associated operators. In Section~\ref{sec:4-pt_tree_unitarity}, we construct the 4-pt on-shell amplitudes recursively and derive the coupling conditions from their high-energy scaling. In Section~\ref{sec:lagrangian_construction}, we present the tree unitarity conditions for 4-pt amplitudes in a unified form, leading to a Lie algebra structure, and simplify the Lagrangian accordingly. Adopting the St\"uckelberg formulation, we derive the coupling conditions for all higher-point amplitudes, thereby obtaining all explicit conditions for diagnosing tree-level unitarity in our setup. In Section~\ref{sec:dark_photon}, we apply the results to dark photon scenarios with two dark matter (DM) particles of spin up to 1. In Section~\ref{sec:higgsless}, we use our results to infer the tree unitarity conditions for theories without scalars. In Section~\ref{sec:conclusion}, we summarize our results.

In Appendix~\ref{appendix:spinor_helicity_formalism}, we review the spinor-helicity formalism and present our conventions. In Appendix~\ref{appendix:CPT}, we show that tree-level Hermiticity imposes reality and Hermiticity on bosonic and fermionic couplings, respectively. In Appendix~\ref{appendix:high-E_limits_spinor}, we present the high-energy limits of spinor contractions used in the analysis of 4-pt amplitudes. In Appendix~\ref{appendix:summary_conditions}, we summarize the tree unitarity conditions for direct use in practical analyses.

\section{Tree unitary theories}
\label{sec:tree_unitarity}

\subsection{Tree unitarity}
\vspace{0.2cm}

\noindent We first review tree unitarity, which we use synonymously with tree-level unitarity. From the $S$-matrix, $S=1+iT$, we define the amplitude $\mathcal{M}_{n-r,r}$ for an $r \to n-r$ scattering process, where $r$ initial particles produce $n-r$ final-state particles as
\begin{align}
\langle n-r | S| r \rangle
=\langle n-r|r\rangle+i(2\pi)^4 \delta^4\big(P_r-P_{n-r}'\big) 
\mathcal{M}_{n-r,r},
\label{eq:s_matrix_element}
\end{align}
where $P_r=p_1+\cdots +p_r$ and $P_{n-r}'=p_{r+1}+\cdots +p_{n}$ represent the total momenta of the initial and final states, respectively, and $\mathcal{M}_{n-r,r}$ has mass dimension $4-n$.\footnote{We follow the state convention in Ref.~\cite{Peskin:1995ev}, determining the mass dimension of $\mathcal{M}$.} In this section, we discuss only the case in which no momenta are parallel to each other, thereby avoiding divergences from collinear configurations.

Let us consider a $2\! \to \!2$ elastic process with the high collision energy $E$ in the center of mass (c.m.) frame. Tree unitarity originates from the unitarity condition [$-i(T-T^\dagger)=TT^\dagger$]:
\begin{align}
&
\int d\Phi_{2,n-2}\,
|\mathcal{M}_{2,n-2}|^2
< \mbox{Im}\mathcal{M}_{2,2}
= \mbox{constant},
\label{eq:unitarity}
\end{align}
for $n\geq 4$,\footnote{At high $E$, the case $n=3$ (i.e., $2 \to 1$) is kinematically forbidden, since the mass of the produced on-shell particle is much smaller than $E$, violating energy conservation.} where $d\Phi_{2,n-2}$ is the Lorentz-invariant phase-space (LIPS) measure for $n-2$ particles. For ease of energy-power counting of the LHS above, we define the dimensionless LIPS measure $d\tilde{\Phi}_{2,n-2}\equiv d\Phi_{2,n-2}/E^{2n-8}$ whose explicit form is
\begin{align}
&\quad d\tilde{\Phi}_{2,n-2}=
(2\pi)^4 \delta^4 \Big(\tilde{p}_{1}+\tilde{p}_{2}-\sum_{i=3}^{n} \tilde{p}_i\Big)
\bigg(\prod_{i=3}^{n}\frac{d^3\vec{\tilde{p}}_i}{(2\pi)^3 2\tilde{p}^0_i}\bigg)
\quad \mbox{with}\quad 
\tilde{p}_a= p_a/E\;\;  (a=1,\cdots\!,n).
\end{align}
Defining the dimensionless amplitude $\tilde{\mathcal{M}}\equiv \mathcal{M}/E^{4-n}$ to cancel $E^{2n-8}$ arising from $d\Phi_{2,n-2}$, we re-express Eq.~\eqref{eq:unitarity} as
\begin{align}
&
\int d\tilde{\Phi}_{2,n-2}\,
|\tilde{\mathcal{M}}_{2,n-2}|^2
= \mbox{constant},
\label{eq:unitarity2}
\end{align}
where the integral being constant constrains the integrand $|\tilde{\mathcal{M}}_{2,n-2}|^2$ not to grow at high energy.

In this work, we consider local and perturbative theories, where any tree-level amplitude is polynomially bounded in energy. Then, to satisfy the unitarity condition in Eq.~\eqref{eq:unitarity2}, a tree-level amplitude must obey the high-$E$ scaling as
\begin{align}
(\tilde{\mathcal{M}}_{2,n-2})_{\rm tree}
\lesssim E^{0} \quad \rightarrow \quad
(\mathcal{M}_{2,n-2})_{\rm tree}
\lesssim E^{4-n}, 
\label{eq:tree_unitarity}
\end{align}
showing that $(\mathcal{M}_{2,n-2})_{\rm tree}$ grows no faster than $E^{4-n}$. This condition is known as {\it tree unitarity}~\cite{Cornwall:1974km}.\footnote{This result is stronger than the Froissart–Martin bound~\cite{Froissart:1961ux,Martin:1965jj}.  
The point is simply that local tree-unitary theories require a slower high-energy growth than what analyticity and unitarity alone would guarantee.}

Tree unitarity imposes strong constraints on theories that generate contributions with positive powers of $E/\Lambda$, with the cutoff scale $\Lambda$. Even higher-point operators without derivatives are forbidden because tree unitarity requires stronger suppression in the high-$E$ scaling. It means that the intrinsic coupling of any operator must have mass dimension $\geq 0$. For notational simplicity, we will use the symbol ``[]" to denote the mass dimension of a quantity throughout this work, unless otherwise specified.

To show other constraints from tree unitarity, we take the high-$E$ expansion of a tree-unitary amplitude $\mathcal{M}_n$$=(\mathcal{M}_{2,n-2})_{\rm tree}$. In this expansion, we omit angular parameters, while specifying dimensionful parameters. 
Note that the only dimensionful parameters appearing in the denominators of tree-level amplitudes are the masses of vector particles. For this reason, we take the expansion in terms of the vector masses divided by $E$, i.e., $m_a/E$; we also extract the vector masses from all dimensionful couplings, leading to a full expansion in terms of the vector masses. Then, at high $E$, we have
\begin{gather}
\mathcal{M}_n\bigg(E;\frac{\{m_a\}}{E}\bigg)=
\, E^{4-n}\sum_{\tau_1,\cdots,\tau_k} a_{\tau_1,\cdots , \tau_k}\bigg(\frac{m_{a_1}}{E}\bigg)^{\tau_1}\cdots \bigg(\frac{m_{a_k}}{E}\bigg)^{\tau_k},
\nonumber
\\
\mbox{with} \quad \tau_1+\cdots+ \tau_k\ge 0
\label{eq:M_n_expansion}
\end{gather}
satisfying $\mathcal{M}_n\lesssim E^{4-n}$, 
where $\{m_a\}=m_{a_1},...\,,m_{a_k}$ denote the collection of vector masses appearing the amplitude. Here, the dimensionless coefficients $a_{\tau_1,\cdots,\tau_k}$ 
are independent of the vector masses, but include angular parameters.

Given the longitudinal mode of massive vectors, one knows that the vector masses are the only dimensionful parameters appearing in the denominator of tree-level amplitudes, apart from propagators. Thus, the condition $\tau_1+\cdots+\tau_k \ge 0$ enforces that each coefficient in the expansion of $\mathcal{M}_n$ in terms of $1/E$ must be expressed as a ratio of vector masses. Since tree unitarity allows only non-negative powers of $1/E$ in the expansion, the vector-mass ratios must have mass dimension $\ge 0$. Thus, {\it tree unitarity allows one to define a smooth massless limit of an amplitude in which all vector masses are taken to zero at the same rate}, implying a spontaneously broken gauge theory (SBGT)~\cite{Englert:1964et,Higgs:1964pj,Guralnik:1964eu}. In this work, we use this fact when deriving the 3-pt on-shell amplitudes allowing smooth massless limits.

We note that theories violating tree-level unitarity generally do not admit a well-defined massless limit. This is primarily because the massless limit involves a discontinuous transition from massive to massless representations, resulting in a different set of physical degrees of freedom and admissible amplitudes. Conversely, amplitudes with smooth massless limits can still grow without bound at high $E$, reflecting the clear mismatch between the high-$E$ and massless limits.

Next, we express the amplitude $\mathcal{M}_n$ in a simple form as~\cite{Ema:2024rss}
\begin{align}
\mathcal{M}_n=\mathcal{G}
\times \prod u \times \prod \eps 
\quad \mbox{with} \quad 
\mathcal{G}=\sum_{\rm diagrams}
g \times F,
\label{eq:amp_F_u_eps}
\end{align}
where $\mathcal{G}$ is the tree-level amputated Green's function with on-shell external momenta (which will be discussed in detail in Section~\ref{sec:lagrangian_construction}), and $\prod u$ and $\prod \epsilon$ denote the products of Dirac spinors and polarization vectors, respectively. For later convenience, we decompose $\mathcal{G}$ into a collection of couplings $g$ and the associated stripped amplitude $F$ for each diagram. Dimensional analysis gives
\begin{gather}
[\mathcal{M}_n]=4-n, \;\;\;[u]=1/2,\;\;\;[\eps]=0\quad\;\; \Rightarrow \quad\;\;
[\mathcal{G}]=4-n-N_F/2.
\label{eq:F_dim}
\end{gather}
At high $E$, the function $\mathcal{G}$ is expected to scale as $\mathcal{G} \rightarrow E^{[\mathcal{G}]-[g]_{\rm min}}$, where $[g]_{\rm min}$ denotes the minimal mass dimension of a coupling collection. Thus, tree unitarity implies
\begin{align}
\begin{gathered}
\mathcal{G}\rightarrow E^{[\mathcal{G}]-[g]_{\rm min}},\;\;\, u \rightarrow E^{1/2},\;\;\, \eps^{\mp}\rightarrow E^{0},\;\;\, \epsilon^{0}\cdot \mathcal{G}\rightarrow E^{[\mathcal{G}]-[g]_{\rm min}} 
\\
\Rightarrow \quad [\mathcal{G}]\leq 4-n+[g]_{\rm min}-N_F/2,
\end{gathered}
\label{eq:G_tree_unitarity}
\end{align}
where $N_F$ is the number of spin-1/2 fermions. Here, to see the origin of scaling $\epsilon^{0}\cdot \mathcal{G}\rightarrow E^{[\mathcal{G}]-[g]_{\rm min}}$, one must notice that $\mathcal{G}$ is the function of momenta. 

Then, defining $\mathcal{G}(E;m/E)\equiv E^{[\mathcal{G}]-[g]_{\rm min}}\tilde{\mathcal{G}}(m/E)$ in terms of the mass $m$ of an external longitudinal vector, we get
\begin{align}
\mathcal{G}\bigg(E;\frac{m}{E}\bigg)
=
\,E^{[\mathcal{G}]-[g]_{\rm min}}
\bigg(\tilde{\mathcal{G}}^{(0)}(0)+\tilde{\mathcal{G}}^{(1)}(0) \frac{m}{E}+\cdots\bigg),
\end{align}
at high $E$, with the high-$E$ scalings $\tilde{\mathcal{G}}^{(i)} \rightarrow E^{0}$. As discussed in Eq.~\eqref{eq:M_n_expansion}, in tree-unitary theories, any coefficient in the $1/E$ expansion of an amplitude must be given by a ratio of vector masses with mass dimension $\geq 0$, thereby protecting the amplitude from divergences arising from inverse mass factors. This requirement forces $\eps^{0}\cdot \tilde{\mathcal{G}}^{(0)}=0$. The most conservative possibility is that $\epsilon^{0}\!\cdot \tilde{\mathcal{G}}^{(1)}\neq 0$, resulting in 
$\epsilon^{0}\!\cdot \mathcal{G} \rightarrow E^{[\mathcal{G}]-[g]_{\rm min}}$ in Eq.~\eqref{eq:G_tree_unitarity}. Such a cancellation of the high-$E$ growth of $\epsilon^0\sim E/m$ upon contraction with $\mathcal{G}$ implies the Goldstone boson equivalence theorem~\cite{Nambu:1960tm,Goldstone:1961eq,Valencia:1990jp}.
Thus, substituting Eq.~\eqref{eq:F_dim} into Eq.~\eqref{eq:G_tree_unitarity}, we obtain the bound:
\begin{align}
[g]_{\rm min}\geq 0 \quad \rightarrow \quad [g]\geq 0,
\label{eq:tree_unitarity_bound_g}
\end{align}
for any collection of couplings $g$ satisfying $[g]\geq [g]_{\rm min}$. This implies that every coupling in a tree-unitary theory has non-negative mass dimension.

\subsection{All-Line Transverse shift}

\noindent Later, we will discuss the on-shell constructibility of tree-unitary amplitudes. For this purpose, we first review the recursive on-shell construction of amplitudes based on the All-Line Transverse (ALT) shift \cite{Elvang:2015rqa,Ema:2024vww,Ema:2024rss}. The key idea is to shift the helicity spinors (see the notations and conventions in Appendix \ref{appendix:spinor_helicity_formalism}) by a complex parameter $z$. For spin-1/2 and transverse vector particles, we shift the spinors as
\begin{align}
\begin{aligned}
&|i^2\rangle \;\rightarrow \;
|\hat{i}^2\rangle=|i^2\rangle+zc_i|i^1\rangle\quad \;
\mbox{for }\; h_i=+,
\\
&|i^1] \;\rightarrow \;
|\hat{i}^1]=|i^1]+zc_i|i^2]\quad \;\;\,
\mbox{for }\; h_i=-,
\end{aligned}
\label{eq:spinor_shift1_1}
\end{align}
for massive particles, and
\begin{align}
\begin{aligned}
&|i \rangle \;\rightarrow \;
|\hat{i} \rangle=|i \rangle+zc_i|r_i \rangle\quad \;
\mbox{for }\; h_i=+,
\\
&|i] \;\rightarrow \;
|\hat{i}]=|i]+zc_i|r_i]\quad \;\;\,
\mbox{for }\; h_i=-,
\end{aligned}
\label{eq:spinor_shift1_2}
\end{align}
for massless particles, where $h_i$ denotes the helicity of the $i$-th particle, and $c_i$ are constants required to satisfy momentum conservation. In the massless case, $|r_i\rangle$ and $|r_i]$ denote reference spinors associated with an arbitrary massless momentum that is not collinear with the original one. Additionally, we have
\begin{align}
\left.
\begin{aligned}
&|i^2\rangle \;\rightarrow \;
|\hat{i}^2\rangle=|i^2\rangle+z\frac{c_i}{2}|i^1\rangle 
\\
&[i^2| \;\rightarrow \;
[\hat{i}^2|=[i^2|-z\frac{c_i}{2}[i^1|
\end{aligned}
\right\}
\quad
\mbox{for }\; h_i=0,
\label{eq:spinor_shift2}
\end{align}
for massive longitudinal vectors and scalars and
\begin{align}
\begin{aligned}
|i\rangle \;\rightarrow \;
|\hat{i}\rangle=|i\rangle+z\frac{c_i}{2}|r_i\rangle 
\quad\mbox{and}\quad
[i| \;\rightarrow \;
[\hat{i}|=[i|-z\frac{c_i}{2}[r_i|
\end{aligned},
\label{eq:spinor_shift2}
\end{align}
for massless scalars.

Under these shifts, all external Dirac spinors and polarization vectors remain invariant, while the momentum of each external particle is shifted as  
\begin{align}
&p_i\to\hat{p}_i=p_i+zq_i,\quad 
\left\{
\begin{array}{ll}
q_i=c_i m_i \epsilon^\pm(p_i)&\mbox{(massive)}
\\
q_i=c_i (\ab{i r_i} \epsilon^+(p_i)\mbox{ or }\sqb{i r_i} \epsilon^-(p_i))&\mbox{(massless)}
\end{array}
\right., 
\label{eq:momentum_shift}
\end{align}
in terms of the associated momentum $q_i$, which is defined by the mass $m_i$ or the contraction, $\ab{i r_i}$ or $\sqb{i r_i}$, of the external particle together with the transverse polarization vectors $\epsilon^{\pm}(p_i)$. In $\epsilon^{\pm}(p_i)$, the sign matches the helicity for spin-1/2 and transverse vector particles, whereas it is taken as $+$ for longitudinal vectors and scalars\footnote{We choose a specific shifting convention; see another one leading to the opposite sign in Ref.~\cite{Ema:2024vww}.}. Here, $q_i$ satisfies
\begin{align}
\sum_i q_i=0 \quad \mbox{and}\quad  p_i\cdot q_i=q_i^2=0,
\end{align}
with the metric signature $(1,-1,-1,-1)$\footnote{Four-vector indices are omitted for simplicity.}, ensuring that the momenta $\hat{p}_i$ satisfy the momentum conservation and on-shell conditions.

Let us consider an $n$-pt tree-level amplitude $\mathcal{M}_n$. Under the ALT shift, it becomes a function $\hat{\mathcal{M}}_n(z)$ of $z$, indicating the physical amplitude $\mathcal{M}_n=\hat{\mathcal{M}}_n(0)$. Using Cauchy's theorem, we relate $\mathcal{M}_n$ to other nonanalytic and boundary parts:
\begin{align}
    \mathcal{M}_n=\frac{1}{2\pi i}\oint_{z=0}\frac{dz}{z}\hat{\mathcal{M}}_n(z)=-\sum_{z_I}{\rm Res}_{z=z_{I}}\frac{\hat{\mathcal{M}}_n(z)}{z}+B_{\infty}.
    \label{eq:contour_int}
\end{align}
The residues at $z\neq 0$ arise from propagators $1/(\hat{P}_I^2 - m_I^2)$, which are singular at $z = z_I$, where $\hat{P}_I^2 - m_I^2 = 0$. Here, $I$ denotes a subset of external legs defining a factorization channel, and $\hat{P}_I = \sum_{i\in I} \hat{p}_i$. Given the leading behavior $\hat{\mathcal{M}}_n(|z|\!\to\!\infty)\!\propto\! z^\tau$ with integer $\tau$, the boundary term $B_\infty$ vanishes for $\tau < 0$, but remains nonzero for $\tau \geq 0$, implying that $\mathcal{M}_n$ is on-shell constructible from lower-point amplitudes only when $\tau < 0$.

Since $u$ and $\eps^{\pm,0}$ remain invariant under the ALT shift 
in the large-$z$ limit, only $F$ in Eq.~\eqref{eq:amp_F_u_eps}\footnote{Here, we discuss the associated stripped amplitude $F$ to examine the on-shell constructibility of each amplitude of each diagram.} contributes to the leading behavior of $\hat{\mathcal{M}}_n(|z|\!\to\!\infty)$. Given the contribution reflecting the momentum dependence of $F$, we have the condition for on-shell constructibility:
\begin{align}
F\rightarrow z^{[F]}  \quad \rightarrow \quad [F]< 0.
\label{eq:F_on-shell_constructible}
\end{align}
In Eq.~\eqref{eq:amp_F_u_eps}, we get $[\mathcal{G}]=[g]+[F]$ and inserting this into Eq.~\eqref{eq:F_dim}, we have
\begin{align}
[F]=4-n-[g]-N_F/2,
\end{align}
with which Eq.~\eqref{eq:F_on-shell_constructible} leads to
\begin{align}
[g]>4-n-N_F/2.
\label{eq:raw_on-shell_coup}
\end{align}
To examine the implication of Eq.~\eqref{eq:raw_on-shell_coup}, we consider the simplest case $n=4$, leading to $[g]>-N_F/2$. Taking $N_F=0$ and $2$, one observes that 4-pt amplitudes arising from the dimension-3 bosonic and dimension-4 fermionic 3-pt operators are on-shell constructible. On the other hand, the $n$-pt amplitudes with  $n\geq 5$ can be constructed recursively even with higher-dimensional operators. From the next section, we discuss the interaction operators for particles of spin up to 1. For notational simplicity, we denote $\phi$ and $V$ as a scalar and a vector, and $\psi$ and $\bar{\psi}$ as a fermion and its conjugate in the following.

\subsection{On-shell constructible tree-unitary amplitudes}

\noindent
Collecting the interaction operators admitting a consistent on-shell construction, one finds that Eq.~\eqref{eq:raw_on-shell_coup} allows only non-derivative 3-pt operators of the types $VV\phi$, $\phi\phi\phi$, $V\bar{\psi}\psi$, and $\phi\bar{\psi}\psi$; any other operator already fails at the level of the 4-pt amplitude construction. The theory with such a set of operators, especially in the presence of massive vectors, is physically inconsistent, as the longitudinal components of massive vectors lead to amplitudes growing in the high-energy limit, in conflict with tree unitarity.

However, the theory space can be enlarged by requiring  smooth massless limits of amplitudes. In this case, the amputated on-shell Green's function $\mathcal{G}$ in Eq.~\eqref{eq:amp_F_u_eps} satisfies the Ward identity:
\begin{align}
p_i \cdot \mathcal{G}(p_1,\cdots,p_4)=0 \quad\mbox{with}\quad p_i^2=0, \;\;\;\sum_{i=1}^4 p_i=0,
\label{eq:ward}
\end{align}
for arbitrary complex momenta $p_i$. Since Eq.~\eqref{eq:momentum_shift} ensures
\begin{align}
\lim_{|z|\to \infty}\hat{p}_i \propto  q_i \;\; \mbox{with}\;\; q_i^2=0,
\end{align}
together with $\eps^{\pm}_i \propto q_i$, the shifted amplitude $\hat{\mathcal{M}}_n(z)$ with transverse vectors must satisfy
\footnote{In general, $[F]$ depends on the diagram, since each diagram can involve coupling collections with arbitrary mass dimensions.}
\begin{align}
\hat{\mathcal{M}}_n(|z|\!\rightarrow\!\infty)
\propto 
\left[\sum_{\rm diagrams}
z^{[F]}\times
g \times F(q_1,\cdots\!,q_n)
\right]\times \prod q \times \prod u = 0,
\end{align}
at the boundary $|z|\rightarrow \infty$. This result improves Eq.~\eqref{eq:F_on-shell_constructible} to $[F]<1$, thereby replacing Eq.~\eqref{eq:raw_on-shell_coup} with
\begin{align}
[g]> 4-n-N_F/2-c_T,
\label{eq:on-shell_coup}
\end{align}
where $c_T=1$ for $\mathcal{M}_n$ including at least one transverse vector and $c_T=0$ otherwise.

To identify on-shell constructible operators,we first note the following point. Although $c_T$ in Eq.~\eqref{eq:on-shell_coup} is defined for amplitudes with transverse vectors, in practice, the same bound can be applied to amplitudes with longitudinal vectors, since such amplitudes can be obtained by replacing transverse polarizations with longitudinal ones after the recursive construction. We therefore present the practical bound on $g$:
\begin{align}
[g]> 4-n-N_F/2-c_V,
\label{eq:on-shell_coup2}
\end{align}
where $c_V = 1$ whenever at least one vector is present, and $c_V = 0$ otherwise.

We start to examine Eq.~\eqref{eq:on-shell_coup2} in the case $n=4$, yielding $[g] > -N_F/2 - c_V$. Taking $N_F = 0, 2,$ and $4$ with $c_V = 0$, we find that the dimension-3 $\phi\phi\phi$, dimension-4 $V\phi\phi$, and dimension-4 $\bar\psi\psi(\phi \text{ and } V)$ operators are allowed whereas the $\phi\phi\phi\phi$ and $\bar\psi\psi\bar\psi\psi$ contact operators are forbidden.\footnote{The $\phi\phi\phi\phi$ and $\bar\psi\psi\bar\psi\psi$ contact operators require couplings of mass dimension $\leq 0$ and $\leq -2$, respectively, which are incompatible with the conditions $[g]>0$ and $[g]>-2$ for $n=4$.}
Applying the same procedure with $c_V = 1$, we find that the dimension-3 $VV\phi$, dimension-4 $VVV$, and dimension-4 contact $VV\phi\phi$ and $VVVV$ operators are allowed.

In the case $n \ge 5$, Eq.~\eqref{eq:on-shell_coup2} appears to allow
$n$-pt contact operators involving vectors. However, since each
such operator carries its own independent coupling, their amplitudes must individually satisfy the Ward identity in the massless limit. This forces vector fields to appear only through their field-strength tensors, which inevitably introduce derivatives and therefore fail to be on-shell constructible, e.g., $[(F_{\mu\nu}^a)^2\phi ]=5$. Thus, any ($n \ge 5$)-pt contact operators are forbidden even in on-shell constructible theories improved by admitting smooth massless limits.

The absence of the $\phi\phi\phi\phi$ contact operator in a set of operators exhibiting on-shell constructible amplitudes~\cite{Elvang:2015rqa,Arkani-Hamed:2017jhn} is a typical example differentiating the set of operators in tree unitary theories. However, by treating the amplitude for such a contact operator as a building block, one can still proceed with the recursive construction of higher-pt amplitudes from a practical perspective. In the same sense, even the amplitude $\phi\phi \to V \to \phi\phi$, which is not on-shell constructible due to its $z^0$ behavior at large $z$~\cite{Elvang:2015rqa,Arkani-Hamed:2017jhn}, can also be treated as a building block. 

However, in this work, in determining the coupling conditions for tree unitarity from amplitudes, it is not necessary to construct such non-constructible amplitudes. This is because, as seen in Eqs.~\eqref{eq:G_tree_unitarity}, amplitudes exhibiting $z^0$ behavior at large $z$ correspond to those exhibiting $E^0$ behavior at high $E$. Thus, they do not yield coupling conditions for tree unitarity. As will be discussed in detail, the conditions are derived by examining the high-$E$ scaling of amplitudes, enforcing relations among couplings to eliminate such high-$E$ contributions.

From the next section, we refer to couplings of mass dimension $\geq 0$ as tree-unitary couplings and operators of mass dimension $\leq 4$ as tree-unitary operators.\footnote{Although operators of mass dimension $\le 4$ are conventionally termed renormalizable, renormalizability concerns loop-level ultraviolet behavior. For concreteness, we base our classification solely on tree unitarity.}

\section{3-pt amplitudes and operators}
\label{sec:3-pt_amps_operators}

In Section~\ref{sec:tree_unitarity}, we showed that tree-unitary theories require only couplings of mass dimension $\geq 0$. This leads to a compact construction of 3-pt on-shell amplitudes and the associated 3-pt operators, as will be shown in detail. In this work, we work with self-conjugate bosonic fields without loss of generality, since any complex field can be decomposed into two real ones. For spin-1/2 fermions, we use a chiral Dirac representation, which can be restricted to a Majorana one by imposing charge-conjugation invariance.

Table~\ref{tab:3-pt} presents all tree-unitary 3-pt on-shell amplitudes and operators, together with the notational conventions used throughout this work \cite{Elvang:2015rqa,Arkani-Hamed:2017jhn}. To construct these amplitudes, using Eq.~\eqref{eq:3pt_massless_amps}, we first obtain the massless 3-pt amplitudes, which are completely fixed by little-group scaling (see the details in Appendix~\ref{appendix:3-pt_massless_on-shell}). Using Eqs.~\eqref{eq:massless_limits1} and~\eqref{eq:massless_limits2}, we then trace back the massive 3-pt amplitudes whose massless limits reproduce the original massless amplitudes. From these massive amplitudes, we derive the massive–massless mixed 3-pt amplitudes.
\begin{table}[htb!]
\vskip 0.1cm
\centering
\scalebox{0.85}{
\begin{tabular}{|c|c|}
\hline
&
\\[-8pt]
3-pt on-shell amplitude
& 3-pt operator
\\[6pt] 
\hline
&\multirow{6}{*}{
\makecell{
\vspace{-0.5cm}
$-C_{\capi{ABC}}\partial_{\nu}V_{\capi{A}\mu} V_{\capi{B}}^{\mu} V_{\capi{C}}^{\nu}$
}
}
\\[-10pt]
$(1^{+1},2^{+1},3^{-1})\mbox{: }
\sqrt{2} i C_{\hat{a}_1 \hat{a}_2 \hat{a}_3}\frac{[  12 ]^3}{[ 31 ][ 23 ]}$
,\;\; $(1^{-1},2^{-1},3^{+1})\mbox{: }
\sqrt{2} i C_{\hat{a}_1 \hat{a}_2 \hat{a}_3}\frac{\langle 12\rangle^3}{\langle 31 \rangle\langle 23\rangle }$
&
\\[6pt]
\cline{1-1}
&
\\[-10pt]
$\sqrt{2} i C_{a_1a_2a_3}
\Big(\frac{\ab{\bs{12}}\sqb{\bs{23}}\sqb{\bs{31}}}{m_1m_2}
+ \mbox{c.p.t.}\Big)$
&
\\[6pt]
\cline{1-1}
&
\\[-10pt]
\makecell{$(1^{+1},\bs{2,3}):$\;\;
$\sqrt{2}iC_{\hat{a}_1a_2a_3}\left(\frac{\ab{\bm{23}}[\bm{3}1][1\bm{2}]}{m_2m_3}+\frac{[1\bm{2}r_1\rangle\ab{\bm{23}}[\bm{32}]}{\ab{1r_1}m_2m_3}\right)$
\\[8pt]
$(1^{-1},\bs{2,3}):$\;\;
$\sqrt{2}iC_{\hat{a}_1a_2a_3}\left(\frac{[\bm{23}]\ab{\bm{3}1}\ab{1\bm{2}}}{m_2m_3}+\frac{\langle1\bm{2}r_1]\ab{\bm{23}}[\bm{32}]}{[1r_1]m_2m_3}\right)$
}
$(m_2=m_3)$
&
%
%
\\[19pt]
\hline
&
\\[-7pt]
$2F_{a_1 a_2 i_3} \frac{[\bs{1}\bs{2}]
\langle \bs{2}\bs{1} \rangle}{m_1m_2}$
&
$F_{\capi{AB}i}V_{\capi{A}\mu}V_{\capi{B}}^{\mu}\phi_i$
%
%
\\[9pt] 
\hline
&\multirow{7}{*}{
$-\bar{\psi}_R \slash\!\!\! V_{\capi{A}} R^{\capi{A}} \psi_R
-\bar{\psi}_L \slash\!\!\! V_{\capi{A}} L^{\capi{A}} \psi_L$}
\\[-10pt]
$(1^{+ 1},2^{-\frac12},3^{+\frac12}):
\sqrt{2} R^{\hat{a}_1}
\frac{[13]^2}{[32]},  \;\;
(1^{- 1},2^{-\frac12},3^{+\frac12}): \sqrt{2} R^{\hat{a}_1}
\frac{\langle 12 \rangle^2}{\langle 32\rangle}$
&
\\[5pt]
$(1^{+ 1},2^{+\frac12},3^{-\frac12}):
\sqrt{2} L^{\hat{a}_1}
\frac{[12]^2}{[23]},
\;\;
(1^{- 1},2^{+\frac12},3^{-\frac12}): \sqrt{2}L^{\hat{a}_1}
\frac{\langle 13 \rangle^2}{\langle 23\rangle}$
&
\\[5pt]
\cline{1-1}
&
\\[-10pt]
$\frac{\sqrt{2}}{m_1} 
(R^{a_1}\langle \bs{1}\bs{2}\rangle
[\bs{1}\bs{3}]
+
L^{a_1}[\bs{1}\bs{2}]
\langle \bs{1}\bs{3}\rangle )$
&
\\[6pt]
\cline{1-1}
&
\\[-10pt]
$- 
R^{\hat{a}_1}
(\langle \bs{2} \epsilon^\pm_1 \bs{3} ]
+\langle \bs{3} \epsilon^\pm_1 \bs{2} ] ) \quad (R=L, m_2=m_3)$ 
&
%
%
\\[6pt]
\hline
&
\multirow{5}{*}{\vspace{-9pt} $-G_{\capi{A}ij}V_{\capi{A}\mu}\partial^{\mu} \phi_i\phi_j$}
\\[-10pt]
$(1^{+ 1},2^0,3^0):
\sqrt{2}iG_{\hat{a}_1i_2 i_3} \frac{[12][13]}{[23]}
,\;\;
(1^{- 1},2^0,3^0): \sqrt{2}iG_{\hat{a}_1i_2i_3} \frac{\langle 12 \rangle \langle 31 \rangle}{\langle 23\rangle } 
$&
\\[6pt]
\cline{1-1}
&
\\[-10pt]
$\frac{iG_{a_1 i_2 i_3}}{\sqrt{2}m_1}  
\langle \bs{1}( \bs{2}- \bs{3}
)\bs{1}]$
&
\\[6pt]
\cline{1-1}
&
\\[-10pt]
$\frac{iG_{\hat{a}_1 i_2 i_3}}{2}  \;\epsilon^{\pm}_1 \cdot ( \bs{2}- \bs{3} ) \quad (m_2=m_3)$
&
%
%
\\[6pt]
\hline
&
\\[-10pt]
$H_{i_1}[\bs{2}\bs{3}]
+H_{i_1}^\dagger \langle \bs{2}\bs{3}\rangle$
&
\!\!\!\!
$-(\bar{\psi}_L H_i \phi_i \psi_R
+\mbox{h.c.})$
\!\!\!\!
%
%
\\[6pt]
\hline
&
\\[-10pt]
$-P_{i_1i_2i_3}$
&
$-\frac16 P_{ijk}\phi_i\phi_j\phi_k$
\\[6pt]
\hline
\end{tabular}
}
\caption{\it Tree-unitary 3-pt on-shell amplitudes and associated  operators involving scalar $\phi_i$, right (left)-chiral spin-1/2 fermion $\psi_R$ $(\psi_L)$, and vector $V_{\capi{A}}$ fields. The vector index $A$ collectively represents the massive  $a$ and massless $\hat{a}$ vector indices. The scalar indices are denoted by italic letters $i,j,k$. For simplicity, the spin-1/2 fermion indices, represented by Roman letters $\rm{i},\rm{j},\rm{k}$, are omitted in the table. For fermionic amplitudes, we adopt the convention $(1_{V/\phi},2_{\bar{\psi}},3_{\psi})$, leading to the couplings $R^{\capi{A_1}}_{\rm{i}_2 \rm{i}_3}$, $L^{\capi{A_1}}_{\rm{i}_2 \rm{i}_3}$, and $(H_{i_1})_{\rm{i}_2 \rm{i}_3}$ in the amplitudes. The dot symbol involving $\epsilon_1^{\pm}$ denotes the spinor contraction. For amplitudes involving massless particles, the helicities are indicated on the left of the amplitudes, respectively. For the amplitudes with one massless and two massive particles, their independence of reference spinors enforces $R=L$ (for the fermionic case) and $m_2=m_3$. The notations “c.p.t.” and ``h.c." denote cyclic permutations of the external particle states and Hermitian conjugation, respectively.}
\label{tab:3-pt}
\end{table}

\noindent
{\large $\ab{VVV}$}
\vspace{0.4cm}

\noindent
Recall that the mass dimension of spinor part in any 3-pt massless on-shell amplitude is solely determined by the sum of helicities (see Appendix~\ref{appendix:3-pt_massless_on-shell}). Since tree-unitary couplings must have mass dimension $\ge 0$, three massless vectors are restricted to carry only the helicity configurations $\{\pm,\pm,\mp\}$. Using Eq.~\eqref{eq:3pt_massless_amps}, we have
\begin{align}
\mathcal{M}_3(1^+_{\hat{a}_1} 2^+_{\hat{a}_2} 3^-_{\hat{a}_3})=\sqrt{2}iC_{\hat{a}_1\hat{a}_2\hat{a}_3}\frac{[12]^3}{[23][13]},\quad\mathcal{M}_3(1^-_{\hat{a}_1} 2^-_{\hat{a}_2} 3^+_{\hat{a}_3})=\sqrt{2}iC_{\hat{a}_1\hat{a}_2\hat{a}_3}\frac{\ab{12}^3}{\ab{23}\ab{13}}
,
\label{eq:3pt_massless_VVV_amps}
\end{align}
for the helicity orderings $(\pm\pm\mp)$ with a conventional overall factor $\sqrt{2}i$ and coupling $C_{\hat{a}_1\hat{a}_2\hat{a}_3}$. Other helicity orderings, such as $(\pm\mp\pm)$ or $(\mp\pm\pm)$, are obtained by exchanging $i_{\hat{a}_i}\leftrightarrow j_{\hat{a}_j}$ for two external legs with opposite helicities in the amplitudes. Then, one finds that $C_{\hat{a}_1 \hat{a}_2 \hat{a}_3}$ must be antisymmetric so that every helicity ordering is invariant under the exchange $i_{a_i}\leftrightarrow j_{a_j}$ for the same helicity, as required by Bose statistics.

Next, we trace back the massive $VVV$ amplitude from the massless ones in Eq.~\eqref{eq:3pt_massless_VVV_amps}. Using Eqs.~\eqref{eq:massless_limits1} and~\eqref{eq:massless_limits2}, we have
\begin{align}
\begin{aligned}
    \mathcal{M}_3(1^+_{\hat{a}_1}2^+_{\hat{a}_2}3^-_{\hat{a}_3})=\sqrt{2}iC_{\hat{a}_1\hat{a}_2\hat{a}_3}\frac{[12]^3}{[23][13]}\quad &\longrightarrow\quad \sqrt{2}iC_{a_1a_2a_3}\frac{[\bm{12]}\ab{\bm{23}}\ab{\bm{31}}}{m_1m_2},\\
    \mathcal{M}_3(1^+_{\hat{a}_1}2^-_{\hat{a}_2}3^+_{\hat{a}_3})=\sqrt{2}iC_{\hat{a}_3\hat{a}_1\hat{a}_2}\frac{[13]^3}{[12][23]}\quad &\longrightarrow\quad \sqrt{2}iC_{a_1a_2a_3}\frac{[\bm{31]}\ab{\bm{12}}\ab{\bm{23}}}{m_1m_3},\\
    \mathcal{M}_3(1^-_{\hat{a}_1}2^+_{\hat{a}_2}3^+_{\hat{a}_3})=\sqrt{2}iC_{\hat{a}_2\hat{a}_3\hat{a}_1}\frac{[23]^3}{[12][13]}\quad &\longrightarrow\quad \sqrt{2}iC_{a_1a_2a_3}\frac{[\bm{23]}\ab{\bm{31}}\ab{\bm{12}}}{m_2m_3},
    \end{aligned}
    \label{eq:2a1s}
\end{align}
and
\begin{align}
\begin{aligned}
    \mathcal{M}_3(1^-_{\hat{a}_1}2^-_{\hat{a}_2}3^+_{\hat{a}_3})=\sqrt{2}iC_{\hat{a}_1\hat{a}_2\hat{a}_3}\frac{\ab{12}^3}{\ab{23}\ab{13}}\quad &\longrightarrow \quad \sqrt{2}iC_{a_1a_2a_3}\frac{\ab{\bm{12}}[\bm{23}][\bm{31}]}{m_1m_2},\\
    \mathcal{M}_3(1^-_{\hat{a}_1}2^+_{\hat{a}_2}3^-_{\hat{a}_3})=\sqrt{2}iC_{\hat{a}_3\hat{a}_1\hat{a}_2}
    \frac{\ab{13}^3}{\ab{12}\ab{23}}\quad &\longrightarrow\quad \sqrt{2}iC_{a_1a_2a_3}\frac{\ab{\bm{31}}[\bm{12}][\bm{23}]}{m_1m_3},\\
    \mathcal{M}_3(1^+_{\hat{a}_1}2^-_{\hat{a}_2}3^-_{\hat{a}_3})=\sqrt{2}iC_{\hat{a}_2\hat{a}_3\hat{a}_1}\frac{\ab{23}^3}{\ab{12}\ab{13}}\quad &\longrightarrow\quad \sqrt{2}iC_{a_1a_2a_3}\frac{\ab{\bm{23}}[\bm{31}][\bm{12}]}{m_1m_2}.
    \end{aligned}
    \label{eq:1a2s}
\end{align}
These results lead us to construct the massive $VVV$ amplitude through their combination. Using an identity in Eq.~\eqref{eq:m1m2m3_identity}, we simplify the structure 
as (see Refs.~\cite{Durieux:2019eor,Durieux:2020gip,Liu:2022alx})
\begin{align}
\mathcal{M}_3(\bs{1}_{a_1} \bs{2}_{a_2} \bs{3}_{a_3})=&
\sqrt{2} iC_{ a_1a_2 a_3} 
\bigg(
\frac{\ab{\bs{12}}\sqb{\bs{23}}\sqb{\bs{31}}}{m_1m_2}
+\rm{c.p.t}
\bigg),
\label{eq:VVV_amp}
\end{align}
only with the three structures in Eq.~\eqref{eq:1a2s}, where the notation “c.p.t.” denotes cyclic permutations of the external particle states. For concreteness, let us confirm the massless limit of Eq.~\eqref{eq:VVV_amp} for the helicity ordering $(++-)$ in Eq.~\eqref{eq:2a1s} as
\begin{align}\begin{split}
\mathcal{M}_3(\bs{1}^+_{a_1}\bs{2}^+_{a_2}\bs{3}^-_{a_3})
&\stackrel{m\to 0}{\longrightarrow}\sqrt{2}iC_{a_1a_2a_3}\frac{\ab{\eta_23}\ab{3\eta_1}\sqb{12}}{m_1m_2}=\sqrt{2}iC_{a_1a_2a_3}\frac{[12]^3}{[23][13]}\ ,
\label{eq:VVV_massless_limit}
\end{split}\end{align}
reproducing the massless amplitude. In this computation, we have neglected the subleading first and third terms (see Eq.~\eqref{eq:leading_term_massless_limit}). 

In the same manner as in Eq.~\eqref{eq:VVV_massless_limit}, one can find that Eq.~\eqref{eq:VVV_amp} vanishes in the massless limit for the other helicity configurations $\{\pm,\pm,\pm\}$, $\{0,\pm,\pm\}$ and $\{0,\pm,\mp\}$. Here, the case of $\{0,\pm,\mp\}$ is subtle to show its vanishment, so we directly present the massless limit for the ordering $(0-+)$ as an example:
\begin{align}
\mathcal{M}_3(\bs{1}^0_{a_1}\bs{2}^-_{a_2}\bs{3}^+_{a_3})
\;\stackrel{m\to 0}{\longrightarrow}\;
iC_{a_1a_2a_3}
\frac{\ab{12}\sqb{\eta_23}\sqb{31}}{\sqrt{2}m_1m_2} \sim\,
iC_{a_1a_2a_3}
\frac{\ab{12}^2}{\ab{31}^2}
\frac{(m_2^2-m_3^2-m_1^2)}{\sqrt{2} m_1}
=0,
\end{align}
vanishing when all masses diminish at the same rate, where we used Eqs.~\eqref{eq:massless_limits1} and~\eqref{eq:massless_limits2}, and Eq.~\eqref{eq:[]_<>_suppress}. On the other hand, the case of $\{0,0,\pm\}$ yields a smooth massless limit~\cite{Christensen:2018zcq,Durieux:2019eor,Liu:2022alx}. For the ordering $(00+)$, we get
\begin{align}
\mathcal{M}_3(\bs{1}^0_{a_1}\bs{2}^0_{a_2}\bs{3}^+_{a_3})
\;\stackrel{m\to 0}{\longrightarrow}\;iC_{a_1a_2a_3}\frac{(m_1^2+m_2^2-m_3^2)}{\sqrt{2}m_1m_2}\frac{[23][31]}{[12]},
\label{eq:}
\end{align}
remaining finite when all masses diminish at the same rate.

In the massive $VVV$ amplitude in Eq.~\eqref{eq:VVV_amp}, taking the massless limit of a transverse vector, we easily derive the one-massless and two-massive $VVV$ amplitude. For the helicity $+1$, using Eq.~\eqref{eq:spinor_dirac_eq} and Eq.~\eqref{eq:massive_schouten}, we get
\begin{align}\begin{split}
\mathcal{M}_3(\bs{1}^{+}_{a_1}\bm{2}_{a_2}\bm{3}_{a_3})&\;\stackrel{m\to 0}{\longrightarrow}\;
\sqrt{2}iC_{a_1a_2a_3}\left(\frac{\ab{\bm{23}}[\bm{3}1][1\bm{2}]}{m_2m_3}+\frac{[1\bm{2}\eta_1\rangle\ab{\bm{23}}[\bm{32}]}{\ab{1\eta_1}m_2m_3}\right)\ ,
\label{eq:1massless_2massive_VVV_0}
\end{split}\end{align}
with the notation in Eq.~\eqref{eq:two_bracket_notations}
where $m_1=\ab{1\eta_1}$. Since any physical amplitude must be independent of the choice of reference spinor, the second term in Eq.~\eqref{eq:1massless_2massive_VVV_0} must be independent of $|\eta_1\rangle$. This can be verified by rewriting the second term as
\begin{align}
\frac{[1\bm{2}\eta_1\rangle\ab{1x}\ab{\bm{23}}[\bm{32}]}{\ab{1\eta_1}\ab{1x}m_2m_3}=\frac{[1\bm{2}r\rangle\ab{\bm{23}}[\bm{32}]}{\ab{1r}m_2m_3}+ \frac{(m_3^2-m_2^2)\ab{\eta_1r}\ab{\bm{23}}[\bm{32}]}{\ab{1\eta_1}\ab{1r}m_2m_3}\ ,
\label{eq:1massless_2massive_VVV_1}
\end{align}
with an arbitrary spinor $|r\rangle$, where we used Eq.~\eqref{eq:spinor_dirac_eq} and Eq.~\eqref{eq:massive_schouten}. Eq.~\eqref{eq:1massless_2massive_VVV_1} shows that the condition $m_2=m_3$ is required for the one-massless and two-massive $VVV$ amplitude. Thus, applying the same argument to the case of the helicity $-1$, we get
\begin{align}
\begin{aligned}
\mathcal{M}_3(1^{+}_{\hat{a}_1}\bm{2}_{a_2}\bm{3}_{a_3})
&=
\sqrt{2}iC_{\hat{a}_1a_2a_3}\left(\frac{\ab{\bm{23}}[\bm{3}1][1\bm{2}]}{m_2m_3}+\frac{[1\bm{2}r_1\rangle\ab{\bm{23}}[\bm{32}]}{\ab{1 r_1}m_2m_3}\right),
\\
\mathcal{M}_3(1^{-}_{\hat{a}_1}\bm{2}_{a_2}\bm{3}_{a_3})
&=\sqrt{2}iC_{\hat{a}_1a_2a_3}\left(\frac{[\bm{23}]\ab{\bm{3}1}\ab{1\bm{2}}}{m_2m_3}+\frac{\langle1\bm{2}r_1]\ab{\bm{23}}[\bm{32}]}{[1r_1]m_2m_3}\right),
\end{aligned}
\label{eq:1massless_2massive_VVV}
\end{align}
with $m_2=m_3$. The condition $m_2=m_3$ also ensures the absence of the two-massless and one-massive $VVV$ amplitude, leading to $C_{\hat{a}\hat{b}c}=0$. Using Eq.~\eqref{eq:massive_polarization}, we trace back the operator yielding all $VVV$ amplitudes derived so far
~\cite{Durieux:2019eor,Li:2022tec}:
\begin{align}
-C_{\capi{ABC}}\partial_{\nu}V_{\capi{A}\mu} V_{\capi{B}}^{\mu} V_{\capi{C}}^{\nu},
\label{eq:VVV_op}
\end{align}
with the capital indices $\{A\}$ collecting both the massive $\{a\}$ and massless $\{\hat{a}\}$ ones. We note that tree-level Hermiticity of the $T$ matrix $(S=1+iT)$ ensures all couplings appearing in this work are real (Hermitian) for bosonic (fermionic) couplings (see Appendix~\ref{appendix:CPT}); hence their reality and Hermiticity will not be mentioned hereafter.

\vspace{0.4cm}
\noindent
{\large $\ab{VV\phi}$}
\vspace{0.4cm}

\noindent
For the massless $VV\phi$ amplitude, the bound on tree-unitary couplings requires only the helicity configurations $\{\pm , \mp,0\}$. Using Eq.~\eqref{eq:3pt_massless_amps}, we obtain
\begin{align}
\mathcal{M}_3(1_{\hat{a}_1}^+ 2_{\hat{a}_2}^- 3_{i_3})
=\frac{\ab{23}^2}{\ab{31}^2}\mbox{ or }\frac{\sqb{31}^2}{\sqb{12}^2}
,\quad 
\mathcal{M}_3(1_{\hat{a}_1}^- 2_{\hat{a}_2}^+ 3_{i_3})
=\frac{\ab{31}^2}{\ab{23}^2}\mbox{ or }\frac{\sqb{23}^2}{\sqb{31}^2},
\label{eq:VVphi_massless}
\end{align}
However, such amplitudes cannot arise from any local $VV\phi$ operator, implying the necessity of non-local description~\cite{Conde:2016izb}. Thus, only in the case of $VV\phi$, we first present the tree-unitary local massive $VV\phi$ operator as
\begin{align}
F_{abi}V_{a\mu}V_{b}^{\mu}\phi_i.
\label{eq:VVphi_op}
\end{align}
which is the unique structure satisfying the tree-unitarity condition discussed in Section~\ref{sec:tree_unitarity}. Bose statistics requires the dimension-1 coupling $F_{abi}$ to be symmetric under the exchange $a\leftrightarrow b$. In Eq.~\eqref{eq:VVphi_op}, using Eq.~\eqref{eq:eps_massive_spinor}, we derive the massive $VV\phi$ amplitude as
\begin{align}
\mathcal{M}_3(\bs{1}_{a_1}\bs{2}_{a_2}\bs{3}_{i_3})=
2F_{ a_1a_2 i_3}\frac{\ab{\bs{12}}\sqb{\bs{21}}}{m_1 m_2 }. 
\label{eq:VVS_amp}
\end{align}
To classify Eq.~\eqref{eq:VVS_amp} as a tree-unitary amplitude, we need to verify whether Eq.~\eqref{eq:VVS_amp} can have a smooth massless limit. Using Eqs.~\eqref{eq:massless_limits1} and~\eqref{eq:massless_limits2}, and Eq.~\eqref{eq:[]_<>_suppress}, we derive the massless limit for the ordering $(-+0)$ as 
\begin{align}
\mathcal{M}_3(\bs{1}_{a_1}^-\bs{2}_{a_2}^+\bs{3}_{i_3})&\;\stackrel{m\to 0}{\longrightarrow}\;
2F_{ a_1a_2 i_3}
\bigg\{\frac{[23]^2}{[31]^2}
\bigg(\frac{m_2^2-m_3^2-m_1^2}{m_1^2-m_2^2-m_3^2}
\bigg)
\;\mbox{or}\;
\frac{\ab{31}^2}{\ab{23}^2}
\bigg(\frac{m_1^2-m_2^2-m_3^2}{m_2^2-m_3^2-m_1^2}\bigg)
\bigg\},
\end{align}
requiring a non-local description as in Eq.~\eqref{eq:VVphi_massless}. However, by imposing $F_{abi} \rightarrow 0$ in the massless limit, we can define a smooth massless limit of Eq.~\eqref{eq:VVS_amp}, valid even for the configurations $\{\pm,\pm,0\}$.

In the case that one of two vectors is massless, it is not possible to construct an amplitude independent of the reference spinor.
Let us consider the case $m_1=0$ in Eq.~\eqref{eq:VVS_amp}. Then, as in Eq.~\eqref{eq:1massless_2massive_VVV_1}, with an arbitrary spinor $|r'\rangle$, we compute
\begin{align}
    \frac{\ab{r_1\bm{2}}[\bm{2}1]}{\ab{1r_1} m_2}=\frac{\ab{r_1\bm{2}}\ab{1r'}[\bm{2}1]}{\ab{1 r_1}\ab{1r'}m_2}=\frac{\ab{r'\bm{2}}[\bm{2}1]}{\ab{1r'}m_2}+\frac{\ab{r_1r'}\ab{1\bm{2}}[\bm{2}1]}{\ab{1r_1}\ab{1r'}m_2},
\end{align}
which leads to an explicit dependence on the reference spinor and hence results in an unphysical amplitude. Thus, defining $F_{\hat{a}\hat{b}i}=F_{\hat{a}bi}=0$ as discussed above, we present the tree-unitary $VV\phi$ operator including both massive and massless vectors as
\begin{align}
F_{\capi{AB}i}V_{\capi{A}\mu}V_{\capi{B}}^{\mu}\phi_i,    
\end{align}
with the capital indices of vectors.

\vspace{0.4cm}
\noindent
{\large $\ab{V\bar{\psi}\psi}$}
\vspace{0.4cm}

Using Eq.~\eqref{eq:3pt_massless_amps}, we present the massless $\ab{V\bar{\psi}\psi}$ amplitudes as
\begin{align}
\mathcal{M}_3(1^{-1}_{\hat{a}_1} \bar{2}^{-\frac12}_{\rm{i}_2} 
3^{+\frac12}_{\rm{i}_3})
=-\sqrt{2}R^{\hat{a}_1}_{\rm{i}_2\rm{i}_3}\frac{\ab{12}^2}{\ab{23}}
,\quad 
\mathcal{M}_3(1^{-1}_{\hat{a}_1} \bar{2}^{+\frac12}_{\rm{i}_2} 
3^{-\frac12}_{\rm{i}_3}) 
=-\sqrt{2}L^{\hat{a}_1}_{\rm{i}_2\rm{i}_3}\frac{\sqb{12}^2}{\sqb{23}},
\label{eq:VFF_massless_amp}
\end{align}
for the orderings $(-\mp \pm)$, where $R^{\hat{a}_1}_{\rm{i}_2\rm{i}_3}$ and $L^{\hat{a}_1}_{\rm{i}_2\rm{i}_3}$ are dimensionless right- and left-chiral couplings, respectively. As discussed in Appendix~\ref{appendix:CPT}, these couplings are Hermitian, i.e., $R^{\hat{a}_1*}_{\rm{i}_3\rm{i}_2}=R^{\hat{a}_1}_{\rm{i}_2\rm{i}_3}$ and $L^{\hat{a}_1*}_{\rm{i}_3\rm{i}_2}=L^{\hat{a}_1}_{\rm{i}_2\rm{i}_3}$. Next, using Eqs.~\eqref{eq:massless_limits1} and~\eqref{eq:massless_limits2}, we trace back the massive $V\bar{\psi}\psi$ amplitude:
\begin{align}
\mathcal{M}_{3}(\bs{1}_{a_1}\bar{\bs{2}}_{\rm{i}_2}\bs{3}_{\rm{i}_3})=-
\frac{\sqrt{2}}{m_1} 
(R^{a_1}_{\rm{i}_2\rm{i}_3}\langle \bs{1}\bs{2}\rangle
[\bs{3}\bs{1}]
+
L^{a_1}_{\rm{i}_2\rm{i}_3}[\bs{1}\bs{2}]
\langle \bs{3}\bs{1}\rangle )\ ,
\label{eq:VFF_amp}
\end{align}
reproducing the massless ones for the orderings $(+\pm\mp)$ and $(-\pm\mp)$ in the massless limit. 

Using Eqs.~\eqref{eq:massless_limits1} and~\eqref{eq:massless_limits2}, and Eq.~\eqref{eq:[]_<>_suppress}, we verify that Eq.~\eqref{eq:VFF_amp} has a smooth massless limit for the vector mass, even in the configurations $\{0,\pm,\mp\}$ and $\{0,\pm,\pm\}$ as
\begin{align}
&\mathcal{M}_{3}(\bs{1}_{a_1}^0\bar{\bs{2}}_{\rm{i}_2}^{-\frac12}\bs{3}_{\rm{i}_3}^{+\frac12})
\stackrel{m_1\to 0}{\longrightarrow}
-R^{a_1}_{\rm{i}_2\rm{i}_3}
\frac{\langle 12\rangle[31]}{\sqrt{2}m_1}
\sim 
-R^{a_1}_{\rm{i}_2\rm{i}_3}
\frac{\ab{12}}{\ab{31}}
\frac{m_2^2-m_3^2-m_1^2}{\sqrt{2}m_1}
=0,
\nonumber
\\
&\mathcal{M}_{3}(\bs{1}_{a_1}^0\bar{\bs{2}}_{\rm{i}_2}^{+\frac12}
\bs{3}_{\rm{i}_3}^{+\frac12})
\stackrel{m_1\to 0}{\longrightarrow}
-\frac{R^{a_1}_{\rm{i}_2\rm{i}_3}m_2-L^{a_1}_{\rm{i}_2\rm{i}_3}m_3}{\sqrt{2} m_1}[23]\ ,
\label{eq:VFF_limit}
\end{align}
which remain finite when $m_2^2-m_3^2$ and $Rm_2-Lm_3$ diminish at the same rate as $m_1^2$ and $m_1$, respectively. In the same manner, one can find the massless limits for the orderings $(0+-)$ and $(0--)$, similar to Eq.~\eqref{eq:VFF_limit}.
Next, we derive the $V\bar{\psi}\psi$ amplitude with a massless vector. Taking the limit $m_1\rightarrow 0$ in Eq.~\eqref{eq:VFF_amp}, we get
\begin{align}
\mathcal{M}_{3}(\bs{1}_{a_1}^+\bar{\bs{2}}_{\rm{i}_2}\bs{3}_{\rm{i}_3})\stackrel{m_1\to 0}{\longrightarrow}
\frac{\sqrt{2}}{\ab{1\eta_1}} 
(R^{a_1}_{\rm{i}_2\rm{i}_3}\langle \eta_1\bs{2}\rangle
[1\bs{3}]
+
L^{a_1}_{\rm{i}_2\rm{i}_3}[1\bs{2}]
\langle \eta_1\bs{3}\rangle ),
\label{eq:VFF_massless_limit}
\end{align}
for the helicity $+1$ of the massless vector. To find the condition of Eq.~\eqref{eq:VFF_massless_limit} of being the physical amplitude, using an arbitrary spinor $|r\rangle$, we compute
\begin{align}
&\frac{\sqrt{2}}{\ab{1\eta_1}} 
(R^{a_1}_{\rm{i}_2\rm{i}_3}\langle \eta_1\bs{2}\rangle
[1\bs{3}]
+
L^{a_1}_{\rm{i}_2\rm{i}_3}[1\bs{2}]
\langle \eta_1\bs{3}\rangle )
\nonumber
\\
&=\frac{\sqrt{2}}{\ab{1r}} (R^{\hat{a}_1}_{\rm{i}_2\rm{i}_3}\langle r\bs{2}\rangle[1\bs{3}]+L^{\hat{a}_1}_{\rm{i}_2\rm{i}_3}[1\bs{2}]\langle r\bs{3}\rangle )
\nonumber
\\
&+\sqrt{2}\ab{\eta_1r}\frac{L^{\hat{a}_1}_{\rm{i}_2\rm{i}_3}(m_2\ab{\bm{23}}+m_3[\bm{23}])-R^{\hat{a}_1}_{\rm{i}_2\rm{i}_3}(m_2[\bm{23}]+m_3\ab{\bm{23}})}{\ab{1\eta_1}\ab{1r}}\ ,
\label{eq:VFF_massless_amp}
\end{align}
which is independent of $|\eta_1\rangle$ and $|r\rangle$ when $R=L$ and $m_2=m_3$. It leads us to obtain the amplitude with a massless vector as
\begin{align}\begin{split}
    \mathcal{M}_{3}(1_{\hat{a}_1}^\pm \bar{\bs{2}}_{\rm{i}_2}\bs{3}_{\rm{i}_3})&=-R^{\hat{a}_1}_{\rm{i}_2\rm{i}_3}(\langle\bm{2}|\epsilon^{\pm}_1|\bm{3}]+[\bm{2}|\epsilon^{\pm}_1|\bm{3}\rangle)\quad\mbox{with}\quad m_2=m_3,
    \label{eq:VFF_V_massless}
\end{split}\end{align}
for the helicities $\pm 1$, where the explicit form of polarization vector $\eps^{\pm}_1$ is present in Eq.~\eqref{eq:eps_massless_spinor}. It is straightforward to verify that  Eq.~\eqref{eq:VFF_V_massless} also has smooth massless limits, leading to Eq.~\eqref{eq:VFF_massless_amp}.

From all tree-unitary $V\bar{\psi}\psi$ amplitudes discussed above, 
using Eq.~\eqref{eq:uv_to_spinors} and Eqs.~\eqref{eq:eps_massless_spinor} and~\eqref{eq:eps_massive_spinor}, we trace back the tree-unitary $V\bar{\psi}\psi$ operator as
\begin{align}
    -\bar{\psi}_R \slash\!\!\! V_{\capi{A}} R^{\capi{A}} \psi_R
-\bar{\psi}_L \slash\!\!\! V_{\capi{A}} L^{\capi{A}} \psi_L, 
\label{eq:VFF_op}
\end{align}
with the capital index $A$ for a vector, where 
$\psi_{R}$ ($\psi_L$) is the right (left)-chiral spin-1/2 fermion field and
we have omitted the fermion indices for simplicity.

\vspace{0.4cm}
\noindent
{\large $\ab{V\phi\phi}$}
\vspace{0.4cm}

Using Eq.~\eqref{eq:3pt_massless_amps}, we present the massless $V\phi\phi$ amplitude for the ordering $(+00)$ as
\begin{align}
\mathcal{M}_3(1^{+ 1}_{\hat{a}_1} 2^0_{i_2} 3^0_{i_3})=
\sqrt{2}iG_{a_1i_2i_3} \frac{\sqb{12}\sqb{31}}{\sqb{23}},
\label{eq:VSS_V_massless_amp}
\end{align}
in terms of the overall factor $\sqrt{2}i$ where the dimensionless coupling $G_{\hat{a}_1 i_2 i_3}$ is antisymmetric under the exchange $2_{i_2}\leftrightarrow 3_{i_3}$, ensured by Bose statistics. Using Eqs.~\eqref{eq:massless_limits1} and~\eqref{eq:massless_limits2}, we trace back the $V\phi\phi$ amplitude as
\begin{align}
\mathcal{M}_3(\bs{1}_{a_1} \bs{2}^0_{i_2} \bs{3}^0_{i_3})
=
\frac{iG_{a_1 i_2 i_3}}{\sqrt{2}m_1}  
\langle \bs{1}( \bs{2}- \bs{3}
)\bs{1}]\ .
\label{eq:VSS_amp}
\end{align}
satisfying Bose statistics under the exchange $\bs{2}_{i_2}\leftrightarrow \bs{3}_{i_3}$. To verify tree unitarity of Eq.~\eqref{eq:VSS_amp}, we take its massless limit for the vector mass in the ordering $(000)$:
\begin{align}
\mathcal{M}_3(\bs{1}_{a_1}^{0} \bs{2}^0_{i_2} \bs{3}^0_{i_3})
\;\stackrel{m_1\rightarrow 0}{\longrightarrow}\; iG_{a_1 i_2 i_3}\frac{m_3^2-m_2^2}{\sqrt{2}m_1}=0,  
\end{align}
which remains finite when $m_1^2$ and $m_3^2-m_2^2$ diminish at the same rate.

To derive the $V\phi\phi$ amplitude with a massless vector, using an arbitrary spinor $|r\rangle$, we compute
\begin{align}
\frac{iG_{a_1 i_2 i_3}\ab{1r}}{\sqrt{2}\ab{1\eta_1}\ab{1r}}  \langle \eta_1( \bm{2}- \bm{3})1]
\;\stackrel{m_1\to 0}{\longrightarrow}\;
\frac{iG_{a_1 i_2 i_3}}{\sqrt{2}\ab{1r}}  \langle r( \bm{2}- \bm{3})1]+\frac{iG_{a_1 i_2 i_3}}{\sqrt{2}} \frac{\ab{\eta_1r} \langle 1( \bm{2}- \bm{3})1]}{\ab{1\eta_1}\ab{1r}}\ ,
\label{eq:VSS_massless_limit}
\end{align}
for the ordering $(+00)$ in Eq.~\eqref{eq:VSS_amp}. Since  $\asb{1\bs{2}1}=m_3^2-m_2^2$ and $\asb{1\bs{3}1}=m_2^2-m_3^2$ (see Eq.~\eqref{eq:p_square}), Eq.~\eqref{eq:VSS_massless_limit} becomes independent of $|\eta_1\rangle$ when $m_2=m_3$, leading to a physical amplitude. Thus, using the polarization vector $\eps^\pm_1$, we obtain the $V\phi\phi$ amplitude with a massless vector as
\begin{align}
    \mathcal{M}_3 (1^{\pm}_{\hat{a}_1} \bm{2}_{i_2} \bm{3}_{i_3})=\frac{iG_{\hat{a}_1 i_2 i_3}}{2}\epsilon^{\pm}_1\cdot(\bm{2}-\bm{3}) \quad\mbox{with}\quad m_2=m_3,
    \label{eq:Vphiphi_V_massless_amp}
\end{align}
for both helicities $\pm 1$, where the symbol ``$\cdot$" denotes the spinor contraction. As in the case of $V\bar{\psi}\psi$, one can easily observe that Eq.~\eqref{eq:Vphiphi_V_massless_amp} has smooth massless limits, reproducing Eq.~\eqref{eq:VSS_V_massless_amp}.

From Eqs.~\eqref{eq:VSS_amp} and~\eqref{eq:Vphiphi_V_massless_amp}, using Eq.~\eqref{eq:uv_to_spinors}, we trace back the tree-unitary $V\phi\phi$ operators as
\begin{align}
    G_{Aij}V_{A}^{\mu}(\partial_{\mu}\phi_{i})\phi_{j},
\end{align}
with the captial index $A$ for a vector.

\vspace{0.4cm}
\noindent
{\large $\ab{\phi\bar{\psi}\psi}$}
\vspace{0.4cm}

Using Eq.~\eqref{eq:3pt_massless_amps}, we obtain the massless $\phi\bar{\psi}\psi$ amplitudes as
\begin{align}
\mathcal{M}_3(1^0_{i_1}2^{+\frac12}_{{\rm i}_2}3^{+\frac12}_{{\rm i}_3})=(H_{i_1})_{{\rm i}_2{\rm i}_3}[23]
,\quad
\mathcal{M}_3(1^0_{i_1}2^{-\frac12}_{{\rm i}_2}3^{-\frac12}_{{\rm i}_3})=(H_{i_3}^{\dagger})_{{\rm i}_2{\rm i}_3}\ab{23},
\label{eq:SFF_massless_amp}
\end{align}
for the orderings $(0\pm\pm)$, with the dimensionless Hermitian coupling $(H_{i})_{\rm{ij}}$ satisfying $(H_{i})^*_{\rm{ji}}=(H_{i})_{\rm{ij}}$. Here, we do not consider the orderings $(0\pm\mp)$ since they cannot arise from any local $\phi\bar{\psi}\psi$ operator, requiring a non-local description~\cite{Conde:2016izb}.

From Eq.~\eqref{eq:SFF_massless_amp}, using Eqs.~\eqref{eq:massless_limits1} and~\eqref{eq:massless_limits2}, we obtain the massive $\phi\bar{\psi}\psi$ amplitude as
\begin{align}
\mathcal{M}_3 (\bs{1}_{i_1} \bm{2}_{\rm{i}_2} \bm{3}_{\rm{i}_3})
=(H_{i_3})_{{\rm i}_1{\rm i}_2}[\bs{12}]
+(H_{i_3}^{\dagger})_{{\rm i}_1{\rm i}_2}\ab{\bs{12}},
\end{align}
Using Eq.~\eqref{eq:uv_to_spinors}, we trace back the tree-unitary $\phi\bar{\psi}\psi$ operator:
\begin{align}
-\bar\psi_LH_{i}\phi_{i}\psi_R-\text{h.c.},
\end{align}
where ``h.c." denotes the Hermitian conjugate and we omitted the fermion indices for simplicity.

\vspace{0.4cm}
\noindent
{\large $\ab{\phi\phi\phi}$}
\vspace{0.4cm}

Eq.~\eqref{eq:3pt_massless_amps} restricts the tree-unitary $\phi\phi\phi$ amplitude to involve only a constant coupling regardless of the scalar masses. Thus, we directly cast the massive $\phi\phi\phi$ amplitude as
\begin{align}
\mathcal{M}_3(\bs{1}_{i_1}\bs{2}_{i_2}\bs{3}_{i_3})=-P_{i_1 i_2 i_3},
\label{eq:SSS_massive_amp}
\end{align}
where the dimension-1 coupling $P_{ijk}$ is symmetric under any exchange of scalar indices. The tree-unitary $\phi\phi\phi$ operator is
\begin{align}
-\frac{1}{6}P_{ijk}\phi_i\phi_j\phi_k,
\end{align}
yielding Eq.~\eqref{eq:SSS_massive_amp}.

In the massless limit, Eq.~\eqref{eq:SSS_massive_amp} takes the same form, but can have different values. Note that we have not discussed the changes of dimensionless couplings in the massless limit. This is because, to realize such changes, one must introduce additional dimensionful parameters independent of those already present in the theory. For example, a dimensionless coupling $c$ may be defined as
\begin{align}
c=\hat{c}+\frac{m}{\Lam},
\label{eq:dimensionless_c_massless}
\end{align}
with the mass $m$ of a massive vector in the theory and an independent dimensionful parameter $\Lambda$. Unlike the dimensionful case, this definition necessarily introduces a new independent scale, so that the independent parameter content differs between the massive theory and its massless limit. It therefore does not represent a continuous limit within a fixed parameter structure. For this reason, we restrict our discussion to the dimensionless couplings leading to the same values even in the massless limit.

\section{Tree unitarity in 4-pt amplitudes}
\label{sec:4-pt_tree_unitarity}
In this section, we recursively construct the 4-pt on-shell amplitudes from the 3-pt on-shell amplitudes based on the ALT shift \cite{Elvang:2015rqa,Ema:2024vww,Ema:2024rss}, discussed in Section~\ref{sec:3-pt_amps_operators}. Afterward, we investigate the high-$E$ scaling of 4-pt amplitudes, and find the conditions on couplings which lead to the elimination of divergent terms violating tree unitarity. For practical applications in future works, we present in Appendix~\ref{appendix:amp_no_high-E} the 4-pt amplitudes that do not lead to any high-$E$ growth.

\subsection{$VVVV$}

We begin by constructing the $VVVV$ amplitude. After presenting the detailed construction for this case, we provide only the explicit expressions for other processes in what follows, deferring the details to the readers.

A 4-pt amplitude generally includes the $s$-, $t$-, and $u$-channel exchange diagrams, as well as contact interactions. However, the recursive construction allows us to consider only the channel-induced contributions; This is because, in the contour integration, non-analyticities appear only at the poles of the amplitude, i.e., the factorized amplitude encodes the contribution of contact interactions. With this in mind, we first examine the massive $VVVV$ amplitude with all helicities $+1$ in a single $s$-channel mediated by a massive vector. As the first step, we shift the amplitude using Eqs.~\eqref{eq:spinor_shift1_1} and~\eqref{eq:spinor_shift2} for massive spinors and Eq.~\eqref{eq:spinor_shift1_2} for massless ones. Then, the contour integration over the complex parameter $z$ allows us to get
\begin{align}
&\mathcal{M}_{4,s}(\bs{1}_{a_1}^+ \bs{2}_{a_2}^+ \bs{3}_{a_3}^+ 
\bs{4}_{a_4}^+)
\nonumber
\\
&= -\underset{z=z_{12}}{\rm Res}
\bigg[\frac{
\hat{\mathcal{M}}_{3, a_1a_2 b}^{\{I I'\}}(\hat{P}_{12})
\,\veps_{IJ}\veps_{I'J'}\,
\hat{\mathcal{M}}_{3, ba_3a_4}^{\{JJ'\}}(-\hat{P}_{12})}{\hat{s}_{12}-m_{b}^2}\bigg]+\text{(other exchanges)},
\label{eq:origin_VVVV_amp}
\end{align}
where $\hat{\mathcal{M}}_{3,a_1a_2 b}^{\{I I'\}}$ $(\hat{\mathcal{M}}_{3,ba_3a_4}^{\{J J'\}})$ is the massive $VVV$ on-shell amplitude involving the spinors of the intermediate vector state with mass $m_b$ and shifted momentum $\hat{P}_{12}$ ($-\hat{P}_{12}$), and $z_{12}$ is the solution for $\hat{s}_{12}=m_b^2$. The summation over the little-group indices implements the polarization sum for the intermediate state. Here, the shifted invariant $\hat{s}_{12}=\hat{P}_{12}^2$ is given by
\begin{align}
\hat{s}_{12}=m_1^2+m_2^2+2p_1\cdot p_2+2z(q_1\cdot p_2+p_1\cdot q_2)+2z^2q_1\cdot q_2\ .
\label{eq:shifted_s}
\end{align}
with the associated momenta $q_i$ (see the definition of $q$ in Eq.~\eqref{eq:momentum_shift}). Then, the first term of the RHS in Eq.~\eqref{eq:origin_VVVV_amp} includes the poles at the following $z$ values,
\begin{align}
z_{12}^{\pm}=\frac{1}{Q^2_{12}}\left( -P_{12}\cdot Q_{12}\pm\sqrt{(P_{12}\cdot Q_{12})^2-(s_{12}-m_b^2)Q^2_{12}}\right),
\label{eq:z_12}
\end{align}
where $Q_{12}=q_1+q_2$. 

Using Eqs.~\eqref{eq:shifted_s} and~\eqref{eq:z_12}, we obtain
\begin{align}
    \frac{1}{\hat{s}_{12}-m_{b}^2}=\frac{1}{s_{12}-m_{b}^2}\frac{z_{12}^+z_{12}^-}{(z-z_{12}^+)(z-z_{12}^-)}.
\end{align}
Defining the numerator in Eq.~\eqref{eq:origin_VVVV_amp} as 
\begin{align}
    \hat{\mathcal{N}}(z)=
\hat{\mathcal{M}}_{3, a_1a_2 b}^{\{I I'\}}(\hat{P}_{12})
\,\veps_{IJ}\veps_{I'J'}\,
\hat{\mathcal{M}}_{3, ba_3a_4}^{\{JJ'\}}(-\hat{P}_{12}),
\label{eq:numerator}
\end{align}
we re-express Eq.~\eqref{eq:origin_VVVV_amp} as
\begin{align}
\mathcal{M}_{4,s}(\bs{1}_{a_1}^+ \bs{2}_{a_2}^+ \bs{3}_{a_3}^+ 
\bs{4}_{a_4}^+)=\frac{z_{12}^+\hat{\mathcal{N}}(z_{12}^-)-z_{12}^-\hat{\mathcal{N}}(z_{12}^+)}{(s_{12}-m^2_b)(z_{12}^+-z_{12}^-)}
+\mbox{(other exchanges)}.
\label{eq:4V_s_pole}
\end{align}
Using the explicit form of the massive $VVV$ on-shell amplitude in Table~\ref{tab:3-pt}, we confirm that $\hat{\mathcal{M}}_{3,abc}^{\{I_1 I_2\}}$ includes terms up to linear order in $z$:
\begin{align}
\hat{\mathcal{M}}_{3}(\hat{P})
=\hat{\mathcal{M}}^{(0)}_3(P)
+z\hat{\mathcal{M}}^{(1)}_3(P),
\end{align}
where we omitted the vector and little group indices for simplicity. This indicates that Eq.~\eqref{eq:numerator} includes terms up to quadratic order in $z$:
\begin{align}
    \hat{\mathcal{N}}(z_{12}^{\pm})=\hat{\mathcal{N}}^{(0)}(0)
    +z_{12}^{\pm}\hat{\mathcal{N}}^{(1)}(0)+(z_{12}^{\pm})^2\hat{\mathcal{N}}^{(2)}(0),
\end{align}
where $\mathcal{N}^{(0)}(0)$ is the product of two massive $VVV$ on-shell amplitudes with unshifted momenta.

In Eq.~\eqref{eq:4V_s_pole}, the contributions from $\hat{\mathcal{N}}^{(1)}$ cancel identically; $(z_{12}^+z_{12}^- - z_{12}^-z_{12}^+)\hat{\mathcal{N}}^{(1)}=0$, while those from $\hat{\mathcal{N}}^{(0)}$ and $\hat{\mathcal{N}}^{(2)}$ remain.
To identify the term associated with $\hat{\mathcal{N}}^{(2)}$, using Eq.~\eqref{eq:z_12}, we compute
\begin{align}
    \frac{z_{12}^+(z_{12}^-)^2-z_{12}^-(z_{12}^+)^2}{z_{12}^+-z_{12}^-}=-\frac{s^2_{12}-m^2_b}{2q_1\cdot q_2},
\end{align}
resulting in the propagator cancellation in Eq.~\eqref{eq:4V_s_pole}. It implies that the term associated with  $\hat{\mathcal{N}}^{(2)}$ corresponds to the contact diagram. Inserting the contributions from $\hat{\mathcal{N}}^{(0,1,2)}$ into Eq.~\eqref{eq:4V_s_pole}, we obtain
\begin{align}
\mathcal{M}_{4,s}(\hat{\bs{1}}_{a_1}^+ \hat{\bs{2}}_{a_2}^+ \hat{\bs{3}}_{a_3}^+ 
\hat{\bs{4}}_{a_4}^+)&=\frac{\mathcal{N}^{(0)}(0)}{s_{12}-m^2_b}-\frac{\mathcal{N}^{(2)}(0)}{2q_1\cdot q_2},
\end{align}
For clarity, we present the explicit forms of $\mathcal{N}^{(0)}$ and $\mathcal{N}^{(2)}/(2q_1\cdot q_2)$:
\begin{align}
    \hat{\mathcal{N}}^{(0)}(0)&=
    \hat{\mathcal{M}}_{3, a_1a_2 b}^{\{I I'\}}(P_{12})
\,\veps_{IJ}\veps_{I'J'}\,
\hat{\mathcal{M}}_{3, ba_3a_4}^{\{JJ'\}}(-P_{12}),
\label{eq:N0}
\\
\frac{\hat{\mathcal{N}}^{(2)}(0)}{2q_1\cdot q_2}
&=C_{a_1a_2b}C_{ba_3a_4}\frac{\langle 1^1 2^1\rangle\langle  3^1 4^1\rangle}{m^2_b}
\nonumber
\\
&\times
\left(\frac{\langle 2^14^1\rangle[1^13^1]}{m_1m_3}+\frac{\langle 1^14^1\rangle[2^13^1]}{m_2m_3}+\frac{\langle 2^13^1\rangle[1^14^1]}{m_1m_4}+\frac{\langle 1^13^1\rangle[2^14^1]}{m_2m_4}\right),
\label{eq:N2}
\end{align}
leading to the explicit expression for the massive $VVVV$ amplitude with all helicities $+1$, mediated by a massive vector in the $s$-channel.

In Eqs.~\eqref{eq:N0} and \eqref{eq:N2}, taking each spinor to be bolded without the little-group index, we derive the massive $VVVV$ amplitude with arbitrary helicities, mediated by a massive vector in the $s$-channel. In this way, we proceed with the computation for all $s$-channel vector and scalar exchanges. Using Bose statistics, the amplitudes for the $t$- and $u$-channels can then be obtained by the exchanges $\bs{2}_{a_2}\leftrightarrow \bs{3}_{a_3}$ and $\bs{2}_{a_2}\leftrightarrow \bs{4}_{a_4}$, respectively. Therefore, with the $s$-channel amplitude $\mathcal{M}_{4,s}
(\bs{1}_{a_1} \bs{2}_{a_2} \bs{3}_{a_3} \bs{4}_{a_4})$ including all possible exchanges, we obtain the massive $VVVV$ amplitude as
\begin{align}
&\mathcal{M}_{4}
(\bs{1}_{a_1} \bs{2}_{a_2} \bs{3}_{a_3} \bs{4}_{a_4})
\nonumber
\\
&=
\mathcal{M}_{4,s}
(\bs{1}_{a_1} \bs{2}_{a_2} \bs{3}_{a_3} \bs{4}_{a_4})
+
\mathcal{M}_{4,s}
(\bs{1}_{a_1} \bs{3}_{a_3} \bs{2}_{a_2} \bs{4}_{a_4})
+
\mathcal{M}_{4,s}
(\bs{1}_{a_1} \bs{4}_{a_4} \bs{3}_{a_3} \bs{2}_{a_2}).
\label{eq:stu_crossing}
\end{align}
Here, the explicit form of $\mathcal{M}_{4,s}
(\bs{1}_{a_1} \bs{2}_{a_2} \bs{3}_{a_3} \bs{4}_{a_4})$ is given by
\begin{align}
&\mathcal{M}_{4,s}
(\bs{1}_{a_1} \bs{2}_{a_2} \bs{3}_{a_3} \bs{4}_{a_4})
\nonumber
\\
&=-\sum_b \frac{C_{a_1a_2 b}C_{a_3a_4b} (N_{V^4}+N_{V^4,b}')}{m_{a_1}m_{a_2}m_{a_3}m_{a_4}(s_{12}-m_{b}^2)}
-\sum_i 
\frac{
F_{a_1a_2 i}F_{a_3a_4 i} 
\ab{\bs{1}\bs{2}}\ab{\bs{3}\bs{4}}
\sqb{\bs{1}\bs{2}}\sqb{\bs{3}\bs{4}}
}{m_{a_1}m_{a_2}m_{a_3}m_{a_4}(s_{12}-m_{i}^2)}
\nonumber
\\
&\quad\;
-\sum_{\hat{b}} \frac{C_{a_1a_2 \hat{b}}C_{a_3a_4 \hat{b}} 
\,N_{V^4}}{m_{a_1}m_{a_2}m_{a_3}m_{a_4} \, s_{12} },
\label{eq:VVVV_s-ch}
\end{align}
where the combinations of spinor contractions, $N_{V^4}$ and $N_{V^4,b}'$, are
\begin{align}
N_{V^4}=&
(\asb{\bs{3}\bs{4}\bs{3}}
\asb{\bs{4}(\bs{2}-\bs{1})\bs{4}}
+
\asb{\bs{4}\bs{3}\bs{4}}
\asb{\bs{3}(\bs{1}-\bs{2})\bs{3}})
\ab{\bs{1}\bs{2}}
\sqb{\bs{1}\bs{2}}
\nonumber
\\
&+
(\asb{\bs{2}\bs{1}\bs{2}}
\asb{\bs{1}(\bs{3}-\bs{4})\bs{1}}
+
\asb{\bs{1}\bs{2}\bs{1}}
\asb{\bs{2}(\bs{4}-\bs{3})\bs{2}}
)
\ab{\bs{3}\bs{4}}
\sqb{\bs{3}\bs{4}}
\nonumber
\\
&+2\asb{\bs{2}\bs{1}\bs{2}}
(
\asb{\bs{3}\bs{4}\bs{3}}
\ab{\bs{1}\bs{4}}
\sqb{\bs{1}\bs{4}}
-
\asb{\bs{4}\bs{3}\bs{4}}
\ab{\bs{1}\bs{3}}
\sqb{\bs{1}\bs{3}}
)
\nonumber
\\
&+2\asb{\bs{1}\bs{2}\bs{1}}
(
\asb{\bs{4}\bs{3}\bs{4}}
\ab{\bs{2}\bs{3}}
\sqb{\bs{2}\bs{3}}
-
\asb{\bs{3}\bs{4}\bs{3}}
\ab{\bs{2}\bs{4}}
\sqb{\bs{2}\bs{4}}
)
\nonumber
\\
&+
\ab{\bs{1}\bs{2}}
\ab{\bs{3}\bs{4}}
\sqb{\bs{1}\bs{2}}
\sqb{\bs{3}\bs{4}}
(s_{13}-s_{14})
\nonumber
\\
&+
(
\ab{\bs{1}\bs{3}}
\ab{\bs{2}\bs{4}}
\sqb{\bs{1}\bs{3}}
\sqb{\bs{2}\bs{4}}
-
\ab{\bs{1}\bs{4}}
\ab{\bs{2}\bs{3}}
\sqb{\bs{1}\bs{4}}
\sqb{\bs{2}\bs{3}})\, s_{12},
\end{align}
and
\begin{align}
N_{V^4,b}'=&
\ab{\bs{1}\bs{2}}
\ab{\bs{3}\bs{4}}
\sqb{\bs{1}\bs{2}}
\sqb{\bs{3}\bs{4}}
\bigg(
\frac{(m_{a_1}^2-m_{a_2}^2)(m_{a_3}^2-m_{a_4}^2)}{m_b^2}
\bigg)
\nonumber
\\
&-
(
\ab{\bs{1}\bs{3}}
\ab{\bs{2}\bs{4}}
\sqb{\bs{1}\bs{3}}
\sqb{\bs{2}\bs{4}}
-
\ab{\bs{1}\bs{4}}
\ab{\bs{2}\bs{3}}
\sqb{\bs{1}\bs{4}}
\sqb{\bs{2}\bs{3}}
)\,m_b^2.
\end{align}

As the next step, we will find the conditions on couplings required to satisfy tree unitarity. The procedure is as follows: using Eq.~\eqref{eq:lam_eta_spinors_explicit}, we parameterize the spinors for the kinematic configuration in the c.m. frame. Expanding the amplitude at high $E$ with the formulas in Appendix~\ref{appendix:high-E_limits_spinor}, we find the coupling relations eliminating the higher-order $E$ growth. Let us first examine the $E^4$, $E^3$, $E^2$, and $E$ growth in the helicity orderings, $(0000)$, $(\pm 000)$, $(\pm\!\pm\! 00)$, and $(\pm\!\pm\!\pm 0)$, respectively. In this procedure, by replacing the transverse polarization vectors with those of massless vectors, we can also examine the higher-order $E$ growth even in the cases of external massless vectors. Thus, by investigating the massive and massless cases together, we obtain
\begin{align}
\sum_{\capi{B}}
(C_{\capi{A_1 A_2 B}}C_{\capi{A_3 a_4 B}}
+C_{\capi{A_1 A_3 B}}C_{\capi{a_4 A_2 B}}
+C_{\capi{A_1 a_4 B}}C_{\capi{A_2 A_3 B}})=0,
\label{eq:relation_VVVV1_0}
\end{align}
where each of capital indices collects the massive and massless vector indices together, as introduced in Table~\ref{tab:3-pt}. Eq.~\eqref{eq:relation_VVVV1_0} 
includes $C_{\hat{a}\hat{b}c}=0$ in Eq.~\eqref{eq:VVV_op}.

For any $VVVV$ amplitude with external massless vectors, it must be independent of reference spinors. Thus, the $VVVV$ amplitude with all external massless vectors satisfies
\begin{align}
\mathcal{M}_{4}
(1^{\pm}_{\hat{a}_1} 2^{-}_{\hat{a}_2} 3^{+}_{\hat{a}_3} 4^{+}_{\hat{a}_4})\; \rightarrow \;0,
\label{eq:VVVV_ward}
\end{align}
under the replacement $\epsilon_1^{\pm} \rightarrow \sqrt{2}\,|1\rangle [1|$. This allows us to get
\begin{align}
\sum_{B}
(C_{\hat{a}_1\hat{a}_2 B}
C_{\hat{a}_3\hat{a}_4 B}
+
C_{\hat{a}_1\hat{a}_3 B}
C_{\hat{a}_4\hat{a}_2 B}
+
C_{\hat{a}_1\hat{a}_4 B}
C_{\hat{a}_2\hat{a}_3 B})
=0.
\label{eq:relation_VVVV_massless}
\end{align}
Combining Eqs.~\eqref{eq:relation_VVVV1_0} and \eqref{eq:relation_VVVV_massless}, we obtain
\begin{align}
\sum_{\capi{B}}
(C_{\capi{A_1 A_2 B}}C_{\capi{A_3 A_4 B}}
+C_{\capi{A_1 A_3 B}}C_{\capi{A_4 A_2 B}}
+C_{\capi{A_1 A_4 B}}C_{\capi{A_2 A_3 B}})=0,
\label{eq:relation_VVVV1}
\end{align}
which is known as the Jacobi identity.

Next, we examine the $E^2$ and $E$ growth in the orderings, $(0000)$ and $(\pm 000)$. To eliminate such growth, the following relation is required:
\begin{align}
&-4
(
F_{\capi{A_1 A_3} i}
F_{\capi{A_2 A_4} i}
-
F_{\capi{A_1 A_4} i}
F_{\capi{A_2 A_3} i}
)
\nonumber
\\
&=\sum_{b} 
\Big\{
(
2C_{\capi{A_1 A_2} b}C_{\capi{A_3 A_4} b}
+C_{\capi{A_1 A_3} b}C_{\capi{A_2 A_4} b}
-C_{\capi{A_1 A_4} b}C_{\capi{A_2 A_3} b}
)m_b^2
\nonumber
\\
&\qquad \quad +\frac{1}{m_b^2}
\big[
C_{\capi{A_1 A_3} b}C_{\capi{A_2 A_4} b}
(m_{\capi{A_1}}^2-m_{\capi{A_3}}^2)
(m_{\capi{A_2}}^2-m_{\capi{A_4}}^2)
\nonumber
\\
&\qquad\qquad
-
C_{\capi{A_1 A_4} b}C_{\capi{A_2 A_3} b}
(m_{\capi{A_1}}^2-m_{\capi{A_4}}^2)
(m_{\capi{A_2}}^2-m_{\capi{A_3}}^2)
\big]\Big\}
\nonumber
\\
&\quad
-\sum_{B}
(C_{\capi{A_1 A_3 B}}C_{\capi{A_2 A_4 B}}
-C_{\capi{A_1 A_4 B}}C_{\capi{A_2 A_3 B}})
(m_{\capi{A_1}}^2+m_{\capi{A_2}}^2+m_{\capi{A_3}}^2+m_{\capi{A_4}}^2),
\label{eq:relation_VVVV2}
\end{align}
where we extended the result such that all couplings include the capital indices for all external vectors. To show the validity of this extension, let us consider the case $\{\hat{a}_1\hat{a}_2a_3a_4\}$. Then, we obtain
\begin{gather}
\begin{aligned}
0&=\sum_{b} 
\Big\{
(
C_{\hat{a}_1 a_3 b}C_{\hat{a}_2a_4 b}
-C_{ \hat{a}_1 a_4 b}C_{\hat{a}_2a_3 b}
)m_b^2
\\
&\qquad \quad +\frac{m_{a_3}^2m_{a_4}^2}{m_b^2}
(
C_{ \hat{a}_1 a_3  b}C_{ \hat{a}_2 a_4  b}
-
C_{  \hat{a}_1 a_4  b}C_{ \hat{a}_2 a_3  b}
)
\big]\Big\}
\\
&\quad
-\sum_{b}
(C_{ \hat{a}_1 a_3  b}C_{ \hat{a}_2 a_4  b}
-
C_{ \hat{a}_1 a_4  b}C_{ \hat{a}_2 a_3  b})
(m_{a_3}^2+m_{a_4}^2)
\end{aligned}
\nonumber
\\
\rightarrow \;\sum_{\capi{B}}
(C_{\hat{a}_1 \hat{a}_2\capi{ B}}C_{a_3 a_4 \capi{B}}
+C_{\hat{a}_1 a_3 \capi{B}}C_{a_4 \hat{a}_2 \capi{B}}
+C_{\hat{a}_1 a_4 \capi{B}}C_{\hat{a}_2 a_3 \capi{B}})=0,
\label{eq:hat_a_hat_a_aa}
\end{gather}
returning to Eq.~\eqref{eq:relation_VVVV1}, where we used $(VV\phi)$ 
$F_{\hat{a}\hat{b}i}=F_{\hat{a}bi}=0$ in Eq.~\eqref{eq:VVS_amp} and $(VVV)$ $m_b=m_c$ for the case $\{\hat{a}bc\}$ in Eq.~\eqref{eq:1massless_2massive_VVV} together with $C_{\hat{a}\hat{b}c}=0$. In a similar way, one can confirm that Eq.~\eqref{eq:relation_VVVV1} also yields Eq.~\eqref{eq:hat_a_hat_a_aa} even for the cases, $\{\hat{a}_1\hat{a}_2\hat{a}_3a_4\}$ and $\{\hat{a}_1\hat{a}_2\hat{a}_3\hat{a}_4\}$, leading to no additional conditions.

\subsection{$VVV\phi$}

The $s$-channel massive $VVV\phi$ amplitude is given by
\begin{align}
&\mathcal{M}_{4,s}
(\bs{1}_{\capi{A_1}} \bs{2}_{\capi{A_2}} \bs{3}_{\capi{A_3}} \bs{4}_{i_4})
\nonumber
\\
&=-i\sqrt{2}
\bigg(\sum_b \frac{C_{\capi{A_1 A_2} b} 
F_{ b\capi{A_3} i_4} N'_{W^3\phi, b}}{
m_{\capi{A_1}}m_{\capi{A_2}}m_{\capi{A_3}}(s_{12}-m_b^2)}
+
\sum_j
\frac{2 F_{ \capi{A_1 A_2} j} G_{ \capi{A_3} j i_4}
\langle \bs{3}\bs{4}\bs{3}]
\langle \bs{1}\bs{2} \rangle
[\bs{1}\bs{2}]
}{
m_{\capi{A_1}}m_{\capi{A_2}}m_{\capi{A_3}}(s_{12}-m_j^2)}
\bigg),
\end{align}
where the combination of spinor products $N'_{W^3\phi,b}$ is
\begin{align}
N'_{W^3\phi,b}
=&\asb{\bs{3}(\bs{1}-\bs{2})\bs{3}}
\ab{\bs{1}\bs{2}}
\sqb{\bs{1}\bs{2}}
+2(\asb{\bs{1}\bs{2}\bs{1}}
\ab{\bs{2}\bs{3}}
\sqb{\bs{2}\bs{3}}
-
\asb{\bs{2}\bs{1}\bs{3}}
\ab{\bs{1}\bs{3}}
\sqb{\bs{1}\bs{3}}
)
\nonumber
\\
&+\asb{\bs{3}\bs{4}\bs{3}}
\ab{\bs{1}\bs{2}}
\sqb{\bs{1}\bs{2}}
\frac{(m^2_{\capi{A_1}}-m^2_{\capi{A_2}})}{m_b^2}.
\label{eq:VVVphi_amp}
\end{align}
Here, the intermediate vector state involving the index $b$ must be massive due to $F_{\hat{a}bi}=F_{\hat{a}\hat{b}i}=0$. Deriving the $t$- and $u$-channel massive $VVV\phi$ amplitudes through the exchanges $\bs{2}_{a_2}\leftrightarrow \bs{3}_{a_3}$ and $\bs{1}_{a_1}\leftrightarrow \bs{3}_{a_3}$ in Eq.~\eqref{eq:VVVphi_amp}, and combining them with the Eq.~\eqref{eq:VVVphi_amp}, we can derive the massive $VVV\phi$ amplitude.

Examining the $E^2$ and $E$ growth in the orderings, $(0000)$ and $(\pm 000)$, respectively, we obtain
\begin{align}
&\sum_{b}\frac{1}{2m_b^2}
\big[
C_{\capi{A_2 A_3} b}(m^2_{\capi{A_2}}-m^2_{\capi{A_3}}+m_b^2)
F_{b \capi{A_1}i_4}
-
C_{\capi{A_2 A_1} b}(m^2_{\capi{A_2}}-m^2_{\capi{A_1}}+m_b^2)
F_{b \capi{A_3}i_4}
\big]
\nonumber
\\
&=
\sum_{j}F_{\capi{A_2 A_3} j} 
G_{\capi{A_1} i j}
-
F_{\capi{A_1 A_2} j} 
G_{\capi{A_3} i j}
-\sum_B C_{\capi{A_1 A_3} B}
F_{B \capi{A_2}i_4},
\label{eq:relation_VVVphi}
\end{align}
where we extended the relation such that all couplings include the capital indices for all external vectors, since this yields no additional conditions for the cases  $\{\hat{a}_1 \hat{a}_2  a_3 i_4\}$ and 
$\{\hat{a}_1 \hat{a}_2  \hat{a}_3 i_4\}$.

\subsection{$VV\phi\phi$, $V\psi V\bar{\psi}$}

We first present the massive $VV\phi\phi$ amplitude directly as
\begin{align}
&\mathcal{M}_{4}
(\bs{1}_{\capi{A_1}} \bs{2}_{\capi{A_2}} 
\bs{3}_{i_3} \bs{4}_{i_4})
\nonumber
\\
&=\sum_b
\Bigg\{
\frac{C_{\capi{A_1 A_2} b}G_{b i_3 i_4} ( N_{V^2\phi^2}+ N_{V^2\phi^2,b}' )}{m_{\capi{A_1}}m_{\capi{A_2}}(s_{12}-m_{b}^2)}
\nonumber
\\
&\quad\;
+\bigg[\frac{F_{\capi{A_1 b}i_3}F_{\capi{A_2 b}i_4}
(2\asb{\bs{1}\bs{3}\bs{1}}\asb{\bs{2}\bs{4}\bs{2}}
-(s_{13}+m_{b}^2)
\ab{\bs{1}\bs{2}}\sqb{\bs{1}\bs{2}})
}{m_{\capi{A_1}}m_{\capi{A_2}}m_{b}^2(s_{13}-m_{b}^2)}
+
(\bs{3}_{i_3}\!\leftrightarrow \bs{4}_{i_4})\bigg]
\Bigg\}
\nonumber
\\
&\quad\;
-\sum_j
\Bigg\{
\frac{F_{\capi{A_1 A_2}j}P_{j i_3 i_4} 
\ab{\bs{1}\bs{2}}
[\bs{1}\bs{2}]
}{m_{\capi{A_1}} m_{\capi{A_2}}(s_{12}-m_{j}^2)}
\nonumber
\\
&\quad\;
-\bigg[\frac{G_{\capi{A_1} i_3j}G_{\capi{A_2} i_4j}
(2\asb{\bs{1}\bs{3}\bs{1}}\asb{\bs{2}\bs{4}\bs{2}}
-(s_{13}-m_j^2)\ab{\bs{1}\bs{2}}\sqb{\bs{1}\bs{2}})}{m_{\capi{A_1}} m_{\capi{A_2}}(s_{13}-m_{j}^2)}
+
(\bs{3}_{i_3}\!\leftrightarrow \bs{4}_{i_4})\bigg]
\Bigg\}
\nonumber
\\
&\quad\;
+\sum_{\hat{b}}
\frac{C_{\capi{A_1}\capi{A_2} \hat{b}}G_{\hat{b} i_3 i_4} N_{V^2\phi^2}}{m_{\capi{A_1}} m_{\capi{A_2}} \, s_{12} },
\label{eq:VVphiphi_amp}
\end{align}
where the combinations of spinor products, $N_{V^2\phi^2}$ and $N'_{V^2\phi^2,b}$, are
\begin{align}
N_{V^2 \phi^2}
=&\asb{\bs{1}\bs{2}\bs{1}}
\asb{\bs{2}(\bs{3}-\bs{4})\bs{2}}
+
\asb{\bs{2}\bs{1}\bs{2}}
\asb{\bs{1}(\bs{4}-\bs{3})\bs{1}}
+
(s_{13}-s_{14})\ab{\bs{1}\bs{2}}
[\bs{1}\bs{2}],
\\
N'_{V^2 \phi^2,b}
=&\bigg(\frac{(m_{\capi{A_1}}^2-m_{i_3}^2)(m_{\capi{A_2}}^2-m_{i_4}^2)}{m_b^2}\bigg)
\ab{\bs{1}\bs{2}}
\sqb{\bs{1}\bs{2}},
\end{align}
In Eq.~\eqref{eq:VVphiphi_amp}, we used Bose statistics for scalars; $(\bs{3}_{i_3} \leftrightarrow \bs{4}_{i_4})$ denotes the term obtained by applying the exchange $\bs{3}_{i_3} \leftrightarrow \bs{4}_{i_4}$ to the preceding expression, thereby yielding the $u$-channel amplitude from the $t$-channel one. 

The $E^2$ and $E$ growth in the orderings, $(0000)$ and $(\pm 000)$, allows us to get
\begin{align}
&-\sum_{b}\frac{1}{m_b^2}
(F_{\capi{A_1} b i_3 }F_{\capi{A_2} b i_4 }
-F_{\capi{A_1} b i_4 }F_{\capi{A_2} b i_3 })
\nonumber
\\
&=G_{\capi{A_1}  i_3 j}G_{\capi{A_2}  i_4 j}
-G_{\capi{A_1}  i_4 j}G_{\capi{A_2}  i_3 j}
+\sum_B C_{\capi{A_1A_2 B}}G_{\capi{B}  i_3i_4},
\label{eq:relation_VVphiphi}
\end{align}
where we extended the relation so that all couplings involve the capital indices for all external vectors. To confirm the validity of the extension, we use 
\begin{align}
\mathcal{M}_{4}(1^{\pm}_{\hat{a}_1} 2^{-}_{\hat{a}_2} 3_{i_3} 4_{i_4}) \; \rightarrow \; 0,
\end{align}
under the replacement $\epsilon_1^{\pm} \rightarrow \sqrt{2}\,|1\rangle [1|$ as in Eq.~\eqref{eq:VVVV_ward}. It leads to
\begin{align}
G_{\hat{a}_1  i_3 j}G_{\hat{a}_2  i_4 j}
-G_{\hat{a}_1  i_4 j}G_{\hat{a}_2  i_3 j}
+\sum_{\hat{b}} 
C_{\hat{a}_1\hat{a}_2 \hat{b}}G_{\hat{b} i_3i_4}
=0,
\label{eq:GC_relation}
\end{align}
consistent with Eq.~\eqref{eq:relation_VVphiphi}.

Next, we present the massive $V\psi V\bar{\psi}$ amplitude as
\begin{align}
&
\mathcal{M}_{4}
(\bs{1}_{\capi{A_1}} \bs{2}_{\mathrm{i}_2} 
\bs{3}_{\capi{A_3}} \bar{\bs{4}}_{\mathrm{i}_4})
\nonumber
\\
&=\sum_b
\frac{i C_{\capi{A_1 A_3} b} (N_{V^2\psi^2}+N_{V^2\psi^2,b}')}{m_{\capi{A_1}}m_{\capi{A_3}}(s_{13}-m_{b}^2)}
-
\sum_j \frac{F_{\capi{A_1A_3}j} 
\ab{\bs{1}\bs{3}}\sqb{\bs{1}\bs{3}}
\Big(
(H_{j})_{\mathrm{i}_4\mathrm{i}_2} 
\sqb{\bs{2}\bs{4}}
+
(H^\dagger_{j})_{\mathrm{i}_4\mathrm{i}_2} 
\langle \bs{2}\bs{4}\rangle
\Big)
}{m_{\capi{A_1}}m_{\capi{A_3}}(s_{13}-m^2_j)}
\nonumber
\\
&\quad +
\frac{2}{m_{\capi{A_1}}m_{\capi{A_3}}}
\sum_{\mathrm{j}}
\bigg(
\frac{M_{W^2 \psi^2 \!, \,\mathrm{j}}}{s_{12}-m^2_{\mathrm{j}}}
+
(\bs{1}_{\capi{A_1}}\!\leftrightarrow \bs{3}_{\capi{A_3}})
\bigg)
+
\sum_{\hat{b}}
\frac{i C_{\capi{A_1 A_3} \hat{b}}  N_{V^2\psi^2} }{m_{\capi{A_1}}m_{\capi{A_3}} s_{13} },
\label{eq:VVpsipsi_amp}
\end{align}
where the combinations of spinor products, $N_{W^2\psi^2}$, $N'_{W^2\psi^2\!,b}$, and $M_{W^2\psi^2\! ,\,\mathrm{j}}$ are
\begin{align}
N_{W^2\psi^2}
=&
R^b_{\mathrm{i}_4 \mathrm{i}_2}
\Big\{\asb{\bs{4}(\bs{3}-\bs{1}) \bs{2}}
\ab{\bs{1}\bs{3}}\sqb{\bs{1}\bs{3}}
+2 (\asb{\bs{3}\bs{1}\bs{3}}
\ab{\bs{1}\bs{4}}\sqb{\bs{1}\bs{2}}
+ 
\asb{\bs{1}\bs{3}\bs{1}}
\ab{\bs{3}\bs{4}}\sqb{\bs{2}\bs{3}}
\Big\}
\nonumber
\\
&+L^b_{\mathrm{i}_4 \mathrm{i}_2}
\Big\{\asb{\bs{2}(\bs{3}-\bs{1}) \bs{2}}
\ab{\bs{1}\bs{3}}
\sqb{\bs{1}\bs{3}}
+2 (\asb{\bs{3}\bs{1}\bs{3}}
\ab{\bs{1}\bs{2}}\sqb{\bs{1}\bs{4}}
+ 
\asb{\bs{1}\bs{3}\bs{1}}
\ab{\bs{2}\bs{3}}\sqb{\bs{3}\bs{4}}
\Big\},
\\
N'_{W^2\psi^2\!,b}
=&
R^b_{\mathrm{i}_4 \mathrm{i}_2}
\ab{\bs{1}\bs{3}}\sqb{\bs{1}\bs{3}}
(
m_{\mathrm{i}_2} \langle \bs{2}\bs{4}\rangle
-
m_{\mathrm{i}_4} [ \bs{2}\bs{4}]
)
\frac{(m_{\capi{A_1}}^2-m_{\capi{A_3}}^2)}{m_b^2}
\nonumber
\\
&+L^b_{\mathrm{i}_4 \mathrm{i}_2}
\ab{\bs{1}\bs{3}}\sqb{\bs{1}\bs{3}}
(
m_{\mathrm{i}_2} [ \bs{2}\bs{4}]
-
m_{\mathrm{i}_4} \langle \bs{2}\bs{4}\rangle
)
\frac{(m_{\capi{A_1}}^2-m_{\capi{A_3}}^2)}{m_b^2},
\\
M_{W^2\psi^2\! ,\,\mathrm{j}}
=&
-R^{\capi{A_1}}_{\mathrm{j} \mathrm{i}_2}
R^{\capi{A_3}}_{\mathrm{i}_4 \mathrm{j}}
\ab{\bs{3}\bs{4}}
\sqb{\bs{1}\bs{2}}
\asb{\bs{1}\bs{(1+2)}\bs{3}}
\nonumber
\\
&+
L^{\capi{A_1}}_{\mathrm{j} \mathrm{i}_2}
L^{\capi{A_3}}_{\mathrm{i}_4 \mathrm{j}}
\ab{\bs{1}\bs{2}}\sqb{\bs{3}\bs{4}}
\asb{\bs{3}(\bs{3+4})\bs{1}}
\nonumber
\\
&+m_{\mathrm{j}}
(
R^{\capi{A_1}}_{\mathrm{j} \mathrm{i}_2}
L^{\capi{A_3}}_{\mathrm{i}_4 \mathrm{j}}
\ab{\bs{1}\bs{3}}
\sqb{\bs{3}\bs{4}}
\sqb{\bs{1}\bs{2}}
+
L^{\capi{A_1}}_{\mathrm{j} \mathrm{i}_2}
R^{\capi{A_3}}_{\mathrm{i}_4 \mathrm{j}}
\ab{\bs{1}\bs{2}}
\ab{\bs{3}\bs{4}}
\sqb{\bs{1}\bs{3}}
).
\label{eq:M_W2phi2j}
\end{align}
To make Eq.~\eqref{eq:VVpsipsi_amp} valid even for the external massless vectors, in Eq.~\eqref{eq:M_W2phi2j}, we do not simplify the terms $\asb{\bs{1(1+2)3}}$ and $\asb{\bs{3(3+4)1}}$ using the Dirac equation in Eq.~\eqref{eq:spinor_dirac_eq}, since the massless-vector spinors necessarily involve reference spinors. Similarly to Eq.~\eqref{eq:VVphiphi_amp}, the exchanged term $(\bs{1}_{\capi{A_1}} \leftrightarrow \bs{3}_{\capi{A_3}})$ in Eq.~\eqref{eq:VVpsipsi_amp} denotes the $u$-channel amplitude derived from the $s$-channel one.

The $E^2$ and $E$ growth in the orderings $(0\!\pm\! 0\mp)$ and $(0\! \pm \!\pm \mp)$ $[\mbox{ or }(0\! \pm\! \mp \mp)]$, respectively, requires 
\begin{align}
[R^{a_1},R^{\capi{A_3}}]=iC_{a_1 \capi{A_3 B}}R^{\capi{B}}
,\quad
[L^{a_1},L^{\capi{A_3}}]=iC_{ a_1 \capi{A_3 B}}L^{\capi{B}},\quad  
\label{eq:relation_VpsiVpsi1_0}
\end{align}
where we omitted the indices for spin-1/2 fermions. For all external massless vectors, Eq.~\eqref{eq:VVpsipsi_amp} does not involve any high-$E$ growth. However, such amplitudes must satisfy
\begin{gather}
\mathcal{M}_{4}
(1^{\pm}_{\hat{a}_1} 2^{\pm}_{\rm{i}_2} 3_{\hat{a}_3}^\pm \bar{4}_{\rm{i}_4}^\mp),\mathcal{M}_{4}
(1^{\pm}_{\hat{a}_1} 2^{\pm}_{\rm{i}_2} 3_{\hat{a}_3}^\mp \bar{4}_{\rm{i}_4}^\mp)\;\rightarrow\;0,
\end{gather}
under the replacement $\epsilon_1^{\pm} \rightarrow \sqrt{2}\,|1\rangle [1|$, implying the independence of reference spinors. It leads to
\begin{align}
[R^{\hat{a}_1},R^{\hat{a}_3}]
=iC_{ \hat{a}_1 \hat{a}_3 \hat{b} }R^{\hat{b}},\quad 
[L^{\hat{a}_1},L^{\hat{a}_3}]=
iC_{\hat{a}_1 \hat{a}_3 \hat{b}}L^{\hat{b}}.
\label{eq:relation_VpsiVpsi1_1}
\end{align}
Combining Eqs.~\eqref{eq:relation_VpsiVpsi1_0} and \eqref{eq:relation_VpsiVpsi1_1}, we obtain
\begin{align}
[R^{\capi{A_1}},R^{\capi{A_3}}]=iC_{\capi{A_1A_3 B}}R^{\capi{B}}
,\quad
[L^{\capi{A_1}},L^{\capi{A_3}}]=iC_{\capi{A_1A_3 B}}L^{\capi{B}},
\label{eq:relation_VpsiVpsi1}
\end{align}
with the capital indices for all external vectors.

Additionally, the $E$ growth in the orderings $(0\! \pm \!0 \pm)$ yields
\begin{align}
&2F_{\capi{A_1 A_3} i}(H_{i})_{\mathrm{i}_4\mathrm{i}_2}
-m_{\mathrm{i}_4}
\{R^{\capi{A_1}},R^{\capi{A_3}}\}_{\mathrm{i}_4\mathrm{i}_2}
-m_{\mathrm{i}_2}
\{L^{\capi{A_1}},L^{\capi{A_3}}\}_{\mathrm{i}_4\mathrm{i}_2}
+\sum_{\mathrm{j}}2m_{\mathrm{j}} 
(
L^{\capi{A_1}}_{\mathrm{i}_4 \mathrm{j}}R^{\capi{A_3}}_{\mathrm{j} \mathrm{i}_2}
+
L^{\capi{A_3}}_{\mathrm{i}_4 \mathrm{j}}R^{\capi{A_1}}_{\mathrm{j} \mathrm{i}_2}
)
\nonumber
\\
&=-\sum_{b}iC_{\capi{A_1 A_3} b}
(
m_{\mathrm{i}_4}R^{b}_{\mathrm{i}_4 \mathrm{i}_2}
-
m_{\mathrm{i}_2}L^{b}_{\mathrm{i}_4 \mathrm{i}_2}
)
\frac{(m_{\capi{A_1}}^2-m_{\capi{A_3}}^2)}{m_b^2},
\label{eq:relation_VpsiVpsi2}
\end{align}
where we extended the relation to involve the capital indices for all external vectors, since no additional conditions appear in the cases, $\{\hat{a}_1\mathrm{i}_2a_3\mathrm{i}_4\}$ and $\{\hat{a}_1\mathrm{i}_2\hat{a}_3\mathrm{i}_4\}$. For concreteness, let us consider Eq.~\eqref{eq:relation_VpsiVpsi2} in the case of $\{\hat{a}_1\mathrm{i}_2a_3\mathrm{i}_4\}$. Then, we have
\begin{align}
&
m_{\mathrm{i}_4}
[R^{\hat{a}_1},R^{a_3}]_{\mathrm{i}_4\mathrm{i}_2}
-m_{\mathrm{i}_2}
[L^{\hat{a}_1},L^{a_3}]_{\mathrm{i}_4\mathrm{i}_2}
=\sum_{b}iC_{\hat{a}_1 a_3 b}
(
m_{\mathrm{i}_4}R^{b}_{\mathrm{i}_4 \mathrm{i}_2}
-
m_{\mathrm{i}_2}L^{b}_{\mathrm{i}_4 \mathrm{i}_2}
).
\label{VFVF_hat_a_&_a}
\end{align}
Using $m_{\mathrm{i}}=m_{\mathrm{j}}$ and $R^{\hat{a}}=L^{\hat{a}}$ for the case $\{\hat{a}\mathrm{i}\mathrm{j}\}$ in Eq.~\eqref{eq:VFF_massless_amp}, one can observe that Eq.~\eqref{VFVF_hat_a_&_a} is guaranteed by  Eq.~\eqref{eq:relation_VpsiVpsi1}.

\subsection{$V\psi \phi \bar{\psi}$, $V\phi\phi\phi$}

The massive $V\psi \phi \bar{\psi}$ amplitude is given by
\begin{align}
&
\mathcal{M}_{4}
(\bs{1}_{\capi{A_1}} \bs{2}_{\mathrm{i}_2} 
\bs{3}_{i_3} \bar{\bs{4}}_{\mathrm{i}_4})
\nonumber
\\
&=-\sum_{\mathrm{j}}
\frac{\sqrt{2}}{m_{\capi{A_1}}}
\Bigg\{
\frac{1}{s_{12}-m^2_{\mathrm{j}}}
\Big[
\sqb{\bs{1}\bs{2}}
\Big(
m_{\mathrm{j}}
\ab{\bs{1}\bs{4}}
(H^\dagger_{i_3})_{\mathrm{i}_4 \mathrm{j}}
+
(m_{i_4}\ab{\bs{1}\bs{4}}+\asb{\bs{1}\bs{3}\bs{4}})
(H_{i_3})_{\mathrm{i}_4 \mathrm{j}}
\Big)
R^{\capi{A_1}}_{\mathrm{j}\mathrm{i}_2}
\nonumber
\\
&\qquad\qquad\qquad\qquad\qquad\;\;
+
\ab{\bs{1}\bs{2}}
\Big(
m_{\mathrm{j}} [\bs{14}]
(H_{i_3})_{\mathrm{i}_4 \mathrm{j}}
+
(m_{i_4} [\bs{14}]
+
\langle \bold{431}|
)
(H_{i_3}^{\dagger})_{\mathrm{i}_4 \mathrm{j}}
\Big)L^{\capi{A_1}}_{\mathrm{j}\mathrm{i}_2}
\Big]
\nonumber
\\
&+\frac{1}{s_{14}-m^2_{\mathrm{j}}}
\Big[
R^{\capi{A_1}}_{\mathrm{i}_4 \mathrm{j}}
\ab{\bs{1}\bs{4}}
\Big(
m_{\mathrm{j}}
\sqb{\bs{1}\bs{2}}
(H_{i_3})_{\mathrm{j} \mathrm{i}_2}
+
(m_{\rm{i}_2}\sqb{\bs{1}\bs{2}}
+\langle \bold{231}])
(H^{\dagger}_{i_3})_{\mathrm{j}\rm{i}_2}
\Big)
\Big]
\nonumber
\\
&\qquad\qquad\quad\;
+L^{\capi{A_1}}_{\mathrm{i}_4 \mathrm{j}}
\sqb{\bs{1}\bs{4}}
\Big(
m_{\mathrm{j}}
\ab{\bs{1}\bs{2}}
(H^\dagger_{i_3})_{\mathrm{j} \mathrm{i}_2}
+
(m_{\rm{i}_2}\ab{\bs{1}\bs{2}}
+\asb{\bs{1}\bs{3}\bs{2}})
(H_{i_3})_{\mathrm{j}\rm{i}_2}
\Big)
\Big]
\Bigg\}
\nonumber
\\
&+\sum_b \frac{\sqrt{2}F_{b\capi{A_1} i_3}}{m_{\capi{A_1}} m_b^2 (s_{13}-m_b^2)}
\Big[
R^{b}_{\mathrm{i}_4 \mathrm{i}_2}
\big(
2m_b^2 
\ab{\bs{1}\bs{4}}
\sqb{\bs{1}\bs{2}}
+\asb{\bs{1}\bs{3}\bs{1}}
(m_{\mathrm{i}_2}\ab{\bs{2}\bs{4}}
-m_{\mathrm{i}_4}\sqb{\bs{2}\bs{4}})
\big)
\nonumber
\\
&\qquad\qquad\qquad\qquad\qquad\;\;
+
L^{b}_{\mathrm{i}_4 \mathrm{i}_2}
\big(
2m_b^2 
\ab{\bs{1}\bs{2}}
\sqb{\bs{1}\bs{4}}
+\asb{\bs{1}\bs{3}\bs{1}}
(m_{\mathrm{i}_2}\sqb{\bs{2}\bs{4}}
-m_{\mathrm{i}_4}\ab{\bs{2}\bs{4}})
\big)
\Big]
\nonumber
\\
%
%
&-\sum_j \frac{\sqrt{2}i G_{\capi{A_1}j i_3}\langle \bold{131}]
\Big(
(H_j)_{\mathrm{i}_4\mathrm{i}_2}
\sqb{\bs{2}\bs{4}}
+
(H_j^\dagger)_{\mathrm{i}_4\mathrm{i}_2}
\ab{\bs{2}\bs{4}}
\Big)
}{m_{\capi{A_1}}(s_{13}-m^2_j)},
\end{align}
where $m_{\capi{A_1}}$ in denominators is necessary to be replaced with $\ab{1\eta_1}$ or $\sqb{1\eta_1}$ for the external massless vector.

The $E$ growth in the ordering $(0\! \pm \!0 \pm)$ requires
\begin{align}
\sum_{b}
\frac{1}{m_b^2}
F_{\capi{A_1}b i_3}
(m_{\mathrm{i}_4} R^b_{\mathrm{i}_4\mathrm{i}_2}
-m_{\mathrm{i}_2} L^b_{\mathrm{i}_4\mathrm{i}_2})
=
-iG_{\capi{A_1} i_3 j}
(H_j)_{\mathrm{i}_4\mathrm{i}_2}
-
(L^{\capi{A_1}} H_{i_3})_{\mathrm{i}_4\mathrm{i}_2}
+
(H_{i_3} R^{\capi{A_1}} )_{\mathrm{i}_4\mathrm{i}_2},
\label{eq:relation_Vpsiphipsi}
\end{align}
where we have extended the relation so that it involves the capital index of the external vector. To verify the validity of this extension, we use
\begin{gather}
\mathcal{M}_{4}
(1^{\pm}_{\hat{a}_1} 2^{\pm}_{\rm{i}_2} 3_{i_3} \bar{4}_{\rm{i}_4}^\pm),\mathcal{M}_{4}
(1^{\mp}_{\hat{a}_1} 2^{\pm}_{\rm{i}_2} 3_{i_3} \bar{4}_{\rm{i}_4}^\pm)\;\rightarrow \;0,
\end{gather}
under the replacement $\epsilon_1^{\pm} \rightarrow \sqrt{2}\,|1\rangle [1|$. It leads to
\begin{align}
iG_{\hat{a}_1 i_3 j}
(H_j)_{\mathrm{i}_4\mathrm{i}_2}
+
(L^{\hat{a}_1} H_{i_3})_{\mathrm{i}_4\mathrm{i}_2}
-
(H_{i_3} R^{\hat{a}_1} )_{\mathrm{i}_4\mathrm{i}_2}=0,
\end{align}
which can be directly derived from Eq.~\eqref{eq:relation_Vpsiphipsi}.

Next, we present the $s$-channel massive $V\phi\phi\phi$ amplitude as
\begin{align}
&\mathcal{M}_{4,s}
(\bs{1}_{\capi{A_1}} \bs{2}_{i_2} \bs{3}_{i_3} \bs{4}_{i_4})
\nonumber
\\
&=i\sqrt{2}
\Bigg(
\sum_b
\frac{F_{b\capi{A_1} i_2}G_{b i_3 i_4}
\big(
2\asb{\bs{1}\bs{3}\bs{1}}m_b^2
+\asb{\bs{1}\bs{2}\bs{1}}(m_b^2+m_{i_3}^2-m_{i_4}^2)
\big)
}{m_b^2 m_{\capi{A_1}}(s_{12}-m_b^2)}
+\sum_j \frac{G_{\capi{A_1} i_2 j}P_{j i_3 i_4}\langle \bold{121}]}{
m_{\capi{A_1}}(s_{12}-m_j^2)}
\Bigg),
\label{eq:Vphiphiphi_s-ch_amp}
\end{align}
where $m_{\capi{A_1}}$ in denominators is necessary to be replaced with $\ab{1\eta_1}$ or $\sqb{1\eta_1}$ for the external massless vector. Deriving the $t$- and $u$-channel massive $V\phi\phi\phi$ amplitudes through the exchanges $\bs{2}_{i_2}\leftrightarrow \bs{3}_{i_3}$ and $\bs{2}_{i_2}\leftrightarrow \bs{4}_{i_4}$ in Eq.~\eqref{eq:Vphiphiphi_s-ch_amp}, and combining them with Eq.~\eqref{eq:Vphiphiphi_s-ch_amp}, we obtain the massive $V\phi\phi\phi$ amplitude.

\section{Constructing tree-unitarity theories}
\label{sec:lagrangian_construction}

In Section~\ref{sec:4-pt_tree_unitarity}, we derived conditions on couplings by requiring tree unitarity of 4-pt amplitudes. Using these conditions, we simplify the Lagrangian in the mass basis. We then employ the Stückelberg formulation~\cite{Stueckelberg:1938hvi,Ruegg:2003ps}, we restore the gauge symmetry of the Lagrangian.

The advantage of such a formulation is that all high-energy divergences of higher-pt amplitudes arise only from higher-pt contact interaction terms of scalar fields, i.e., from the scalar potential~\cite{Cornwall:1974km}. Thus, to derive the Lagrangian of tree-unitary theories, we eliminate these higher-pt scalar contact interaction terms. As a result, the manifestly gauge-invariant structure of the scalar potential is no longer explicit. The tree-unitarity conditions for higher-pt amplitudes can therefore be extracted from the requirement of gauge invariance of the scalar potential.

Building on this result, we further specify the additional coupling conditions, thereby finalizing the derivation of all explicit conditions for diagnosing tree-level unitarity in our setup. For this purpose, we keep the original coupling basis, maintaining an intuitive understanding of the derivation.

\subsection{Lie algebra from the coupling conditions}
\label{sec:lie_algebra}

We first show that the coupling conditions from tree unitarity of 4-pt amplitudes organize the couplings into a Lie algebra of a group $K$. Referring to Table~\ref{tab:3-pt}, we define
\begin{gather}
T^{\capi{(A)}}_{bc}=i C_{\capi{A}bc}
\frac{m_{\capi{A}}^2-m_{b}^2-m_{c}^2}{2m_{b} m_{c}},
\nonumber
\\
T^{(a)}_{\hat{b}c}=T^{(a)}_{c\hat{b}}=
T^{(\hat{a})}_{\hat{b}c}=T^{(\hat{a})}_{c\hat{b}}=
T^{(a)}_{\hat{b}\hat{c}}=0
,\quad
T^{(\hat{a})}_{\hat{b}\hat{c}}=-i C_{\hat{a}\hat{b}\hat{c}},
\label{eq:C_to_T}
\\
T^{(a)}_{ib}=
-T^{(a)}_{bi}=
\frac{i}{m_b}
F_{ab i}
,\quad
T^{(\hat{a})}_{ib}=T^{(\hat{a})}_{bi}=
T^{(a)}_{i\hat{b}}=T^{(a)}_{\hat{b}i}=
T^{(\hat{a})}_{i\hat{b}}=
T^{(\hat{a})}_{\hat{b}i}=0,
\label{eq:F_to_T}
\\
T^{\capi{(A)}}_{ij}
=iG_{\capi{A}ij},
\label{eq:G_to_T}
\end{gather}
for the $VVV$, $VV\phi$, and $V\phi\phi$ couplings, respectively. Using the definitions above, we recast the coupling relations in the following simple forms,
\begin{align}
T^{\capi{(A)}}_{\capi{C \mathcal{K}}}
T^{\capi{(B)}}_{\capi{\mathcal{K} D}}
-
T^{\capi{(B)}}_{\capi{C \mathcal{K}}}
T^{\capi{(A)}}_{\capi{\mathcal{K} D}}
&=
iC_{\capi{ABE}}T^{\capi{(E)}}_{\capi{CD}},
\\
T^{\capi{(A)}}_{\capi{i \mathcal{K}}}
T^{\capi{(B)}}_{\capi{\mathcal{K} C}}
-
T^{\capi{(B)}}_{\capi{i \mathcal{K}}}
T^{\capi{(A)}}_{\capi{\mathcal{K} C}}
&=
iC_{\capi{ABD}}T^{\capi{(D)}}_{i\capi{C}},
\\
T^{\capi{(A)}}_{\capi{i \mathcal{K}}}
T^{\capi{(B)}}_{\capi{\mathcal{K} j}}
-
T^{\capi{(B)}}_{\capi{i\mathcal{K}}}
T^{\capi{(A)}}_{\capi{\mathcal{K} j}}
&=
iC_{\capi{ABE}}T^{\capi{(E)}}_{\capi{ij}},
\end{align}
for Eqs.~\eqref{eq:relation_VVVV2}, \eqref{eq:relation_VVVphi}, and \eqref{eq:relation_VVphiphi}, respectively, where the calligraphic index $\mathcal{K}$ collects both the vector $A$ and scalar $i$ indices.
By treating the vector and scalar indices collectively, we present the unified form of the relations above:
\begin{align}
[T^{\capi{(A)}},T^{\capi{(B)}}]_{\capi{\mathcal{I}\mathcal{J}}}=
iC_{\capi{ABC}}T^{\capi{(C)}}_{\capi{\mathcal{I}\mathcal{J}}},
\label{eq:lie_algebra_T}
\end{align}
with the indices $\mathcal{I},\mathcal{J}$ collecting the vector and scalar indices together. Together with Eq.~\eqref{eq:lie_algebra_T}, from the fermionic conditions in Eq.~\eqref{eq:relation_VpsiVpsi1}, we obtain
\begin{align}
[R^{\capi{A}},R^{\capi{B}}]=iC_{\capi{ABC}}R^{\capi{C}}
,\quad
[L^{\capi{A}},L^{\capi{B}}]=iC_{\capi{ABC}}L^{\capi{C}},
\label{eq:lie_algebra_RL}
\end{align}
for the $V\bar{\psi}\psi$ couplings.

Next, we introduce a coupling involving a vector and two fermion indices as 
\begin{gather}
(H_a)_{\mathrm{i}\mathrm{j}}
\equiv \frac{i}{m_a}(L^a \mu -\mu R^a)_{\mathrm{i}\mathrm{j}}
,\quad\;\;
(H_{\hat{a}})_{\mathrm{i}\mathrm{j}}
\equiv 0,
\label{eq:H_a}
\end{gather}

Using Eq.~\eqref{eq:H_a}, we recast the fermionic relations in Eqs.~\eqref{eq:relation_VpsiVpsi2} and \eqref{eq:relation_Vpsiphipsi} as
\begin{gather}
H_b R^a- L^a H_b- T^{(a)}_{bi}H_i -T^{(a)}_{bc}H_c=0,
\\
H_i R^a- L^a H_i- T^{(a)}_{ij}H_j -T^{(a)}_{ib}H_b=0,
\end{gather}
of which the unified form reads
\begin{align}
H_{\capi{\mathcal{I}}}R^{\capi{A}}
-L^{\capi{A}}H_{\capi{\mathcal{I}}}
-T^{\capi{A}}_{\capi{\mathcal{I}\mathcal{J}}}H_{\capi{\mathcal{J}}}=0,
\label{eq:lie_algebra_H}
\end{align}
with the indices $\mathcal{I,J}$ collecting both the vector and scalar indices.

\subsection{Stückelberg Formulation}

Using the local 3-pt interaction operators in Table~\ref{tab:3-pt} and referring to the 4-pt amplitudes in Section~\ref{sec:4-pt_tree_unitarity}, we construct the Lagrangian in terms of fields that are mass eigenstates as
\begingroup
\small
\begin{align}
\mathcal{L}&=-\frac{1}{4}
(\partial_{\mu} V_{\capi{A}\nu}-\partial_{\nu} V_{\capi{B}\mu}
-C_{\capi{ABC}} V_{\capi{B}\mu}V_{\capi{C}\mu})^2
+
\bar{\psi}_R i(\slash \!\!\! \partial +i \slash \!\!\! V_{\capi{A}}
R^{\capi{A}})\psi_R
+
\bar{\psi}_L i(\slash \!\!\! \partial +i \slash \!\!\! V_{\capi{A}}
L^{\capi{A}})\psi_L
\nonumber
\\
&\quad
+\frac{1}{2}V_{\capi{A}\mu}V_{\capi{B}}^\mu
\bigg[ m_{\capi{A}}^2\delta_{\capi{AB}}
+2F_{\capi{AB}i}\phi_i+
\bigg(\frac{1}{m_c^2}F_{\capi{A}ci}F_{\capi{B}cj}
+G_{\capi{A}ki}G_{\capi{B}kj}\bigg)
\phi_i \phi_j\bigg]
\nonumber
\\
&\quad +\frac{1}{2}(\partial_{\mu}\phi_i)^2
-G_{\capi{A}ij}V_{\capi{A}\mu}\partial^\mu\phi_i \phi_j
-\frac{1}{2}
\bigg[m_i^2\phi_i^2+\frac{1}{3}P_{ijk}\phi_i\phi_j\phi_k 
+\frac{1}{12}Q_{ijkl}\phi_i\phi_j\phi_k\phi_l\bigg]
\nonumber
\\
&\quad 
-\big[\bar{\psi}_L  (\mu+ H_i \phi_i) \psi_R +\mbox{h.c.}\big],  \qquad (\mu_{\mathrm{ij}}=m_{\mathrm{i}}\delta_{\mathrm{ij}})
\label{eq:mass_L}
\end{align}
\endgroup
with the kinetic term of each particle, where we defined $\slash \!\!\!\! A=\gamma^\mu A_{\mu}$ for a four-vector quantity. Here, we traced back the 4-pt contact $VVVV$ and $VV\phi\phi$ operators in Eq.~\eqref{eq:VVVV_s-ch} and Eq.~\eqref{eq:VVphiphi_amp}, respectively. In addition, we introduced the scalar quartic term with coupling $Q$, leading to no tree unitarity violation. In the following, we refer to Eq.~\eqref{eq:mass_L} as the mass-basis Lagrangian for simplicity. 

The Lie algebra in Section~\ref{sec:lie_algebra} allows us to reduce the structure of Eq.~\eqref{eq:mass_L}. To see this, let us first re-express Eq.~\eqref{eq:mass_L} by replacing the couplings $F$ and $G$ with the generators $T$ of the Lie algebra in Section~\ref{sec:lie_algebra} as
\begingroup
\small
\begin{align}
\mathcal{L}&=-\frac{1}{4}
(\partial_{\mu} V_{\capi{A}\nu}-\partial_{\nu} V_{\capi{B}\mu}
-C_{\capi{ABC}} V_{\capi{B}\mu}V_{\capi{C}\mu})^2
+
\bar{\psi}_R i(\slash \!\!\! \partial +i \slash \!\!\! V_{\capi{A}}
R^{\capi{A}})\psi_R
+
\bar{\psi}_L i(\slash \!\!\! \partial +i \slash \!\!\! V_{\capi{A}}
L^{\capi{A}})\psi_L
\nonumber
\\
&\quad
+\frac{1}{2}V_{\capi{A} \mu}V_{\capi{B}}^\mu
\bigg[m_{\capi{A}}^2\delta_{\capi{AB}}+m_{\capi{A}}(iT^{\capi{(B)}}_{\capi{A}\capmi{I}})\phi_{\capmi{I}}+
m_{\capi{B}}(iT^{\capi{(A)}}_{\capi{B}\capmi{I}})\phi_{\capmi{I}}+
(iT^{\capi{(A)}}_{\capmi{K}\capmi{I}})(iT^{\capi{(B)}}_{\capmi{K}\capmi{J}})
\phi_{\capmi{I}}\phi_{\capmi{J}}\bigg]
\nonumber
\\
&\quad +\frac{1}{2}(\partial_{\mu}\phi_{\capmi{I}})^2
+iV_{\capi{A}\mu}\partial^{\mu}\phi_{\capmi{I}}
T^{\capi{(A)}}_{\capmi{I}\capmi{J}}
\phi_{\capmi{J}}
-\mathcal{V}(\phi)
-\big[\bar{\psi}_L \mathcal{Y}(\phi) \psi_R + \mbox{h.c.}\big].
\label{eq:L_generator_rep1}
\end{align}
\endgroup
Here, we extend the scalar index as $\phi_i \to \phi_{\capmi{I}}$ with $\phi_{\capi{A}}=0$ for the vector index $A$. Here, the scalar potential $\mathcal{V}(\phi)$ and Yukawa term $\mathcal{Y}(\phi)$ are
\begin{align}
\mathcal{V}(\phi)&=
\frac12 M_{\capmi{IJ}}^2\phi_{\capmi{I}}\phi_{\capmi{I}}+\frac{1}{6}P_{\capmi{IJK}}\phi_{\capmi{I}}\phi_{\capmi{J}}\phi_{\capmi{K}}
+\frac{1}{24}Q_{\capmi{IJKL}}\phi_{\capmi{I}}\phi_{\capmi{J}}\phi_{\capmi{K}}\phi_{\capmi{L}},
\label{eq:first_V}
\\
\mathcal{Y}(\phi)&=\mu+H_{\capmi{I}}\phi_{\capmi{I}},
\label{eq:first_Y}
\end{align}
with $M^2_{ij}=m_i^2$, where the couplings $M^2$, $P$, $Q$, and $H$ are extended to include the vector index. Since $\phi_{\capi{A}}=0$ forbids their contributions, Eq.~\eqref{eq:L_generator_rep1} reduces to the original form in Eq.~\eqref{eq:mass_L}. However, we will determine the explicit structures of the couplings with vector indices, which are determined by tree unitarity. Note that we have already defined the fermionic coupling $H_{\capi{A}}$ in Eq.~\eqref{eq:H_a} to derive the Lie algebra in Eq.~\eqref{eq:lie_algebra_H}.

Next, let us define
\begin{gather}
\lambda^{\capi{(A)}}_{\capmi{I}}
\equiv \delta_{\capmi{I}\capi{A}}m_{\capi{A}},
\label{eq:lambda_def}
\end{gather}
where the constant vector $\lambda^{\capi{(A)}}$ vanishes for massless vector indices. Inserting \eqref{eq:lambda_def} into Eq.~\eqref{eq:L_generator_rep1}, we reduce the Lagrangian as
\begingroup
\small
\begin{align}
\mathcal{L}&=-\frac{1}{4}
(\partial_{\mu} V_{\capi{A}\nu}-\partial_{\nu} V_{\capi{A}\mu}
-C_{\capi{ABC}} V_{\capi{B}\mu}V_{\capi{C}\mu})^2
+
\bar{\psi}_R i(\slash \!\!\! \partial +i \slash \!\!\! V_{\capi{A}}
R^{\capi{A}})\psi_R
+
\bar{\psi}_L i(\slash \!\!\! \partial +i \slash \!\!\! V_{\capi{A}}
L^{\capi{A}})\psi_L
\nonumber
\\
&\quad
+\frac{1}{2}V_{\capi{A} \mu}V_{\capi{B}}^\mu
\bigg\{
\big[(iT^{\capi{(A)}} \phi )+\lambda^{\capi{(A)}}\big]_{\capmi{I}}
\big[(iT^{\capi{(B)}} \phi )+\lambda^{\capi{(B)}}\big]_{\capmi{I}}
\bigg\}
\nonumber
\\
&\quad +\frac{1}{2}(\partial_{\mu}\phi_{\capmi{I}})^2
+iV_{\capi{A}\mu}\partial^{\mu}\phi_{\capmi{I}}
T^{\capi{(A)}}_{\capmi{IJ}}
\phi_{\capmi{J}}
-\mathcal{V}(\phi)
-\big[\bar{\psi}_L \mathcal{Y}(\phi) \psi_R + \mbox{h.c.}\big].
\label{eq:L_generator_rep2}
\end{align}
\endgroup

For the further analysis of the constant vector $\lambda^{\capi{(A)}}$, using Eqs.~\eqref{eq:C_to_T} and~\eqref{eq:F_to_T}, we derive the following relation:
\begin{gather}
T^{\capi{(a)}}_{cb} m_{b}-
T^{\capi{(b)}}_{ca} m_{a}
=iC_{abc}m_c,\quad 
T^{\capi{(a)}}_{ib} m_{b}-
T^{\capi{(b)}}_{ia} m_{a}
=0
\nonumber
\\
\rightarrow \;\; T^{\capi{(A)}}_{\capmi{I}\capi{B}} m_{\capi{B}}-
T^{\capi{(B)}}_{\capmi{I}\capi{A}} m_{\capi{A}}
=iC_{\capi{ABC}}\delta_{\capmi{I}\capi{C}}m_{\capi{C}}.
\label{eq:vector_scalar_index_relation1}
\end{gather}
Using Eq.~\eqref{eq:lambda_def}, we recast Eq.~\eqref{eq:vector_scalar_index_relation1} as
\begin{align}
T^{\capi{(A)}} \lambda^{\capi{(B)}}
-
T^{\capi{(B)}} \lambda^{\capi{(A)}}
=iC_{\capi{ABC}}\lambda^{\capi{(C)}}.
\label{eq:vector_scalar_index_relation2}
\end{align}
Defining a constant vector $v$ satisfying 
\begin{align}
(iT^{\capi{(A)}}v)_{\capmi{I}}=\lam^{\capi{(A)}}_{\capmi{I}}=
\delta_{\capmi{I}\capi{A}}m_{\capi{A}},
\label{eq:iTv=lam}
\end{align}
one can observe that Eq.~\eqref{eq:vector_scalar_index_relation2} can be rewritten as
\begin{align}
([T^{\capi{A}},T^{\capi{B}}]-iC_{\capi{ABC}}T^{\capmi{C}})v=0,
\end{align}
which is valid due to the Lie algebra in Eq.~\eqref{eq:lie_algebra_T}. The constant vector $v$ does not involve nonzero vector indices:
\begin{align}
v_{\capmi{I}}(iT^{\capi{(A)}}_{\capmi{IJ}})v_{\capmi{J}}
=v_{\capmi{J}}(iT^{\capi{(A)}}_{\capmi{JI}})v_{\capmi{I}}
=-v_{\capmi{I}}(iT^{\capi{(A)}}_{\capmi{IJ}})v_{\capmi{J}}
=v_{\capi{A}}m_{\capi{A}}=0\quad \to \quad v_{\capi{A}}=0,
\end{align}
due to the antisymmetry of generators $T^{\capi{(A)}}$ (see Section~\ref{sec:lie_algebra}). Using Eq.~\eqref{eq:lambda_def}, we find the explicit expression of $v$:
\begin{align}
-\Big[\sum_{a}T^{(a)}T^{(a)}v\Big]_{\capmi{I}}
=\sum_{a}(iT^{a}_{\capmi{I}a})m_{a}
\;\;\; \to \;\;\; 
v_{\capmi{I}}
=-\Big(\sum_{b}T^{(b)}T^{(b)}\Big)^{-1}_{\capi{IJ}}
\sum_{a}(iT^{a}_{\capmi{J}a})m_{a},
\label{eq:explicit_v}
\end{align}
where the operator $\sum_{a}T^{(a)} T^{(a)}$ with the sum taken only over massive vector indices is invertible on the subspace spanned by $T^{(a)}v$, reducing to a nonzero multiple of the identity in each irreducible component.

For several Abelian vector indices $A$ and $B$, we may have $T^{\capi{(A)}}\lam^{\capi{(B)}}-T^{\capi{(B)}}\lam^{\capi{(A)}}=0$ due to $C_{\capi{ABC}}=0$. However, this does not generally imply that $\lambda^{\capi{(A)}}$ and $\lambda^{\capi{(B)}}$ vanish. Let the bolded symbol $\bs{a}$ denote the massive Abelian vector index. Since the vectors $iT^{(\bs{a})}v$ span a vector subspace, any $\lambda^{(\bs{a})}$ can be decomposed into components parallel and orthogonal to this subspace. Thus, after an appropriate redefinition of the Abelian basis within the parallel subspace, we can write
\begin{align}
\lambda^{(\bs{a})}=iT^{(\bs{a})}v+\lam_{\bot}^{(\bs{a})}
\quad \mbox{with}\quad
(\lam_{\bot}^{(\bs{a})})_{\capmi{I}}(T^{(\bs{b})}v)_{\capmi{I}}=0,
\end{align}
for an arbitrary Abelian index $\bs{b}$.

Thus, extracting such vectors with Abelian vector indices from the second line of Eq.~\eqref{eq:L_generator_rep2}, we obtain
\begin{align}
&\frac{1}{2}V_{\capi{A} \mu}V_{\capi{B}}^\mu
\bigg\{
\big[(iT^{\capi{(A)}} \phi )+\lambda^{\capi{(A)}}\big]_{\capmi{I}}
\big[(iT^{\capi{(B)}} \phi )+\lambda^{\capi{(B)}}\big]_{\capmi{I}}
\bigg\}
\nonumber
\\
&=
\frac{1}{2}V_{\capi{A} \mu}V_{\capi{B}}^\mu
\bigg[
(iT^{\capi{(A)}}_{\capmi{KI}})(iT^{\capi{(B)}}_{\capmi{KJ}})
(\phi+v)_{\capmi{I}}(\phi+v)_{\capmi{J}}\bigg]
+\frac{1}{2}
m_{\perp \bs{a}}^2
V_{\bs{a} \mu}^2,
\label{eq:second_line}
\end{align}
with $(\lam^{(\bs{a})}_{\bot})_{ \capmi{I}}(\lam^{(\bs{b})}_{\bot})_{ \capmi{I}}
=m_{\perp\bs{a}}^2 \delta_{\bs{ab}}$\footnote{In general, the vectors $\lambda_{\bot}^{(\bs{a})}$ need not remain mutually orthogonal. Nevertheless, since their inner-product matrix is real and symmetric, one may diagonalize it through an appropriate orthogonal transformation acting only on the Abelian index space. Thus, we assume the orthogonality of $\lambda_{\bot}^{(\bs{a})}$ for simplicity.}.

Next, we restore the gauge symmetry using the Stückelberg formulation~\cite{Stueckelberg:1938hvi,Ruegg:2003ps}. As the first step, we define 
\begin{align}
V_{\capi{A}\mu} (iT^{\capi{(A)}}_{\capmi{IJ}})
&\equiv \big[e^{-i\sigma \cdot T}
W_{\capi{A}\mu}(iT^{\capi{(A)}})e^{i\sigma \cdot T}
+e^{-i\sigma \cdot T}\partial_{\mu}e^{i\sigma \cdot T}
\big]_{\capmi{IJ}},
\nonumber
\\
\psi_R&\equiv e^{-i\sigma\cdot R}q_R,
\nonumber
\\
\psi_L&\equiv e^{-i\sigma\cdot L}q_L,
\label{eq:stuckelberg_VF}
\end{align}
for the vector and fermion fields, where the Stückelberg fields $\sigma_A$ vanish for massless vector indices $\hat{a}$. Here, for Abelian vectors, Eq.~\eqref{eq:stuckelberg_VF} leads to
\begin{align}
V_{\capi{A}\mu}\equiv W_{\capi{A}\mu}+\partial_\mu \sigma_{\capi{A}}.
\label{eq:abelian_VtoW}
\end{align}
Then, the Lagrangian in Eq.~\eqref{eq:L_generator_rep2} is invariant under the following gauge transformations:
\begin{align}
W_{\capi{A}\mu} (iT^{\capi{(A)}}_{\capmi{IJ}})
&\;\;\rightarrow \;\; \big[e^{i\Lam \cdot T}
W_{\capi{A}\mu}(iT^{\capi{(A)}})e^{-i\Lam \cdot T}
+e^{i\sigma \cdot T}\partial_{\mu}e^{-i\Lam \cdot T}
\big]_{\capmi{IJ}}, \quad \mbox{(non-Abelian)}
\nonumber
\\
W_{\capi{A}\mu}& \;\; \rightarrow \;\; 
W_{\capi{A}\mu}+\partial_{\mu}\Lam_{\capi{A}}, 
\qquad\qquad\qquad\qquad\qquad\qquad\qquad \mbox{(Abelian)}
\nonumber
\\
q_R&\;\;\rightarrow \;\; e^{i\Lam\cdot R}q_R,
\nonumber
\\
q_L&\;\;\rightarrow \;\; e^{i\Lam\cdot L}q_L,
\label{eq:gauge_physical}
\end{align}
for the physical fields and
\begin{align}
e^{i\sigma\cdot T}&\;\;\rightarrow \;\;
e^{i\Lam\cdot T}e^{i\sigma\cdot T}, \quad \mbox{(non-Abelian)}
\nonumber
\\
\sigma_{\capi{A}}&\;\;\rightarrow \;\; \sigma_{\capi{A}}+\Lam_{\capi{A}},
\quad\quad \mbox{(Abelian)}
\label{eq:gauge_stuckel}
\end{align}
for the Stückelberg fields, respectively. Gauge invariance is ensured by the Lie algebra in Eqs.~\eqref{eq:lie_algebra_T}, \eqref{eq:lie_algebra_RL}, and~\eqref{eq:lie_algebra_H}, which are collections of coupling conditions derived from the tree unitarity of 4-pt amplitudes.

Using the Lie algebra in Eqs.~\eqref{eq:lie_algebra_T} and~\eqref{eq:lie_algebra_RL}, we observe that the first line of Eq.~\eqref{eq:L_generator_rep2} does not involve the Stückelberg fields:
\begin{align}
&-\frac{1}{4}
(\partial_{\mu} V_{\capi{A}\nu}-\partial_{\nu} V_{\capi{A}\mu}
-C_{\capi{ABC}} V_{\capi{B}\mu}V_{\capi{C}\mu})^2
+
\bar{\psi}_R i(\slash \!\!\! \partial +i \slash \!\!\! V_{\capi{A}}
R^{\capi{A}})\psi_R
+
\bar{\psi}_L i(\slash \!\!\! \partial +i \slash \!\!\! V_{\capi{A}}
L^{\capi{A}})\psi_L
\nonumber
\\
&=
-\frac{1}{4}
(\partial_{\mu} W_{\capi{A}\nu}-\partial_{\nu} W_{\capi{A}\mu}
-C_{\capi{ABC}} W_{\capi{B}\mu}W_{\capi{C}\mu})^2
+
\bar{q}_R i(\slash \!\!\! \partial +i \slash \!\!\! W_{\capi{A}}
R^{\capi{A}})q_R
+
\bar{q}_L i(\slash \!\!\! \partial +i \slash \!\!\! W_{\capi{A}}
L^{\capi{A}})q_L,
\end{align}
in terms of the definitions in Eq.~\eqref{eq:stuckelberg_VF}. Using Eq.~\eqref{eq:stuckelberg_VF}, we recast the first term of Eq.~\eqref{eq:second_line} as
\begin{align}
&\frac12 V_{\capi{A}\mu} V_{\capi{B}\mu} 
(iT^{\capi{(A)}}_{\capmi{KI}})
(iT^{\capi{(B)}}_{\capmi{KJ}})(v+\phi)_{\capmi{I}}
(v+\phi)_{\capmi{J}}
\nonumber
\\
&=\frac12 
\Big\{(\partial_{\mu}+iW_{\capi{A}\mu} T^{\capi{(A)}})
\big[e^{i\sigma \cdot T}(v+\phi)\big]
\Big\}_{\capmi{I}}
\Big\{(\partial^{\mu}+iW^{\mu}_{\capi{B}} T^{\capi{(B)}})
\big[e^{i\sigma \cdot T}(v+\phi)\big]
\Big\}_{\capmi{I}}
\nonumber
\\
&-\frac12 \partial_{\mu}\phi_{\capmi{I}} \partial^{\mu}\phi_{\capmi{I}}
-\partial_{\mu}\phi_{\capmi{I}} 
\Big\{\big[e^{-i\sigma \cdot T}W^{\mu}_{\capi{B}} (iT^{\capi{(B)}})
e^{i\sigma \cdot T}+e^{-i\sigma \cdot T} \partial^{\mu}e^{i\sigma \cdot T}\big](v+\phi)\Big\}_{\capmi{I}}
\label{eq:WW_varphivarphi_int}
\end{align}
On the other hand, using Eq.~\eqref{eq:abelian_VtoW}, we re-express the second term in Eq.~\eqref{eq:second_line} as
\begin{align}
\frac{1}{2}
m_{\bot\bs{a}}^2
V_{\bs{a} \mu}^2
\equiv
\frac{1}{2}
m_{\bot\bs{a}}^2
(W_{\bs{a}\mu}+\partial_{\mu}\sigma_{\bs{a}})^2,
\label{eq:abel_mass_term}
\end{align}
Additionally, for the second term in the third line of Eq.~\eqref{eq:L_generator_rep2}, using Eq.~\eqref{eq:stuckelberg_VF}, we obtain
\begin{align}
&V_{\capi{A}\mu}\partial^{\mu}\phi_{\capmi{I}}
(iT^{\capi{(A)}}_{\capmi{IJ}})
\phi_{\capmi{J}}
\nonumber
\\
&=
\partial^{\mu}\phi_{\capmi{I}}
\big[e^{-i\Lam \cdot T}
W_{\capi{A}\mu}(iT^{\capi{(A)}})
e^{i\sigma \cdot T}
+
e^{-i\Lam \cdot T}\partial_{\mu}e^{i\Lam \cdot T}
\big]_{\capmi{IJ}}
\phi_{\capmi{J}}.
\label{eq:W_varphivarphi_int}
\end{align}
Summing Eqs.~\eqref{eq:WW_varphivarphi_int} and \eqref{eq:W_varphivarphi_int}, we have
\begin{align}
&\frac12 V_{\capi{A}\mu} V_{\capi{B}\mu} 
(iT^{\capi{(A)}}_{\capmi{KI}})
(iT^{\capi{(B)}}_{\capmi{KJ}})(v+\phi)_{\capmi{I}}
(v+\phi)_{\capmi{J}}
+
iV_{\capi{A}\mu}\partial^{\mu}\phi_{\capmi{I}}
T^{\capi{(A)}}_{\capmi{IJ}}
\phi_{\capmi{J}}
\nonumber
\\
&\equiv \frac12 
\Big\{(\partial_{\mu}+iW_{\capi{A}\mu} T^{\capi{(A)}})
\big[e^{i\sigma \cdot T}(v+\phi)\big]
\Big\}_{\capmi{I}}
\Big\{(\partial^{\mu}+iW^{\mu}_{\capi{B}} T^{\capi{(B)}})
\big[e^{i\sigma \cdot T}(v+\phi)\big]
\Big\}_{\capmi{I}}
\nonumber
\\
&-\frac12 \partial_{\mu}\phi_{\capmi{I}} \partial^{\mu}\phi_{\capmi{I}}
-\partial^{\mu}\phi_{\capmi{I}}
\big[e^{-i\sigma \cdot T}(\partial_\mu +iW_{\capi{A}\mu}T^{\capi{(A)}})(e^{i\sigma \cdot T}v)\big]_{\capmi{I}}.
\label{eq:WWpp_p_Wpp_int}
\end{align}
Using Eq.~\eqref{eq:stuckelberg_VF}, we verify that the last term in Eq.~\eqref{eq:WWpp_p_Wpp_int} vanishes:
\begin{align}
&-\partial^{\mu}\phi_{\capmi{I}}
\big[\big(e^{-i\sigma \cdot T}
iW_{\capi{A}\mu}T^{\capi{(A)}}e^{i\sigma \cdot T}
+e^{-i\sigma \cdot T}\partial_{\mu}e^{i\sigma \cdot T}
\big)v\big]_{\capmi{I}}
\nonumber
\\
&= -\partial^{\mu}\phi_{\capmi{I}}
V_{\capi{A}\mu}(iT^{\capi{(A)}}v)_{\capmi{I}}
=
-\partial^{\mu}\phi_{\capmi{I}}
V_{\capi{A}\mu}\lam^{\capi{(A)}}_{\capmi{I}}
=
-\partial^{\mu}\phi_i V_{\capi{A}\mu}
\delta_{i\capi{A}}m_{\capi{A}}=0,
\label{eq:vanishing_term}
\end{align}
due to $\delta_{\capi{A}i}=0$ with the vector and scalar indices.

Next, with the Stückelberg fields, we re-express the fermionic term in the last line of Eq.~\eqref{eq:L_generator_rep2}. As the first step, we decompose the fermionic mass matrix $\mu=\mu_0+\tilde{\mu}$, so that the Yukawa term $\mathcal{Y}(\phi)$ in Eq.~\eqref{eq:L_generator_rep2} becomes
\begin{align}
\mathcal{Y}(\phi)= \mu_0+\tilde{\mu} +H_{\capmi{I}}\phi_{\capmi{I}},
\label{eq:Y_mu0}
\end{align}
where the matrix $\mu_0$ satisfies
\begin{align}
L\mu_0-\mu_0R=0.
\label{eq:mu0}
\end{align}
Note that Eq.~\eqref{eq:mu0} enables us to recast Eq.~\eqref{eq:H_a} as
\begin{align}
(H_a)_{\mathrm{i}\mathrm{j}}
\equiv \frac{i}{m_a}(L^a \tilde{\mu} -\tilde{\mu} R^a)_{\mathrm{i}\mathrm{j}},
\label{eq:H_a2}
\end{align}
in terms of $\tilde{\mu}$. Then, with a continuous parameter $x$, we define the following quantity,
\begin{align}
J(x)\equiv e^{xi\sigma \cdot L}(\tilde{\mu}+H_{\capmi{I}} \phi_{\capmi{I}})e^{-xi\sigma \cdot R}.
\label{eq:J_explicit}
\end{align}

To investigate the coefficients of $x^n$ with $n\geq 0$, we expand $J(x)$ as 
\begin{align}
J(x)= \sum_{n=0}^{\infty} \frac{J_n}{n !} x^n.
\label{eq:J_expand}
\end{align}
Using Eq.~\eqref{eq:J_explicit} and Eq.~\eqref{eq:J_expand}, we compute the first derivative of $J(x)$:
\begin{align}
\frac{dJ(x)}{dx}&= -i\sigma_a\big(J(x)R^{a}-L^{a}J(x) \big)
=\sum_{n=0}^\infty \frac{-i\sigma_a\big(J_n R^{a}-L^{a}J_n \big)}{n!}x^n
=\sum_{n=0}^\infty \frac{J_{n+1}}{n!}x^n,
\end{align}
leading to the recursion relation,
\begin{align}
J_{n+1}=
-i\sigma_a\big(J_n R^{a}-L^{a}J_n \big).
\label{eq:J_recursion}
\end{align}
Extracting the coefficient $J_0=\tilde{\mu}+H_{\capmi{I}}\phi_{\capmi{I}}$ in Eq.~\eqref{eq:J_explicit} and using Eq.~\eqref{eq:H_a2} and Eq.~\eqref{eq:J_recursion}, we obtain the following iteration:
\begin{align}
J_{1}&=H_am_a\sigma_a 
-i\sigma_a (H_{\capmi{I}}R^a-L^a H_{\capmi{I}})\phi_{\capmi{I}}=
H_{\capmi{I}}\big[(i\sigma \cdot T)(v+\phi)\big]_{\capmi{I}},
\nonumber
\\
J_{2}&=H_{\capmi{I}}\big[(i\sigma \cdot T)^2(v+\phi)\big]_{\capmi{I}},
\nonumber
\\
\vdots&
\label{eq:J_iteration}
\end{align}
where
$H_a m_a \sigma_a=H_{\capmi{I}}\delta_{\capmi{I}\capi{A}}m_{\capi{A}}\sigma_{\capi{A}}=H_{\capmi{I}}\big[(i\sigma \cdot T)v\big]_{\capmi{I}}$ (see Eq.~\eqref{eq:lambda_def}). Taking $x=1$ in Eq.~\eqref{eq:J_expand} and using Eq.~\eqref{eq:J_iteration}, we obtain
\begin{align}
J(1)=
e^{i\sigma \cdot L}(\tilde{\mu}+H_{\capmi{I}}\phi_{\capmi{I}})
e^{-i\sigma \cdot R}&
=
H_{\capmi{I}} \big[e^{i\sigma \cdot T}(v+\phi)\big]_{\capmi{I}}.
\label{eq:fermion_yukawa}
\end{align}
Then, using Eq.~\eqref{eq:fermion_yukawa} together with Eqs.~\eqref{eq:Y_mu0} and~\eqref{eq:mu0}, we find 
\begin{align}
e^{i\sigma \cdot L}\mathcal{Y}(\phi)
e^{-i\sigma \cdot R}
=\mu_0 +H_{\capmi{I}} \big[e^{i\sigma \cdot T}(v+\phi)\big]_{\capmi{I}}.
\end{align}

Next, we perform a point transformation of scalar fields as
\begin{gather}
\pi \equiv e^{i\sigma \cdot T}(v+\phi)-v,
\label{eq:pi_to_phi_sigma}
\end{gather}
which combines the degrees of freedom of the physical and St\"uckelberg scalars. For later convenience, we compute
\begin{align}
\phi_i&=[e^{-i\sigma \cdot T}(\pi +v)-v]_i=\pi_i+\mathcal{O}(\pi^2),
\nonumber
\\
m_{\capi{A}}\sigma_{\capi{A}}&=\pi_{\capi{A}}+\mathcal{O}(\pi^2),
\label{eq:phi_sigma_to_pi}
\end{align}
from the definition  in Eq.~\eqref{eq:pi_to_phi_sigma}, where one can verify that all higher-order terms involve the generators $T^{\capi{(A)}}$.

Using Eqs.~\eqref{eq:abel_mass_term} and~\eqref{eq:WWpp_p_Wpp_int}, and Eq.~\eqref{eq:fermion_yukawa}, the definition in Eq.~\eqref{eq:pi_to_phi_sigma} allows us to recast the Lagrangian in Eq.~\eqref{eq:L_generator_rep2} with the scalar fields $\pi_{\capmi{I}}$ as
\begingroup
\small
\begin{align}
\mathcal{L}&=-\frac{1}{4}
(\partial_{\mu} W_{\capi{A}\nu}-\partial_{\nu} W_{\capi{A}\mu}
-C_{\capi{ABC}} W_{\capi{B}\mu}W_{\capi{C}\mu})^2
+
\bar{q}_R i(\slash \!\!\! \partial +i \slash \!\!\! W_{\capi{A}}
R^{\capi{A}})q_R
+
\bar{q}_L i(\slash \!\!\! \partial +i \slash \!\!\! W_{\capi{A}}
L^{\capi{A}})q_L
\nonumber
\\
&\quad
+\frac12 
\Big\{(\partial_{\mu}+iW_{\capi{A}\mu} T^{\capi{(A)}})
(v+\pi)
\Big\}^2
+
\frac{1}{2}
m_{\bot\bs{a}}^2
(W_{\bs{a}\mu}+\partial_{\mu}\sigma_{\bs{a}})^2
-\tilde{\mathcal{V}}(\pi)
\nonumber
\\
&\quad
-\big\{\bar{q}_L \tilde{\mathcal{Y}}(\pi)
q_R + \mbox{h.c.}\big\}-\frac{1}{2\xi}(\partial^\mu W_{\capi{A}\mu}-\xi m_{\capi{A}} \pi_{\capi{A}})^2.
\label{eq:stuckel_L}
\end{align}
\endgroup
where we have employed an $R_\xi$-type gauge defined in terms of $\pi_A$ for later convenience\footnote{This gauge-fixing is admissible since $\pi_A$ parametrizes the same Goldstone directions as $\sigma_A$, and thus provides an equivalent description.}. Here, the scalar potential and Yukawa terms are given by
\begin{align}
\tilde{\mathcal{V}}(\pi)&\equiv \mathcal{V}
\big[e^{-i\sigma(\pi)\cdot T}(v+\pi)-v\big],
\\
\tilde{\mathcal{Y}}(\pi)&\equiv \mu_0+H_{\capmi{I}} (v+\pi)_{\capmi{I}}.
\end{align}
Note that the scalar potential $\tilde{\mathcal{V}}(\pi)$ generally contains higher-pt contact interaction terms. In the following, we refer to Eq.~\eqref{eq:stuckel_L} as the Stückelberg Lagrangian, and introduce the Lagrangian notations $\mathcal{L}(V,\psi_R,\psi_L,\phi)$ in Eq.~\eqref{eq:mass_L} and $\mathcal{L}_S(W,q_R,q_L,\pi)$ in Eq.~\eqref{eq:stuckel_L}, which denote the mass-basis and Stückelberg Lagrangians, respectively. Here, we do not denote the St\"uckelberg fields $\sigma_{\bs{a}}$ for the Abelian massive vectors because they are eliminated by the gauge-fixing term.

\subsection{Tree unitarity of higher-pt amplitudes}
\label{sec:tree_unitarity_higher-pt}

In this section, after showing the equivalence between the tree-level amplitudes arising from the Lagrangians, $\mathcal{L}(V,\psi_R,\psi_L,\phi) $ and $\mathcal{L}_S(W,q_R,q_L,\pi)$, we examine the high-energy behavior of the tree-level amplitudes for $\mathcal{L}_S(W,q_R,q_L,\pi)$. 

As the first step, we define the amputated Green's function as
\begin{align}
&G_{\rm amp}[V_{\capi{A}\mu}(p_1)\cdots \psi_{\mathrm{i}\alpha}(p_2)
\cdots \phi_{i}(p_3)] (2\pi)^4 \delta(p_1+p_2+p_3+\cdots)
\nonumber
\\
&=\int dx_1dx_2dx_3\cdots \,e^{-i(p_1\cdot x_1+p_2\cdot x_2+p_3\cdot x_3+\cdots)}
\nonumber
\\
&\times\;
\langle (p_1^2-m_{\capi{A}}^2)V_{\capi{A}\mu}(x_1)
\cdots [(\slash\!\!\! p_2-m_{\mathrm{i}})\psi_{\mathrm{i}}(x_2)]_{\alpha}
\cdots (p_3^2-m_{i}^2)\phi_{i}(x_3)\cdots
\rangle,
\end{align}
in terms of the correlation function $\langle \cdots \rangle$. For the amputated Green's function $G_{\rm amp(tree)}$ including only tree-level diagrams, by taking the on-shell momenta for external particles, we obtain
\begin{gather}
G_{\rm amp(tree)}[V_{\capi{A}\mu}(p_1)\cdots \psi_{\mathrm{i}}(p_2)
\cdots \phi_{i}(p_3)\cdots]
\;\overset{p^2 \to m^2}{\longrightarrow}
\mathcal{G}[V_{\capi{A}\mu}(p_1)\cdots \psi_{\mathrm{i}}(p_2)
\cdots \phi_{i}(p_3)\cdots].
\label{eq:G_tree}
\end{gather}
Then, multiplying Eq.~\eqref{eq:G_tree} by the wave functions of external particles, we can construct the tree-level amplitude as
\begin{gather}
\mathcal{M}=
\mathcal{G}[\eps_{\capi{A}} \!\cdot \!V_{\capi{A}}(p_1)\cdots \bar{u}_{\mathrm{i}} \psi_{\mathrm{i}}(p_2)
\cdots \phi_{i}(p_3) \cdots ].
\end{gather}
In the mass basis, the propagators of each field are given by
\begin{gather}
\langle \phi_{i} \phi_{j}\rangle
=\frac{i \delta_{ij}}{p^2-m_i^2}
,\quad 
\langle \psi_{\mathrm{i}} \bar{\psi}_{\mathrm{j}}\rangle
=\frac{i(\slash \!\!\! p+m_{\mathrm{i}})}{p^2-m_{\rm i}^2}\delta_{\mathrm{ij}},
\\
\langle V_{\hat{a}\mu} V_{\hat{b}\nu}\rangle=\frac{i(-g_{\mu\nu})}{p^2}\delta_{\hat{a}\hat{b}}
,\quad 
\langle V_{a\mu} V_{b\nu}\rangle=\frac{i\big(-g_{\mu\nu}+\frac{p_{\mu}p_{\nu}}{m_a^2}\big)}{p^2-m^2_a}\delta_{ab},
\end{gather}
we use a compact notation in which propagators are denoted by angle brackets. Here, the Feynman and unitary gauges are employed for massless and massive vectors, respectively.

To see how the amputated Green's function in Eq.~\eqref{eq:G_tree} is related to that for the contents $(W,q_R,q_L,\pi)$, we first compute
\begin{gather}
e^{-i\sigma \cdot T}(iT^{\capi{A}})e^{i\sigma \cdot T}
=(iT^{\capi{A}})\Omega_{\capi{BA}}(\sigma)
,\quad e^{-i\sigma \cdot T}\partial_{\mu}e^{i\sigma \cdot T}=(iT^{\capi{A}})\Omega'_{\capi{BA}}(\sigma),
\nonumber
\\
\mbox{with} \quad
\Omega(\sigma)=e^{-i\sigma \cdot t},\quad 
\Omega'(\sigma)=\frac{e^{-i\sigma \cdot t}-1}{-i\sigma \cdot t},
\end{gather}
where we have introduced the generators $t^{\capi{A}}_{\capi{BC}}=-iC_{\capi{ABC}}$ in the adjoint representation of the group $K$. Then, the St\"uckelberg formulation of vectors in Eq.~\eqref{eq:stuckelberg_VF} can be written as
\begin{gather}
V_{\capi{A}\mu} 
= \Omega_{\capi{AB}}(\sigma)W_{\capi{B}\mu}+\Omega'_{\capi{AB}}(\sigma)\frac{\partial_{\mu}\tilde{\sigma}_{\capi{B}}}{m_{\capi{B}}},
\label{eq:V_Omega_W}
\end{gather}
with the dimensionful St\"uckelberg field $\tilde{\sigma}_{\capi{A}}=m_{\capi{A}}\sigma_{\capi{A}}$.

Next, let us take the following replacements with a parameter $y$:
\begin{gather}
t^{\capi{A}},T^{\capi{(A)}},R^{\capi{(A)}},L^{\capi{(A)}}
\;\to \; y \,t^{\capi{A}},y\,T^{\capi{(A)}},y\,R^{\capi{(A)}},y\,L^{\capi{(A)}},
\label{eq:generators_y}
\end{gather}
which leave the Lie algebra unchanged in Section~\ref{sec:lie_algebra}.
Inserting Eq.~\eqref{eq:generators_y} into Eq.~\eqref{eq:stuckelberg_VF} and Eq.~\eqref{eq:V_Omega_W} and using Eq.~\eqref{eq:phi_sigma_to_pi}, we obtain 
\begin{align}
\mathcal{G}[V_{\capi{A}}\cdots \psi_{R\mathrm{i}}
\cdots \phi_{i} \cdots ]
&=
\bigg[\bigg(\Omega_{\capi{AB}}(y\sigma)W_{\capi{B}}+\Omega'_{\capi{AB}}(y\sigma)\frac{\partial \tilde{\sigma}_{\capi{B}}}{m_{\capi{B}}}\bigg)\cdots (e^{yi\sigma \cdot R}q_{R})_{\mathrm{i}}
\cdots \phi_{i} \cdots \bigg]
\nonumber
\\
&=
\mathcal{G}
\bigg[\bigg(W_{\capi{A}}+\frac{\partial \pi_{\capi{A}}}{m_{\capi{A}}}\bigg)\cdots q_{R\rm i}
\cdots \phi_{i} \cdots \bigg]+\mathcal{O}(y),
\label{eq:y_expansion}
\end{align}
in terms of the tree-level amputated Green's function. The linear independence on the parameter $y$ ensures that the first term of the RHS in Eq.~\eqref{eq:y_expansion} is equivalent to the term of the LHS:
\begin{gather}
\mathcal{G}[V_{\capi{A}}\cdots \psi_{R\mathrm{i}}
\cdots \phi_{i} \cdots ]
=
\mathcal{G}\bigg[\bigg(W_{\capi{A}}+\frac{\partial \pi_{\capi{A}}}{m_{\capi{A}}}\bigg)\cdots q_{R\rm i}
\cdots \pi_{i} \cdots \bigg].
\end{gather}

In the St\"uckelberg Lagrangian in Eq.~\eqref{eq:stuckel_L}, we take the gauge parameter as $\xi=1$, leading to the following propagators:
\begin{gather}
\langle \pi_{i} \pi_{j}\rangle
=\frac{i \delta_{ij}}{p^2-m_i^2}
,\quad 
\langle q_{\mathrm{i}} \bar{q}_{\mathrm{j}}\rangle
=\frac{i(\slash \!\!\! p+m_{\mathrm{i}})}{p^2-m_{\rm i}^2}\delta_{\mathrm{ij}},
\nonumber
\\
\langle W_{\hat{a}\mu} W_{\hat{b}\nu}\rangle=\frac{i(-g_{\mu\nu})}{p^2}\delta_{\hat{a}\hat{b}},\quad
\ab{W_{a\mu}W_{b\nu}}=
\frac{i(-g_{\mu\nu})}{p^2-m^2_a}\delta_{ab}
,\quad
\ab{\pi_{a}\pi_{b}}=
\frac{i}{p^2-m^2_a}\delta_{ab},
\end{gather}
in terms of the particle contents $(W,q_R,q_L,\pi)$. As shown in Eq.~\eqref{eq:y_expansion}, due to the equivalence between the amputated tree-level Green's functions for the mass-basis and St\"uckelberg Lagrangians, we derive the following tree unitarity condition on the amplitude with the contents $(W,q_R,q_L,\pi)$:
\begin{gather}
\mathcal{G} 
\bigg[\eps_{a}\cdot \bigg(W_{a}+\frac{\partial \pi_{a}}{m_{a}}\bigg)\cdots \bar{u}_{\rm i} \, q_{R\rm i}
\cdots \phi_{i} \cdots \bigg]\leq E^{4-n}.
\label{eq:tree_unitarity_G}
\end{gather}
We note that the function $\mathcal{G}$ is proportional to the factor $-g_{\mu\nu}+p_{a\mu}p_{a\nu}/m_{a}^2$, leading to the on-shell condition:
\begin{gather}
\mathcal{G}
\bigg[\bigg(W_{a}+\frac{\partial \pi}{m_{a}}\bigg)\cdots  q_{R\rm i}
\cdots \phi_{i} \cdots \bigg]\propto 
\bigg(-g_{\mu\nu}+\frac{p_{a\mu}p_{a\nu}}{m_{a}^2}\bigg)(\cdots)_{\nu}
\nonumber
\\
\to \;\;
\mathcal{G}
\bigg[p_{a}\cdot \bigg(W_{a}+\frac{\partial \pi_{a}}{m_{a}}\bigg)\cdots  q_{R\rm i}
\cdots \phi_{i} \cdots \bigg]=0.
\label{eq:p_W+sigma=0}
\end{gather}
For the longitudinal mode of a massive vector, the leading term of polarization vector at high $E$ is proportional to its momentum, and therefore contributes to the high-$E$ scaling of amplitudes. However, the St\"uckelberg formulation eliminates such contributions, so that the high-$E$ scaling can arise only from vertices and propagators, allowing us to evaluate the leading high-$E$ scaling by examining only the vertices and propagators.

To show this explicitly, we first compute
\begin{gather}
\eps^0_{a}
=\frac{p_{a}}{m_{a}}+\eta_{a}, 
\nonumber
\\
\eps^0_{a}\cdot \bigg(W_{a}+\frac{i p_{a} \pi_{a}}{m_{a}}\bigg)
-
\frac{p_{a}}{m_{a}}\cdot \bigg(W_{a}+\frac{i p_{a} \pi_{a}}{m_{a}}\bigg)
=-i(\pi_{a}+i\eta_{a}\cdot W_{a}),
\label{eq:eps_W+p sigam=-isigma+xi_W}
\end{gather}
in terms of the longitudinal polarization vector $\eps^0_{a}$ with $\eta_{a} \to E^{-1}$ at high $E$. Then, using Eq.~\eqref{eq:p_W+sigma=0} and Eq.~\eqref{eq:eps_W+p sigam=-isigma+xi_W}, we obtain
\begin{gather}
\mathcal{G}
[\eps^0_{a}\cdot W_{a}\cdots  q_{R\rm i}
\cdots \pi_{i} \cdots ]=\mathcal{G}
[-i(\pi_{a}+i\eta_{a}\cdot W_{a})\cdots  q_{R\rm i}
\cdots \pi_{i} \cdots ],
\label{eq:no_longitudinal}
\end{gather}
leading to the vanishing of high-$E$ contributions from the longitudinal polarization.

To examine the high-$E$ behavior of amplitudes for the St\"uckelberg Lagrangian, we recast Eq.~\eqref{eq:stuckel_L} with the gauge fixing $(\xi=1)$ as
\begingroup
\small
\begin{align}
\mathcal{L}&=-\frac{1}{4}
(\partial_{\mu} W_{\capi{A}\nu}-\partial_{\nu} W_{\capi{A}\mu}
-C_{\capi{ABC}} W_{\capi{B}\mu}W_{\capi{C}\mu})^2
+
\bar{q}_R i(\slash \!\!\! \partial +i \slash \!\!\! W_{\capi{A}}
R^{\capi{A}})q_R
+
\bar{q}_L i(\slash \!\!\! \partial +i \slash \!\!\! W_{\capi{A}}
L^{\capi{A}})q_L
\nonumber
\\
&\quad
+\frac12 
\Big\{(\partial_{\mu}+iW_{\capi{A}\mu} T^{\capi{(A)}})
\pi
\Big\}^2
+\frac12 m_{\capi{A}}^2 W_{\capi{A}\mu}^2
+m_{\capi{B}}W_{\capi{B}\mu} W_{\capi{A}}^\mu (iT_{\capi{B}\capmi{I}}^{\capi{(A)}})\pi_{\capmi{I}}
+
\frac{1}{2}
m_{\bot\bs{a}}^2
(W_{\bs{a}\mu}+\partial_{\mu}\sigma_{\bs{a}})^2
\nonumber
\\
&\quad
-\tilde{\mathcal{V}}(\pi)
-\big\{\bar{q}_L \tilde{\mathcal{Y}}(\pi)
q_R + \mbox{h.c.}\big\}-\frac{1}{2}(\partial^\mu W_{\capi{A}\mu})^2-\frac{1}{2}m_{\capi{A}}^2\pi_{\capi{A}}^2.
\label{eq:stuckel_L2}
\end{align}
\endgroup
Next, let us consider an amplitude $\mathcal{G}[\eps_{\capi{A}} \cdot W_{\capi{A}}\cdots \bar{u}_{\mathbf{i}} q_{\mathbf{i}}\cdots \pi_{i}\cdots]$. In this amplitude, we denote the numbers of $W^n$-type vertices by $\mathcal{B}_{nW}$ $(n=1,...,4)$, vertices arising from the scalar potential $\tilde{\mathcal{V}}$ by $\mathcal{B}_{\tilde{\mathcal{V}}}$, and fermionic vertices by $\mathcal{F}$. The total number of internal lines is always one less than the total number of vertices. Denoting the numbers of bosonic and fermionic internal lines by $I_B$ and $I_F$, respectively, we obtain
\begin{gather}
I_B+I_F+1=\mathcal{B}_{4W}+\mathcal{B}_{3W}+\mathcal{B}_{2W}+\mathcal{B}_{W}+\mathcal{B}_{\tilde{\mathcal{V}}}+\mathcal{F}.
\label{eq:I_B_I_F}
\end{gather}
For external fermions, some fermions form pairs, directly yielding bosonic propagators. The remaining unpaired fermions are connected through internal fermion lines. Combining the internal fermion lines with half the number of unpaired fermions gives the total number of fermionic vertices minus half the number of paired fermions:
\begin{gather}
I_F+\frac{N_{F\mbox{\scriptsize (unpair)}}}{2}=\mathcal{F}-\frac{N_{F\mbox{\scriptsize (pair)}}}{2}
\quad \to \quad 
I_F=\mathcal{F}-N_F/2,
\label{eq:I_F}
\end{gather}
with the number of external fermions $N_F$.

By examining the high-$E$ scaling of propagators and vertices, one can determine the exponent $\mathcal{P}$ in $\mathcal{G}[\eps_{\capi{A}} \cdot W_{\capi{A}}\cdots \bar{u}_{\mathbf{i}} q_{\mathbf{i}}\cdots \pi_{i}\cdots]\leq E^\mathcal{P}$, given by $\mathcal{P}=-2I_B-I_F+\mathcal{B}_{3W}+\mathcal{B}_{W}+N_F/2$. Then, using Eqs.~\eqref{eq:I_B_I_F} and~\eqref{eq:I_F}, we obtain
\begin{gather}
\mathcal{P}=2-2\mathcal{B}_{4W}-\mathcal{B}_{3W}-2\mathcal{B}_{2W}-\mathcal{B}_{W}-2\mathcal{B}_{\tilde{\mathcal{V}}}-\mathcal{F}.
\label{eq:exp_P}
\end{gather}
The bosonic $W$ and $W^3$-type, as well as fermionic, vertices generate 3-pt interactions, while the $W^2$ and $W^4$ types yield 4-pt contact interactions. Together with the tree-unitarity condition in Eq.~\eqref{eq:tree_unitarity_G}, one can then observe that only higher-pt contact operators in the scalar potential violate the tree unitarity of higher-pt amplitudes. Thus, we conclude that the scalar potential $\tilde{V}(\pi)$ in Eq.~\eqref{eq:stuckel_L2} consists of terms up to quartic order:
\begin{align}
\frac{\partial^5 \tilde{V}(\pi)}{\partial \pi_{\capmi{I}_1}\partial \pi_{\capmi{I}_2}\partial \pi_{\capmi{I}_3}\partial \pi_{\capmi{I}_4}\partial \pi_{\capmi{I}_5}}=0.
\end{align}
This result translates the tree unitarity conditions for higher-pt amplitudes into a nontrivial gauge invariance condition. Recall that we adopt the Stückelberg formulation to restore gauge symmetry, which automatically ensures the gauge invariance of the Lagrangian. However, truncating the scalar potential at quartic order forces gauge invariance to be realized with only terms up to quartic order, thereby imposing additional conditions on the couplings.

\subsection{Constructing the Lagrangian of tree-unitary theories}
\label{sec:tree_unitary_L}

By the definition of scalar fields $\pi_{\capmi{I}}$ in Eq.~\eqref{eq:pi_to_phi_sigma}, their gauge transformations take a complicated form. However, tree unitarity restricts the scalar potential to consist of only terms up to quartic order. Thus, one can derive the gauge invariance condition by redefining the scalar fields as
\begin{gather}
\bar{\pi}\equiv v+\pi=e^{i\sigma\cdot T}(v+\phi),
\label{eq:bar_pi_phi}
\end{gather}
of which the gauge transformation is $\bar{\pi} \to e^{i\Lam\cdot T}\bar{\pi}$ for arbitrary gauge parameters $\Lam_{\capi{A}}$. Then, we see that the local minimum of $\bar{V}(\bar{\pi})$ is shifted at $\bar{\pi}=v$:
\begin{gather}
\frac{\partial \tilde{\mathcal{V}}(\pi)}{\partial \pi_{\capmi{I}}}
=
\frac{\partial \tilde{\mathcal{V}}(\bar{\pi}-v)}{\partial \bar{\pi}_{\capmi{I}}}
\;\;\;\to \;\;\;
\frac{\partial \tilde{\mathcal{V}}(\bar{\pi}-v)}{\partial \bar{\pi}_{\capmi{I}}}
\bigg|_{\bar{\pi}=v}=0.
\label{eq:shift_minimum}
\end{gather}
Taking the replacement $\pi \to \bar{\pi}-v$ in Eq.~\eqref{eq:stuckel_L}, we obtain
\begingroup
\small
\begin{align}
\mathcal{L}&=-\frac{1}{4}
(\partial_{\mu} W_{\capi{A}\nu}-\partial_{\nu} W_{\capi{A}\mu}
-C_{\capi{ABC}} W_{\capi{B}\mu}W_{\capi{C}\mu})^2
+
\bar{q}_R i(\slash \!\!\! \partial +i \slash \!\!\! W_{\capi{A}}
R^{\capi{A}})q_R
+
\bar{q}_L i(\slash \!\!\! \partial +i \slash \!\!\! W_{\capi{A}}
L^{\capi{A}})q_L
\nonumber
\\
&\quad
+\frac12 
\big[(\partial_{\mu}+iW_{\capi{A}\mu} T^{\capi{(A)}})
\bar{\pi}
\big]^2
+
\frac{1}{2}
m_{\bot\bs{a}}^2
(W_{\bs{a}\mu}+\partial_{\mu}\sigma_{\bs{a}})^2
-\bar{\mathcal{V}}(\bar{\pi})
-\big\{\bar{q}_L \bar{\mathcal{Y}}(\bar{\pi})
q_R + \mbox{h.c.}\big\},
\label{eq:sbgt_L}
\end{align}
\endgroup
with no gauge-fixing term, where the scalar potential and Yukawa terms are
\begin{align}
\bar{\mathcal{V}}(\bar{\pi})
&=\frac12 \bar{M}^2_{\capmi{IJ}}\bar{\pi}_{\capmi{I}}\bar{\pi}_{\capmi{J}}
+\frac{1}{6}\bar{P}_{\capmi{IJK}}
\bar{\pi}_{\capmi{I}}\bar{\pi}_{\capmi{J}}\bar{\pi}_{\capmi{K}}
+\frac{1}{24}\bar{Q}_{\capmi{IJKL}}
\bar{\pi}_{\capmi{I}}\bar{\pi}_{\capmi{J}}\bar{\pi}_{\capmi{K}}\bar{\pi}_{\capmi{L}},
\nonumber
\\
\tilde{\mathcal{Y}}(\bar\pi)&= \mu_0+H_{\capmi{I}} \bar{\pi}_{\capmi{I}}.
\label{eq:barV_barY}
\end{align}
In the potential $\bar{\mathcal{V}}(\bar{\pi})$, the term linear in $\bar{\pi}$ is forbidden by gauge symmetry.

Note that the Lagrangian in Eq.~\eqref{eq:sbgt_L} is invariant under the following gauge transformations: 
\begin{align}
W_{\capi{A}\mu} (iT^{\capi{(A)}}_{\capmi{IJ}})
&\;\;\rightarrow \;\; \big[e^{i\Lam \cdot T}
W_{\capi{A}\mu}(iT^{\capi{(A)}})e^{-i\Lam \cdot T}
+e^{i\sigma \cdot T}\partial_{\mu}e^{-i\Lam \cdot T}
\big]_{\capmi{IJ}}, \quad \mbox{(non-Abelian)}
\nonumber
\\
W_{\capi{A}\mu}& \;\; \rightarrow \;\; 
W_{\capi{A}\mu}+\partial_{\mu}\Lam_{\capi{A}}, 
\qquad\qquad\qquad\qquad\qquad\qquad\qquad \mbox{(Abelian)}
\nonumber
\\
q_R&\;\;\rightarrow \;\; e^{i\Lam\cdot R}q_R,
\nonumber
\\
q_L&\;\;\rightarrow \;\; e^{i\Lam\cdot L}q_L,
\nonumber
\\
\bar{\pi}& \;\;\rightarrow \;\; e^{i\Lambda \cdot T}\bar{\pi},
\nonumber
\\
\sigma_{\bs{a}}& \;\;\rightarrow \;\; \sigma_{\bs{a}}+\Lam_{\bs{a}}.
\label{eq:SBGT_gauge}
\end{align}
The scalar potential attains its local minimum at $\bar{\pi}=v$, i.e. $\langle \bar{\pi} \rangle = v$ is the vacuum expectation value (VEV). This indicates that tree-unitary theories correspond to spontaneously broken gauge theories supplemented by additional Abelian vector mass terms~\cite{Cornwall:1974km}.

\section{Tree unitarity in higher-pt amplitudes}
\label{sec:higher-pt_tree_unitarity}

Using the results derived in the previous section, we find the coupling conditions arising from the tree unitarity of all higher-pt amplitudes. As discussed above, these are realized as gauge-invariance conditions on the couplings in the scalar potential $\bar{\mathcal{V}}(\bar\pi)$. We therefore examine in detail the couplings in $\bar{\mathcal{V}}(\bar\pi)$ and present all additional conditions in terms of the mass-basis couplings.

The condition for the shifted local minimum in Eq.~\eqref{eq:shift_minimum} constrains the couplings in Eq.~\eqref{eq:barV_barY} to satisfy
\begin{gather}
2\bar{M}^2_{\capmi{IJ}}v_{\capmi{J}}+
\bar{P}_{\capmi{IJK}}v_{\capmi{J}}v_{\capmi{K}}+
\frac13 \bar{Q}_{\capmi{IJKL}} v_{\capmi{J}} v_{\capmi{K}}v_{\capmi{L}}=0. 
\label{eq:tadpole}
\end{gather}
The gauge invariance of the Lagrangian in Eq.~\eqref{eq:sbgt_L} requires the scalar potential to satisfy
\begin{align}
\frac{\partial \bar{\mathcal{V}}}{\partial \bar{\pi}_{\capmi{I}}}
[(i \omega\cdot T)\bar{\pi}]_{\capmi{I}}
=0\;\; \to \;\; 
\frac{\partial \bar{\mathcal{V}}}{\partial \bar{\pi}_{\capmi{I}}}
[(i T^{\capi{(A)}})\bar{\pi}]_{\capmi{I}}
=0,
\label{eq:V_gauge_invariance}
\end{align}
in terms of infinitesimal gauge parameters $\omega_{\capi{A}}$. Then, using Eq.~\eqref{eq:V_gauge_invariance}, we obtain
\begin{align}
&\bar{Q}_{\capmi{I'JKL}}(iT^{\capi{(A)}}_{\capmi{I'I}})
+
\bar{Q}_{\capmi{IJ'KL}}(iT^{\capi{(A)}}_{\capmi{J'J}})
+
\bar{Q}_{\capmi{IJK'L}}(iT^{\capi{(A)}}_{\capmi{K'K}})
+
\bar{Q}_{\capmi{IJKL'}}(iT^{\capi{(A)}}_{\capmi{L'L}})
=0, \label{eq:QbarConstraint}
\\
&\bar{P}_{\capmi{I'JK}}(iT^{\capi{(A)}}_{\capmi{I'I}})
+
\bar{P}_{\capmi{IJ'K}}(iT^{\capi{(A)}}_{\capmi{J'J}})
+
\bar{P}_{\capmi{IJK'}}(iT^{\capi{(A)}}_{\capmi{K'K}})
=0, \label{eq:PbarConstraint}
\\
&\bar{M}^2_{\capmi{I'J}}(iT^{\capi{(A)}}_{\capmi{I'I}})
+
\bar{M}^2_{\capmi{IJ'}}(iT^{\capi{(A)}}_{\capmi{J'J}})
=0. \label{eq:MbarConstraint}
\end{align}
To obtain an additional constraint on the couplings, we take the second derivatives of the scalar potential:
\begin{align}
&\frac{\partial^2 \bar{\mathcal{V}}}{\partial \bar{\pi}_{\capmi{I}}\partial \bar{\pi}_{\capmi{J}}}
[(i T^{\capi{(A)}})\bar{\pi}]_{\capmi{I}}
+
\frac{\partial \bar{\mathcal{V}}}{\partial \bar{\pi}_{\capmi{I}}}
(iT_{\capmi{IJ}}^{\capi{(A)}})
=0.
\label{eq:V_second_d}
\end{align}
By evaluating Eq.~\eqref{eq:V_second_d} at $\bar{\pi}=v$, we obtain
\begin{gather}
\bar{M}^2_{\capmi{I}\capi{A}}+
\bar{P}_{\capmi{I}\capi{A}\capmi{K}}v_{\capmi{K}}+
\frac12 \bar{Q}_{\capmi{I}\capi{A}\capmi{KL}} v_{\capmi{K}}v_{\capmi{L}}=0,
\label{eq:M_{IA}_vanish}
\end{gather}
implying that all components of this rank-2 tensor with any vector index vanish.

To re-express the conditions in Eqs.~\eqref{eq:QbarConstraint}-\eqref{eq:MbarConstraint} in terms of the couplings in the mass basis, we first need to find the relations between the couplings in $\mathcal{V}(\phi)$ and $\bar{\mathcal{V}}(\bar\pi)$. This can be achieved by taking the gauge transformation in Eq.~\eqref{eq:SBGT_gauge} with the gauge parameters $\Lam_{\capi{A}}\equiv \sigma_{\capi{A}}$. Then, using the definitions in Eqs.~\eqref{eq:stuckelberg_VF} and~\eqref{eq:bar_pi_phi}, we can recast Eq.~\eqref{eq:sbgt_L} as
\begingroup
\small
\begin{align}
\mathcal{L}&=-\frac{1}{4}
(\partial_{\mu} V_{\capi{A}\nu}-\partial_{\nu} V_{\capi{A}\mu}
-C_{\capi{ABC}} V_{\capi{B}\mu}V_{\capi{C}\mu})^2
+
\bar{\psi}_R i(\slash \!\!\! \partial +i \slash \!\!\! V_{\capi{A}}
R^{\capi{A}})\psi_R
+
\bar{\psi}_L i(\slash \!\!\! \partial +i \slash \!\!\! V_{\capi{A}}
L^{\capi{A}})\psi_L
\nonumber
\\
&\quad
+\frac12 
\Big\{(\partial_{\mu}+iV_{\capi{A}\mu} T^{\capi{(A)}})
(v+\phi)
\Big\}^2
+
\frac{1}{2}
m_{\bot\bs{a}}^2
V_{\bs{a}\mu}^2
-\bar{\mathcal{V}}[v+\phi]
-\big\{\bar{\psi}_L \bar{\mathcal{Y}}[v+\phi]
\psi_R + \mbox{h.c.}\big\},
\label{eq:sbgt_L2}
\end{align}
\endgroup
in terms of the mass-basis fields. Since the Lagrangian above must be equal to Eq.~\eqref{eq:L_generator_rep1}, we obtain
\begin{align}
\bar{\mathcal{V}}[v+\phi]&=
\frac12 \bar{M}^2_{\capmi{IJ}}\tilde{\phi}_{\capmi{I}}\tilde{\phi}_{\capmi{J}}
+\frac16
\bar{P}_{\capmi{IJK}}\tilde{\phi}_{\capmi{I}}\tilde{\phi}_{\capmi{J}}
\tilde{\phi}_{\capmi{K}}
+\frac{1}{24}
\bar{Q}_{\capmi{IJKL}}\tilde{\phi}_{\capmi{I}}\tilde{\phi}_{\capmi{J}}
\tilde{\phi}_{\capmi{K}}\tilde{\phi}_{\capmi{L}} \quad (\tilde{\phi}=v+\phi),
\nonumber
\\
&=
\frac12 M^2_{\capmi{IJ}}\phi_{\capmi{I}}\phi_{\capmi{J}}
+\frac16
P_{\capmi{IJK}}\phi_{\capmi{I}}\phi_{\capmi{J}}
\phi_{\capmi{K}}
+\frac{1}{24}
Q_{\capmi{IJKL}}\phi_{\capmi{I}}\phi_{\capmi{J}}
\phi_{\capmi{K}}\phi_{\capmi{L}},
\label{eq:V_mass_basis}
\\
\bar{\mathcal{Y}}[v+\phi]&=\mu_0+H_i(v+\phi)_i=\mu +H_i\phi_i
\quad \to \quad \mu=\mu_0+ H_i v_i,
\label{eq:Y_mass_basis}
\end{align}
for the scalar potential and Yukawa term, respectively.

In Eq.~\eqref{eq:V_mass_basis}, we see that the tadpole term for $\phi_{\capmi{I}}$ vanishes due to the condition in Eq.~\eqref{eq:tadpole} and the quartic, cubic, and quadratic couplings are related as
\begin{align}
Q_{\capmi{IJKL}}&=\bar{Q}_{\capmi{IJKL}}
\nonumber
\\
P_{\capmi{IJK}}
&=\bar{P}_{\capmi{IJK}}+\bar{Q}_{\capmi{IJKL}}v_{\capmi{L}}
\nonumber
\\
M_{\capmi{IJ}}^2&=\bar{M}^2_{\capmi{IJ}}+
\bar{P}_{\capmi{IJK}}
v_{\capmi{K}}+\frac12 \bar{Q}_{\capmi{IJKL}}
v_{\capmi{K}}v_{\capmi{L}},
\end{align}
and inversely,
\begin{align}
\bar{Q}_{\capmi{IJKL}}&=Q_{\capmi{IJKL}}
\nonumber
\\
\bar{P}_{\capmi{IJK}}
&=P_{\capmi{IJK}}-Q_{\capmi{IJKL}}v_{\capmi{L}}
\nonumber
\\
\bar{M}^2_{\capmi{IJ}}&=M_{\capmi{IJ}}^2-
P_{\capmi{IJK}}
v_{\capmi{K}}+\frac12 Q_{\capmi{IJKL}}
v_{\capmi{K}}v_{\capmi{L}}.
\label{eq:bar_to_nonbar}
\end{align}
Then, in terms of the mass-basis couplings $M^2$, $P$, and $Q$, we recast the conditions in Eqs.~\eqref{eq:tadpole} and~\eqref{eq:M_{IA}_vanish} as
\begin{gather}
2M^2_{\capmi{IJ}}v_{\capmi{J}}-
P_{\capmi{IJK}}v_{\capmi{J}}v_{\capmi{K}}+
\frac13 Q_{\capmi{IJKL}} v_{\capmi{J}} v_{\capmi{K}}v_{\capmi{L}}=0
\quad\mbox{and}\quad
M^2_{\capmi{I}\capi{A}}=0.
\label{eq:tadpole_&_M^2_{IA}=0}
\end{gather}
Using Eq.~\eqref{eq:bar_to_nonbar}, we find another expression for the first condition in Eq.~\eqref{eq:tadpole_&_M^2_{IA}=0} as
\begin{gather}
\bar{M}^2_{\capmi{IJ}}v_{\capmi{J}}=-M^2_{\capmi{IJ}}v_{\capmi{J}}.
\end{gather}

Next, we identify the couplings $M^2$, $P$, and $Q$ involving vector indices. Multiplying Eq.~\eqref{eq:QbarConstraint} by the VEV $v$, the relation in Eq.~\eqref{eq:iTv=lam} allows us to get
\begin{align}
&Q_{\capi{A}\capmi{JKL}}m_{\capi{A}}
+
Q_{\capmi{IJ'KL}}v_{\capmi{I}}(iT^{\capi{(A)}}_{\capmi{J'J}})
+
Q_{\capmi{IJK'L}}v_{\capmi{I}}(iT^{\capi{(A)}}_{\capmi{K'K}})
+
Q_{\capmi{IJKL'}}v_{\capmi{I}}(iT^{\capi{(A)}}_{\capmi{L'L}})
=0,
\label{eq:QTv}
\end{align}
where we have replaced $\bar{Q}$ by $Q$, ensured by Eq.~\eqref{eq:bar_to_nonbar}. To identify the coupling $Q$ with vector indices, we use the gauge invariance condition for the cubic coupling in Eq.~\eqref{eq:PbarConstraint}:
\begin{align}
Q_{\capi{A}\capmi{IJK}}
=-
P_{\capmi{I'JK}}\frac{(iT^{\capi{(A)}}_{\capmi{I'I}})}{m_{\capi{A}}}
-
P_{\capmi{IJ'K}}\frac{(iT^{\capi{(A)}}_{\capmi{J'J}})}{m_{\capi{A}}}
-
P_{\capmi{IJK'}}\frac{(iT^{\capi{(A)}}_{\capmi{K'K}})}{m_{\capi{A}}}.
\label{eq:Q_AIJK}
\end{align}
This expression, however, requires us to identify the cubic coupling $P$ with vector indices. To achieve this, 
by multiplying Eq.~\eqref{eq:Q_AIJK} and Eq.~\eqref{eq:PbarConstraint} by one additional VEV $v$, respectively, we obtain:
\begin{align}
&
P_{\capi{A}\capmi{IJ}}+Q_{\capi{A}\capmi{IJK}}v_{\capmi{K}}
=-
P_{\capmi{I'JK}}v_{\capi{K}}\frac{(iT^{\capi{(A)}}_{\capmi{I'I}})}{m_{\capi{A}}}
-
P_{\capmi{IJ'K}}v_{\capi{K}}\frac{(iT^{\capi{(A)}}_{\capmi{J'J}})}{m_{\capi{A}}},
\\
&P_{\capi{A}\capmi{IJ}}-Q_{\capi{A}\capmi{IJK}}v_{\capmi{K}}
=-\bigg[P_{\capmi{I'JK}}v_{\capmi{K}}-Q_{\capmi{I'JKL}}v_{\capmi{K}}v_{\capmi{L}}\bigg]
\frac{(iT^{\capi{(A)}}_{\capmi{I'I}})}{m_{\capi{A}}}.
\label{eq:bQTvs}
\end{align}
Combining these two relations and using the gauge-invariance condition for the quadratic coupling in Eq.~\eqref{eq:MbarConstraint}, we obtain
\begin{align}
P_{\capi{A}\capmi{IJ}}
&=-M^2_{\capmi{I'J}}
\frac{(iT^{\capi{(A)}}_{\capmi{I'I}})}{m_{\capi{A}}}
-M^2_{\capmi{IJ'}}
\frac{(iT^{\capi{(A)}}_{\capmi{J'J}})}{m_{\capi{A}}}
=\frac{[iT^{\capi{(A)}},M^2]_{\capmi{IJ}}}{m_{\capi{A}}},
\label{eq:P_AIJ}
\end{align}
described solely by the mass-basis couplings. Note that we have used the conditions for $\bar{M}^2$ and $\bar{P}$ in Eqs.~\eqref{eq:PbarConstraint}-\eqref{eq:MbarConstraint} to identify the couplings $P$ and $Q$ involving vector indices $A$. It implies that the condition in Eq.~\eqref{eq:QbarConstraint} is the unique coupling condition required for tree unitarity of all higher-pt amplitudes. Using Eqs.~\eqref{eq:Q_AIJK} and~\eqref{eq:P_AIJ}, we find the following expressions:
\begin{align}
Q_{\capi{A_1A_2A_3A_4}}
&=\frac{[iT^{\capi{(A_1)}},[iT^{\capi{(A_4)}},M^2]]_{\capi{A_2A_3}}}{2m_{\capi{A_1}}m_{\capi{A_4}}}+\mbox{cyclic}(A_2,A_3,A_4),
\nonumber
\\
Q_{\capi{A_1A_2A_3}i_4}
&=\frac{[iT^{\capi{(A_1)}},[iT^{\capi{(A_3)}},M^2]]_{\capi{A_2}i_4}}{m_{\capi{A_1}}m_{\capi{A_3}}}+(A_2\leftrightarrow A_3)
-
\frac{(iT^{\capi{(A_1)}}[iT^{\capi{(A_3)}},M^2])_{i_4\capi{A_2}}}{m_{\capi{A_1}}m_{\capi{A_3}}},
\nonumber
\\
Q_{\capi{A_1A_2}i_3i_4}
&=\frac{[iT^{\capi{(A_1)}},[iT^{\capi{(A_2)}},M^2]]_{i_3i_4}}{m_{\capi{A_1}}m_{\capi{A_2}}}+\frac{(iT^{\capi{(A_1)}}_{\capi{A_2B}})[iT^{\capi{(B)}},M^2]_{i_3i_4}}{m_{\capi{A_1}}m_{\capi{B}}}
+\frac{(iT^{\capi{(A_1)}}_{\capi{A_2}j})}{m_{\capi{A_1}}}P_{ji_3i_4},
\nonumber
\\
Q_{\capi{A_1}i_2i_3i_4}
&=\frac{(iT^{\capi{(A_1)}}_{i_2\capi{B}})[iT^{\capi{(B)}},M^2]_{i_3i_4}}{m_{\capi{A_1}}m_{\capi{B}}}
+\frac{(iT^{\capi{(A_1)}}_{i_2j})}{m_{\capi{A_1}}}P_{ji_3i_4}
+\mbox{cyclic}(i_2,i_3,i_4),
\label{eq:Q_As}
\end{align}
described only by the mass-basis couplings, where cyclic$(\cdots)$ denotes the terms obtained by cyclic permutations of the indices in $(\cdots)$. 

Inserting the expressions in Eq.~\eqref{eq:Q_As} into  Eq.~\eqref{eq:QbarConstraint}, one can obtain the coupling conditions required for tree unitarity of all higher-pt amplitudes. However, the five free indices in Eq.~\eqref{eq:QbarConstraint} imply that this equation can be derived from tree unitarity of bosonic 5-pt amplitudes. To confirm this, let us construct the 5-pt $W\pi\pi\pi\pi$ amplitude in the St\"uckelberg Lagrangain in Eq.~\eqref{eq:stuckel_L}. For a single channel, we obtain
\begin{align}
&\mathcal{M}_{5,\mbox{\scriptsize 1-ch}}
(\bs{1}_{\capi{A_1}} \bs{2}_{\capmi{I}_2} \bs{3}_{\capmi{I}_3} \bs{4}_{\capmi{I}_4}
\bs{5}_{\capmi{I}_5})
=-\sqrt{2}
\Bigg(
\sum_j \frac{T^{\capi{(A_1)}}_{\capmi{I}_2\capmi{J}}Q_{\capmi{J}\capmi{I}_3\capmi{I}_4\capmi{I}_5}\langle \bold{121}]}{
m_{\capi{A_1}}(s_{12}-m_{\capmi{J}}^2)}
\Bigg)
+
(\mbox{two propagators}),
\label{eq:W4pi}
\end{align}
where the second term encodes the diagrams including two propagators in the single channel. By summing over all amplitudes whose indices ${\cal I}_2, ..., {\cal I}_5$ are cyclically permuted, one can get the full amplitude.

Recall that we used energy power counting when identifying the tree-unitarity-violating higher-pt amplitudes from the St\"uckelberg Lagrangian. In this procedure, we used the fact that the high-$E$ contribution of the longitudinal mode vanishes due to Eq.~\eqref{eq:no_longitudinal}. In Eq.~\eqref{eq:W4pi}, however, we intentionally restore one vector contribution to explicitly see the origin of Eq.~\eqref{eq:QbarConstraint}. The first term in Eq.~\eqref{eq:W4pi} leads to $E^0$ scaling, while the others yield $E^k$ $(k \le -1)$ scaling. This implies that only the terms associated with the first one in Eq.~\eqref{eq:W4pi} generate the tree-unitarity violation, leading to the condition in Eq.~\eqref{eq:QbarConstraint}.

Given the equivalence between the tree-level amplitudes from the mass-basis and St\"uckelberg Lagrangians, we conclude that, {\it apart from the tree unitarity conditions for 4-pt amplitudes, any additional ones arise only from bosonic 5-pt amplitudes}, implying no necessity to examine tree unitarity of $(n \ge 6)$-pt amplitudes. To illustrate whether we indeed get Eq.~\eqref{eq:QbarConstraint} from the 5-pt amplitudes in the mass basis, in Table.~\ref{tab:Vphi5pt}, we present all diagrams and associated coupling configurations appearing in the $VV\phi\phi\phi$ process as an example.
\begin{table}[htbp!]
    \centering
    \begin{tabular}{cc||cc||cc}
        Diagrams & Couplings & Diagrams & Couplings & Diagrams & Couplings \\
        \hline
\tikzset{every picture/.style={line width=0.75pt}} 

\begin{tikzpicture}[x=0.75pt,y=0.75pt,yscale=-1,xscale=1]

\draw  [fill={rgb, 255:red, 0; green, 0; blue, 0 }  ,fill opacity=1 ] (71.35,73.59) .. controls (71.12,71.32) and (72.71,69.3) .. (74.92,69.06) .. controls (77.12,68.82) and (79.1,70.45) .. (79.33,72.72) .. controls (79.57,74.98) and (77.97,77.01) .. (75.77,77.25) .. controls (73.56,77.49) and (71.58,75.85) .. (71.35,73.59) -- cycle ;
\draw    (75.34,73.15) -- (120.49,73.57) ;
\draw  [fill={rgb, 255:red, 0; green, 0; blue, 0 }  ,fill opacity=1 ] (116.5,74) .. controls (116.26,71.74) and (117.86,69.71) .. (120.07,69.47) .. controls (122.27,69.23) and (124.25,70.87) .. (124.48,73.13) .. controls (124.72,75.4) and (123.12,77.42) .. (120.91,77.66) .. controls (118.71,77.9) and (116.73,76.26) .. (116.5,74) -- cycle ;
\draw    (120.49,73.57) -- (150.75,72.98) ;
\draw    (120.49,73.57) -- (120.49,105.93) ;
\draw    (120.49,41.21) -- (120.49,73.57) ;
\draw    (47.2,53.22) .. controls (56.21,46.93) and (47.46,64.67) .. (56.8,58.61) ;
\draw    (56.8,58.61) .. controls (65.81,52.32) and (57.06,70.06) .. (66.4,64) ;
\draw    (66.4,64) .. controls (75.41,57.71) and (66.66,75.45) .. (76,69.39) ;
\draw    (37.6,47.83) .. controls (46.61,41.54) and (37.86,59.28) .. (47.2,53.22) ;
\draw    (48.81,92.07) .. controls (47,80.88) and (58.87,96.49) .. (57.4,85.06) ;
\draw    (57.4,85.06) .. controls (55.59,73.86) and (67.47,89.47) .. (65.99,78.04) ;
\draw    (65.99,78.04) .. controls (64.19,66.85) and (76.06,82.46) .. (74.59,71.03) ;
\draw    (40.21,99.09) .. controls (38.41,87.89) and (50.28,103.5) .. (48.81,92.07) ;

\end{tikzpicture} & \raisebox{0.7cm}{$FQ$} & 

\tikzset{every picture/.style={line width=0.75pt}} 

\begin{tikzpicture}[x=0.75pt,y=0.75pt,yscale=-1,xscale=1]

\draw  [fill={rgb, 255:red, 0; green, 0; blue, 0 }  ,fill opacity=1 ] (86.65,81.28) .. controls (86.42,79.02) and (88.02,76.99) .. (90.22,76.75) .. controls (92.43,76.51) and (94.4,78.15) .. (94.64,80.42) .. controls (94.87,82.68) and (93.27,84.71) .. (91.07,84.95) .. controls (88.87,85.19) and (86.89,83.55) .. (86.65,81.28) -- cycle ;
\draw    (90.65,80.85) -- (116.01,97.1) ;
\draw  [fill={rgb, 255:red, 0; green, 0; blue, 0 }  ,fill opacity=1 ] (112.01,97.53) .. controls (111.78,95.27) and (113.38,93.24) .. (115.58,93) .. controls (117.79,92.76) and (119.76,94.4) .. (120,96.67) .. controls (120.23,98.93) and (118.63,100.96) .. (116.43,101.2) .. controls (114.22,101.44) and (112.25,99.8) .. (112.01,97.53) -- cycle ;
\draw    (90.65,48.49) -- (125.53,28.85) ;
\draw    (116.01,97.1) -- (147.39,115.93) ;
\draw    (142.9,71.21) -- (116.01,97.1) ;
\draw    (61.85,32.32) .. controls (70.85,26.04) and (62.11,43.78) .. (71.45,37.71) ;
\draw    (71.45,37.71) .. controls (80.45,31.42) and (71.7,49.16) .. (81.05,43.1) ;
\draw    (81.05,43.1) .. controls (90.05,36.81) and (81.3,54.55) .. (90.65,48.49) ;
\draw    (52.25,26.94) .. controls (61.25,20.65) and (52.51,38.39) .. (61.85,32.32) ;
\draw    (65.13,101.18) .. controls (63.33,89.99) and (75.2,105.6) .. (73.73,94.17) ;
\draw    (73.73,94.17) .. controls (71.92,82.97) and (83.79,98.58) .. (82.32,87.15) ;
\draw    (82.32,87.15) .. controls (80.51,75.96) and (92.39,91.57) .. (90.91,80.14) ;
\draw    (56.54,108.2) .. controls (54.73,97) and (66.61,112.61) .. (65.13,101.18) ;
\draw    (90.65,48.49) -- (90.65,80.85) ;
\draw  [fill={rgb, 255:red, 0; green, 0; blue, 0 }  ,fill opacity=1 ] (86.65,48.92) .. controls (86.42,46.66) and (88.02,44.63) .. (90.22,44.39) .. controls (92.43,44.15) and (94.4,45.79) .. (94.64,48.06) .. controls (94.87,50.32) and (93.27,52.35) .. (91.07,52.59) .. controls (88.87,52.83) and (86.89,51.19) .. (86.65,48.92) -- cycle ;

\end{tikzpicture} & \raisebox{1cm}{$GGP$} &

\tikzset{every picture/.style={line width=0.75pt}} 

\begin{tikzpicture}[x=0.75pt,y=0.75pt,yscale=-1,xscale=1]

\draw  [fill={rgb, 255:red, 0; green, 0; blue, 0 }  ,fill opacity=1 ] (81.07,70.34) .. controls (80.83,68.08) and (82.43,66.05) .. (84.63,65.81) .. controls (86.84,65.57) and (88.82,67.21) .. (89.05,69.47) .. controls (89.28,71.74) and (87.69,73.76) .. (85.48,74) .. controls (83.28,74.24) and (81.3,72.6) .. (81.07,70.34) -- cycle ;
\draw    (85.06,102.27) -- (123.87,103.81) ;
\draw  [fill={rgb, 255:red, 0; green, 0; blue, 0 }  ,fill opacity=1 ] (81.07,102.7) .. controls (80.83,100.44) and (82.43,98.41) .. (84.63,98.17) .. controls (86.84,97.93) and (88.82,99.57) .. (89.05,101.83) .. controls (89.28,104.1) and (87.69,106.12) .. (85.48,106.36) .. controls (83.28,106.6) and (81.3,104.96) .. (81.07,102.7) -- cycle ;
\draw    (85.06,37.55) -- (123.87,35.56) ;
\draw    (125.55,69.68) -- (85.06,69.91) ;
\draw    (52.75,36.18) .. controls (57.87,26.18) and (58.12,46.15) .. (63.64,36.18) ;
\draw    (63.64,36.18) .. controls (68.76,26.18) and (69.01,46.15) .. (74.52,36.18) ;
\draw    (74.52,36.18) .. controls (79.64,26.18) and (79.89,46.15) .. (85.41,36.18) ;
\draw    (41.87,36.18) .. controls (46.99,26.18) and (47.24,46.15) .. (52.75,36.18) ;
\draw    (51.66,103.23) .. controls (56.3,92.97) and (57.5,112.91) .. (62.54,102.66) ;
\draw    (62.54,102.66) .. controls (67.17,92.39) and (68.38,112.33) .. (73.41,102.08) ;
\draw    (73.41,102.08) .. controls (78.04,91.82) and (79.25,111.76) .. (84.28,101.51) ;
\draw    (40.79,103.8) .. controls (45.43,93.54) and (46.63,113.48) .. (51.66,103.23) ;
\draw    (85.06,69.91) -- (85.06,102.27) ;
\draw  [fill={rgb, 255:red, 0; green, 0; blue, 0 }  ,fill opacity=1 ] (81.07,37.98) .. controls (80.83,35.72) and (82.43,33.69) .. (84.63,33.45) .. controls (86.84,33.21) and (88.82,34.85) .. (89.05,37.11) .. controls (89.28,39.38) and (87.69,41.4) .. (85.48,41.64) .. controls (83.28,41.88) and (81.3,40.24) .. (81.07,37.98) -- cycle ;
\draw    (85.06,37.55) -- (85.06,69.91) ;

\end{tikzpicture} & \raisebox{1cm}{$GGP$} \\

\tikzset{every picture/.style={line width=0.75pt}} 

\begin{tikzpicture}[x=0.75pt,y=0.75pt,yscale=-1,xscale=1]

\draw  [fill={rgb, 255:red, 0; green, 0; blue, 0 }  ,fill opacity=1 ] (72.41,71.59) .. controls (72.17,69.32) and (73.77,67.3) .. (75.98,67.06) .. controls (78.18,66.82) and (80.16,68.45) .. (80.39,70.72) .. controls (80.63,72.98) and (79.03,75.01) .. (76.82,75.25) .. controls (74.62,75.49) and (72.64,73.85) .. (72.41,71.59) -- cycle ;
\draw    (76.4,71.15) -- (108.84,71.57) ;
\draw  [fill={rgb, 255:red, 0; green, 0; blue, 0 }  ,fill opacity=1 ] (104.85,72) .. controls (104.62,69.74) and (106.22,67.71) .. (108.42,67.47) .. controls (110.62,67.23) and (112.6,68.87) .. (112.83,71.13) .. controls (113.07,73.4) and (111.47,75.42) .. (109.27,75.66) .. controls (107.06,75.9) and (105.09,74.26) .. (104.85,72) -- cycle ;
\draw    (140.72,72.59) -- (148.07,100.4) ;
\draw    (108.84,71.57) -- (108.84,103.93) ;
\draw    (144.71,72.16) -- (108.84,71.57) ;
\draw  [fill={rgb, 255:red, 0; green, 0; blue, 0 }  ,fill opacity=1 ] (136.72,73.02) .. controls (136.49,70.76) and (138.09,68.73) .. (140.29,68.49) .. controls (142.5,68.25) and (144.47,69.89) .. (144.71,72.16) .. controls (144.94,74.42) and (143.34,76.45) .. (141.14,76.69) .. controls (138.94,76.93) and (136.96,75.29) .. (136.72,73.02) -- cycle ;
\draw    (140.72,42) -- (140.72,72.59) ;
\draw    (47.88,51.22) .. controls (56.89,44.93) and (48.14,62.67) .. (57.48,56.61) ;
\draw    (57.48,56.61) .. controls (66.49,50.32) and (57.74,68.06) .. (67.08,62) ;
\draw    (67.08,62) .. controls (76.09,55.71) and (67.34,73.45) .. (76.68,67.39) ;
\draw    (38.28,45.83) .. controls (47.29,39.54) and (38.54,57.28) .. (47.88,51.22) ;
\draw    (49.49,90.07) .. controls (47.68,78.88) and (59.55,94.49) .. (58.08,83.06) ;
\draw    (58.08,83.06) .. controls (56.27,71.86) and (68.15,87.47) .. (66.68,76.04) ;
\draw    (66.68,76.04) .. controls (64.87,64.85) and (76.74,80.46) .. (75.27,69.03) ;
\draw    (40.89,97.09) .. controls (39.09,85.89) and (50.96,101.5) .. (49.49,90.07) ;

\end{tikzpicture} & \raisebox{0.7cm}{$FPP$} &

\tikzset{every picture/.style={line width=0.75pt}} 

\begin{tikzpicture}[x=0.75pt,y=0.75pt,yscale=-1,xscale=1]

\draw  [fill={rgb, 255:red, 0; green, 0; blue, 0 }  ,fill opacity=1 ] (100.06,77.28) .. controls (99.83,75.02) and (101.42,72.99) .. (103.63,72.75) .. controls (105.83,72.51) and (107.81,74.15) .. (108.04,76.42) .. controls (108.28,78.68) and (106.68,80.71) .. (104.47,80.95) .. controls (102.27,81.19) and (100.29,79.55) .. (100.06,77.28) -- cycle ;
\draw  [fill={rgb, 255:red, 0; green, 0; blue, 0 }  ,fill opacity=1 ] (125.42,93.53) .. controls (125.18,91.27) and (126.78,89.24) .. (128.99,89) .. controls (131.19,88.76) and (133.17,90.4) .. (133.4,92.67) .. controls (133.64,94.93) and (132.04,96.96) .. (129.83,97.2) .. controls (127.63,97.44) and (125.65,95.8) .. (125.42,93.53) -- cycle ;
\draw    (104.05,44.49) -- (138.94,24.85) ;
\draw    (129.41,93.1) -- (160.79,111.93) ;
\draw    (156.31,67.21) -- (129.41,93.1) ;
\draw    (75.25,28.32) .. controls (84.26,22.04) and (75.51,39.78) .. (84.85,33.71) ;
\draw    (84.85,33.71) .. controls (93.86,27.42) and (85.11,45.16) .. (94.45,39.1) ;
\draw    (94.45,39.1) .. controls (103.46,32.81) and (94.71,50.55) .. (104.05,44.49) ;
\draw    (65.65,22.94) .. controls (74.66,16.65) and (65.91,34.39) .. (75.25,28.32) ;
\draw    (78.54,97.18) .. controls (76.73,85.99) and (88.6,101.6) .. (87.13,90.17) ;
\draw    (87.13,90.17) .. controls (85.33,78.97) and (97.2,94.58) .. (95.73,83.15) ;
\draw    (95.73,83.15) .. controls (93.92,71.96) and (105.79,87.57) .. (104.32,76.14) ;
\draw    (69.94,104.2) .. controls (68.14,93) and (80.01,108.61) .. (78.54,97.18) ;
\draw  [fill={rgb, 255:red, 0; green, 0; blue, 0 }  ,fill opacity=1 ] (100.06,44.92) .. controls (99.83,42.66) and (101.42,40.63) .. (103.63,40.39) .. controls (105.83,40.15) and (107.81,41.79) .. (108.04,44.06) .. controls (108.28,46.32) and (106.68,48.35) .. (104.47,48.59) .. controls (102.27,48.83) and (100.29,47.19) .. (100.06,44.92) -- cycle ;
\draw    (103.37,44.54) .. controls (113.13,49.44) and (94.14,50.65) .. (103.88,55.96) ;
\draw    (103.88,55.96) .. controls (113.65,60.85) and (94.65,62.07) .. (104.4,67.38) ;
\draw    (104.4,67.38) .. controls (114.16,72.27) and (95.17,73.48) .. (104.92,78.79) ;
\draw    (104.05,76.85) -- (129.41,93.1) ;

\end{tikzpicture} & \raisebox{1cm}{$FFP$} &

\tikzset{every picture/.style={line width=0.75pt}} 

\begin{tikzpicture}[x=0.75pt,y=0.75pt,yscale=-1,xscale=1]

\draw  [fill={rgb, 255:red, 0; green, 0; blue, 0 }  ,fill opacity=1 ] (104.75,70.34) .. controls (104.51,68.08) and (106.11,66.05) .. (108.32,65.81) .. controls (110.52,65.57) and (112.5,67.21) .. (112.73,69.47) .. controls (112.96,71.74) and (111.37,73.76) .. (109.16,74) .. controls (106.96,74.24) and (104.98,72.6) .. (104.75,70.34) -- cycle ;
\draw    (108.74,102.27) -- (147.55,103.81) ;
\draw  [fill={rgb, 255:red, 0; green, 0; blue, 0 }  ,fill opacity=1 ] (104.75,102.7) .. controls (104.51,100.44) and (106.11,98.41) .. (108.32,98.17) .. controls (110.52,97.93) and (112.5,99.57) .. (112.73,101.83) .. controls (112.96,104.1) and (111.37,106.12) .. (109.16,106.36) .. controls (106.96,106.6) and (104.98,104.96) .. (104.75,102.7) -- cycle ;
\draw    (108.74,37.55) -- (147.55,35.56) ;
\draw    (149.23,69.68) -- (108.74,69.91) ;
\draw    (76.43,36.18) .. controls (81.55,26.18) and (81.8,46.15) .. (87.32,36.18) ;
\draw    (87.32,36.18) .. controls (92.44,26.18) and (92.69,46.15) .. (98.2,36.18) ;
\draw    (98.2,36.18) .. controls (103.32,26.18) and (103.57,46.15) .. (109.09,36.18) ;
\draw    (65.55,36.18) .. controls (70.67,26.18) and (70.92,46.15) .. (76.43,36.18) ;
\draw    (75.35,103.23) .. controls (79.98,92.97) and (81.19,112.91) .. (86.22,102.66) ;
\draw    (86.22,102.66) .. controls (90.85,92.39) and (92.06,112.33) .. (97.09,102.08) ;
\draw    (97.09,102.08) .. controls (101.72,91.82) and (102.93,111.76) .. (107.96,101.51) ;
\draw    (64.47,103.8) .. controls (69.11,93.54) and (70.31,113.48) .. (75.35,103.23) ;
\draw    (108.74,69.91) -- (108.74,102.27) ;
\draw  [fill={rgb, 255:red, 0; green, 0; blue, 0 }  ,fill opacity=1 ] (104.75,37.98) .. controls (104.51,35.72) and (106.11,33.69) .. (108.32,33.45) .. controls (110.52,33.21) and (112.5,34.85) .. (112.73,37.11) .. controls (112.96,39.38) and (111.37,41.4) .. (109.16,41.64) .. controls (106.96,41.88) and (104.98,40.24) .. (104.75,37.98) -- cycle ;
\draw    (107.98,36.43) .. controls (117.74,41.32) and (98.75,42.53) .. (108.49,47.84) ;
\draw    (108.49,47.84) .. controls (118.26,52.73) and (99.26,53.95) .. (109.01,59.26) ;
\draw    (109.01,59.26) .. controls (118.77,64.15) and (99.78,65.36) .. (109.53,70.67) ;

\end{tikzpicture} & \raisebox{1cm}{$FGG$} \\

\tikzset{every picture/.style={line width=0.75pt}} 

\begin{tikzpicture}[x=0.75pt,y=0.75pt,yscale=-1,xscale=1]

\draw  [fill={rgb, 255:red, 0; green, 0; blue, 0 }  ,fill opacity=1 ] (79.65,75.59) .. controls (79.42,73.32) and (81.01,71.3) .. (83.22,71.06) .. controls (85.42,70.82) and (87.4,72.45) .. (87.63,74.72) .. controls (87.87,76.98) and (86.27,79.01) .. (84.07,79.25) .. controls (81.86,79.49) and (79.88,77.85) .. (79.65,75.59) -- cycle ;
\draw  [fill={rgb, 255:red, 0; green, 0; blue, 0 }  ,fill opacity=1 ] (112.09,76) .. controls (111.86,73.74) and (113.46,71.71) .. (115.66,71.47) .. controls (117.87,71.23) and (119.84,72.87) .. (120.08,75.13) .. controls (120.31,77.4) and (118.71,79.42) .. (116.51,79.66) .. controls (114.3,79.9) and (112.33,78.26) .. (112.09,76) -- cycle ;
\draw    (147.96,76.59) -- (155.31,104.4) ;
\draw    (116.09,75.57) -- (116.09,107.93) ;
\draw    (151.95,76.16) -- (116.09,75.57) ;
\draw  [fill={rgb, 255:red, 0; green, 0; blue, 0 }  ,fill opacity=1 ] (143.97,77.02) .. controls (143.73,74.76) and (145.33,72.73) .. (147.53,72.49) .. controls (149.74,72.25) and (151.71,73.89) .. (151.95,76.16) .. controls (152.18,78.42) and (150.59,80.45) .. (148.38,80.69) .. controls (146.18,80.93) and (144.2,79.29) .. (143.97,77.02) -- cycle ;
\draw    (147.96,46) -- (147.96,76.59) ;
\draw    (84.07,76.98) .. controls (88.36,66.56) and (90.23,86.44) .. (94.92,76.01) ;
\draw    (94.92,76.01) .. controls (99.21,65.58) and (101.08,85.46) .. (105.77,75.03) ;
\draw    (105.77,75.03) .. controls (110.05,64.61) and (111.93,84.49) .. (116.61,74.06) ;
\draw    (55.12,55.07) .. controls (64.13,48.78) and (55.38,66.52) .. (64.72,60.46) ;
\draw    (64.72,60.46) .. controls (73.73,54.17) and (64.98,71.91) .. (74.32,65.85) ;
\draw    (74.32,65.85) .. controls (83.33,59.56) and (74.58,77.3) .. (83.92,71.24) ;
\draw    (45.52,49.68) .. controls (54.53,43.39) and (45.78,61.13) .. (55.12,55.07) ;
\draw    (56.73,93.92) .. controls (54.92,82.73) and (66.8,98.33) .. (65.32,86.91) ;
\draw    (65.32,86.91) .. controls (63.52,75.71) and (75.39,91.32) .. (73.92,79.89) ;
\draw    (73.92,79.89) .. controls (72.11,68.7) and (83.98,84.31) .. (82.51,72.88) ;
\draw    (48.13,100.94) .. controls (46.33,89.74) and (58.2,105.35) .. (56.73,93.92) ;

\end{tikzpicture} & \raisebox{0.7cm}{$CGP$} &

\tikzset{every picture/.style={line width=0.75pt}} 

\begin{tikzpicture}[x=0.75pt,y=0.75pt,yscale=-1,xscale=1]

\draw  [fill={rgb, 255:red, 0; green, 0; blue, 0 }  ,fill opacity=1 ] (108.46,79.28) .. controls (108.22,77.02) and (109.82,74.99) .. (112.02,74.75) .. controls (114.23,74.51) and (116.21,76.15) .. (116.44,78.42) .. controls (116.67,80.68) and (115.08,82.71) .. (112.87,82.95) .. controls (110.67,83.19) and (108.69,81.55) .. (108.46,79.28) -- cycle ;
\draw  [fill={rgb, 255:red, 0; green, 0; blue, 0 }  ,fill opacity=1 ] (133.82,95.53) .. controls (133.58,93.27) and (135.18,91.24) .. (137.38,91) .. controls (139.59,90.76) and (141.56,92.4) .. (141.8,94.67) .. controls (142.03,96.93) and (140.44,98.96) .. (138.23,99.2) .. controls (136.03,99.44) and (134.05,97.8) .. (133.82,95.53) -- cycle ;
\draw    (112.45,46.49) -- (147.33,26.85) ;
\draw    (137.81,95.1) -- (169.19,113.93) ;
\draw    (164.7,69.21) -- (137.81,95.1) ;
\draw    (83.65,30.32) .. controls (92.66,24.04) and (83.91,41.78) .. (93.25,35.71) ;
\draw    (93.25,35.71) .. controls (102.26,29.42) and (93.51,47.16) .. (102.85,41.1) ;
\draw    (102.85,41.1) .. controls (111.86,34.81) and (103.11,52.55) .. (112.45,46.49) ;
\draw    (74.05,24.94) .. controls (83.06,18.65) and (74.31,36.39) .. (83.65,30.32) ;
\draw    (86.93,99.18) .. controls (85.13,87.99) and (97,103.6) .. (95.53,92.17) ;
\draw    (95.53,92.17) .. controls (93.72,80.97) and (105.6,96.58) .. (104.12,85.15) ;
\draw    (104.12,85.15) .. controls (102.32,73.96) and (114.19,89.57) .. (112.72,78.14) ;
\draw    (78.34,106.2) .. controls (76.53,95) and (88.41,110.61) .. (86.93,99.18) ;
\draw  [fill={rgb, 255:red, 0; green, 0; blue, 0 }  ,fill opacity=1 ] (108.46,46.92) .. controls (108.22,44.66) and (109.82,42.63) .. (112.02,42.39) .. controls (114.23,42.15) and (116.21,43.79) .. (116.44,46.06) .. controls (116.67,48.32) and (115.08,50.35) .. (112.87,50.59) .. controls (110.67,50.83) and (108.69,49.19) .. (108.46,46.92) -- cycle ;
\draw    (111.1,78.98) .. controls (120.16,72.78) and (111.26,90.43) .. (120.65,84.46) ;
\draw    (120.65,84.46) .. controls (129.72,78.26) and (120.81,95.91) .. (130.21,89.94) ;
\draw    (130.21,89.94) .. controls (139.27,83.74) and (130.36,101.4) .. (139.76,95.42) ;
\draw    (112.45,46.49) -- (112.45,78.85) ;

\end{tikzpicture} & \raisebox{1cm}{$FGG$} &

\tikzset{every picture/.style={line width=0.75pt}} 

\begin{tikzpicture}[x=0.75pt,y=0.75pt,yscale=-1,xscale=1]

\draw  [fill={rgb, 255:red, 0; green, 0; blue, 0 }  ,fill opacity=1 ] (86.43,73.34) .. controls (86.2,71.08) and (87.79,69.05) .. (90,68.81) .. controls (92.2,68.57) and (94.18,70.21) .. (94.41,72.47) .. controls (94.65,74.74) and (93.05,76.76) .. (90.84,77) .. controls (88.64,77.24) and (86.66,75.6) .. (86.43,73.34) -- cycle ;
\draw    (90.42,105.27) -- (129.23,106.81) ;
\draw  [fill={rgb, 255:red, 0; green, 0; blue, 0 }  ,fill opacity=1 ] (86.43,105.7) .. controls (86.2,103.44) and (87.79,101.41) .. (90,101.17) .. controls (92.2,100.93) and (94.18,102.57) .. (94.41,104.83) .. controls (94.65,107.1) and (93.05,109.12) .. (90.84,109.36) .. controls (88.64,109.6) and (86.66,107.96) .. (86.43,105.7) -- cycle ;
\draw    (90.42,40.55) -- (129.23,38.56) ;
\draw    (130.91,72.68) -- (90.42,72.91) ;
\draw    (58.11,39.18) .. controls (63.23,29.18) and (63.48,49.15) .. (69,39.18) ;
\draw    (69,39.18) .. controls (74.12,29.18) and (74.37,49.15) .. (79.88,39.18) ;
\draw    (79.88,39.18) .. controls (85,29.18) and (85.25,49.15) .. (90.77,39.18) ;
\draw    (47.23,39.18) .. controls (52.35,29.18) and (52.6,49.15) .. (58.11,39.18) ;
\draw    (57.03,106.23) .. controls (61.66,95.97) and (62.87,115.91) .. (67.9,105.66) ;
\draw    (67.9,105.66) .. controls (72.53,95.39) and (73.74,115.33) .. (78.77,105.08) ;
\draw    (78.77,105.08) .. controls (83.4,94.82) and (84.61,114.76) .. (89.64,104.51) ;
\draw    (46.16,106.8) .. controls (50.79,96.54) and (52,116.48) .. (57.03,106.23) ;
\draw  [fill={rgb, 255:red, 0; green, 0; blue, 0 }  ,fill opacity=1 ] (86.43,40.98) .. controls (86.2,38.72) and (87.79,36.69) .. (90,36.45) .. controls (92.2,36.21) and (94.18,37.85) .. (94.41,40.11) .. controls (94.65,42.38) and (93.05,44.4) .. (90.84,44.64) .. controls (88.64,44.88) and (86.66,43.24) .. (86.43,40.98) -- cycle ;
\draw    (89.66,39.43) .. controls (99.42,44.32) and (80.43,45.53) .. (90.18,50.84) ;
\draw    (90.18,50.84) .. controls (99.94,55.73) and (80.95,56.95) .. (90.69,62.26) ;
\draw    (90.69,62.26) .. controls (100.45,67.15) and (81.46,68.36) .. (91.21,73.67) ;
\draw    (89.66,71.78) .. controls (99.42,76.68) and (80.43,77.89) .. (90.18,83.2) ;
\draw    (90.18,83.2) .. controls (99.94,88.09) and (80.95,89.31) .. (90.69,94.62) ;
\draw    (90.69,94.62) .. controls (100.45,99.51) and (81.46,100.72) .. (91.21,106.03) ;

\end{tikzpicture} & \raisebox{1cm}{$FFF$} \\

\raisebox{0.3cm}{
\tikzset{every picture/.style={line width=0.75pt}} 

\begin{tikzpicture}[x=0.75pt,y=0.75pt,yscale=-1,xscale=1]

\draw  [fill={rgb, 255:red, 0; green, 0; blue, 0 }  ,fill opacity=1 ] (78.01,69.59) .. controls (77.78,67.32) and (79.38,65.3) .. (81.58,65.06) .. controls (83.78,64.82) and (85.76,66.45) .. (86,68.72) .. controls (86.23,70.98) and (84.63,73.01) .. (82.43,73.25) .. controls (80.22,73.49) and (78.25,71.85) .. (78.01,69.59) -- cycle ;
\draw  [fill={rgb, 255:red, 0; green, 0; blue, 0 }  ,fill opacity=1 ] (110.46,70) .. controls (110.22,67.74) and (111.82,65.71) .. (114.02,65.47) .. controls (116.23,65.23) and (118.2,66.87) .. (118.44,69.13) .. controls (118.67,71.4) and (117.08,73.42) .. (114.87,73.66) .. controls (112.67,73.9) and (110.69,72.26) .. (110.46,70) -- cycle ;
\draw    (146.32,70.59) -- (153.67,98.4) ;
\draw    (114.45,69.57) -- (114.45,101.93) ;
\draw    (117.87,69.74) -- (82,69.15) ;
\draw  [fill={rgb, 255:red, 0; green, 0; blue, 0 }  ,fill opacity=1 ] (142.33,71.02) .. controls (142.09,68.76) and (143.69,66.73) .. (145.9,66.49) .. controls (148.1,66.25) and (150.08,67.89) .. (150.31,70.16) .. controls (150.54,72.42) and (148.95,74.45) .. (146.74,74.69) .. controls (144.54,74.93) and (142.56,73.29) .. (142.33,71.02) -- cycle ;
\draw    (146.32,40) -- (146.32,70.59) ;
\draw    (113.82,72.16) .. controls (118.11,61.73) and (119.98,81.61) .. (124.66,71.18) ;
\draw    (124.66,71.18) .. controls (128.95,60.76) and (130.82,80.64) .. (135.51,70.21) ;
\draw    (135.51,70.21) .. controls (139.8,59.78) and (141.67,79.66) .. (146.35,69.24) ;
\draw    (52.93,50.25) .. controls (61.93,43.96) and (53.18,61.7) .. (62.53,55.64) ;
\draw    (62.53,55.64) .. controls (71.53,49.35) and (62.78,67.09) .. (72.13,61.02) ;
\draw    (72.13,61.02) .. controls (81.13,54.74) and (72.38,72.47) .. (81.73,66.41) ;
\draw    (43.33,44.86) .. controls (52.33,38.57) and (43.58,56.31) .. (52.93,50.25) ;
\draw    (54.53,89.1) .. controls (52.72,77.9) and (64.6,93.51) .. (63.12,82.09) ;
\draw    (63.12,82.09) .. controls (61.32,70.89) and (73.19,86.5) .. (71.72,75.07) ;
\draw    (71.72,75.07) .. controls (69.91,63.87) and (81.79,79.48) .. (80.31,68.06) ;
\draw    (45.94,96.11) .. controls (44.13,84.92) and (56,100.53) .. (54.53,89.1) ;

\end{tikzpicture}} & \raisebox{1cm}{$FGG$} &

\tikzset{every picture/.style={line width=0.75pt}} 

\begin{tikzpicture}[x=0.75pt,y=0.75pt,yscale=-1,xscale=1]

\draw  [fill={rgb, 255:red, 0; green, 0; blue, 0 }  ,fill opacity=1 ] (88.7,78.28) .. controls (88.46,76.02) and (90.06,73.99) .. (92.27,73.75) .. controls (94.47,73.51) and (96.45,75.15) .. (96.68,77.42) .. controls (96.91,79.68) and (95.32,81.71) .. (93.11,81.95) .. controls (90.91,82.19) and (88.93,80.55) .. (88.7,78.28) -- cycle ;
\draw  [fill={rgb, 255:red, 0; green, 0; blue, 0 }  ,fill opacity=1 ] (114.06,94.53) .. controls (113.82,92.27) and (115.42,90.24) .. (117.62,90) .. controls (119.83,89.76) and (121.81,91.4) .. (122.04,93.67) .. controls (122.27,95.93) and (120.68,97.96) .. (118.47,98.2) .. controls (116.27,98.44) and (114.29,96.8) .. (114.06,94.53) -- cycle ;
\draw    (92.69,45.49) -- (127.57,25.85) ;
\draw    (118.05,94.1) -- (149.43,112.93) ;
\draw    (144.95,68.21) -- (118.05,94.1) ;
\draw    (63.89,29.32) .. controls (72.9,23.04) and (64.15,40.78) .. (73.49,34.71) ;
\draw    (73.49,34.71) .. controls (82.5,28.42) and (73.75,46.16) .. (83.09,40.1) ;
\draw    (83.09,40.1) .. controls (92.1,33.81) and (83.35,51.55) .. (92.69,45.49) ;
\draw    (54.29,23.94) .. controls (63.3,17.65) and (54.55,35.39) .. (63.89,29.32) ;
\draw    (67.18,98.18) .. controls (65.37,86.99) and (77.24,102.6) .. (75.77,91.17) ;
\draw    (75.77,91.17) .. controls (73.96,79.97) and (85.84,95.58) .. (84.36,84.15) ;
\draw    (84.36,84.15) .. controls (82.56,72.96) and (94.43,88.57) .. (92.96,77.14) ;
\draw    (58.58,105.2) .. controls (56.78,94) and (68.65,109.61) .. (67.18,98.18) ;
\draw  [fill={rgb, 255:red, 0; green, 0; blue, 0 }  ,fill opacity=1 ] (88.7,45.92) .. controls (88.46,43.66) and (90.06,41.63) .. (92.27,41.39) .. controls (94.47,41.15) and (96.45,42.79) .. (96.68,45.06) .. controls (96.91,47.32) and (95.32,49.35) .. (93.11,49.59) .. controls (90.91,49.83) and (88.93,48.19) .. (88.7,45.92) -- cycle ;
\draw    (92,45.54) .. controls (101.77,50.44) and (82.77,51.65) .. (92.52,56.96) ;
\draw    (92.52,56.96) .. controls (102.28,61.85) and (83.29,63.07) .. (93.04,68.38) ;
\draw    (93.04,68.38) .. controls (102.8,73.27) and (83.81,74.48) .. (93.56,79.79) ;
\draw    (91.34,77.98) .. controls (100.41,71.78) and (91.5,89.43) .. (100.89,83.46) ;
\draw    (100.89,83.46) .. controls (109.96,77.26) and (101.05,94.91) .. (110.45,88.94) ;
\draw    (110.45,88.94) .. controls (119.51,82.74) and (110.6,100.4) .. (120,94.42) ;

\end{tikzpicture} & \raisebox{1cm}{$FCG$} & & \\

\tikzset{every picture/.style={line width=0.75pt}} 

\begin{tikzpicture}[x=0.75pt,y=0.75pt,yscale=-1,xscale=1]

\draw  [fill={rgb, 255:red, 0; green, 0; blue, 0 }  ,fill opacity=1 ] (78.57,70.59) .. controls (78.34,68.32) and (79.94,66.3) .. (82.14,66.06) .. controls (84.34,65.82) and (86.32,67.45) .. (86.56,69.72) .. controls (86.79,71.98) and (85.19,74.01) .. (82.99,74.25) .. controls (80.78,74.49) and (78.81,72.85) .. (78.57,70.59) -- cycle ;
\draw  [fill={rgb, 255:red, 0; green, 0; blue, 0 }  ,fill opacity=1 ] (111.02,71) .. controls (110.78,68.74) and (112.38,66.71) .. (114.58,66.47) .. controls (116.79,66.23) and (118.76,67.87) .. (119,70.13) .. controls (119.23,72.4) and (117.64,74.42) .. (115.43,74.66) .. controls (113.23,74.9) and (111.25,73.26) .. (111.02,71) -- cycle ;
\draw    (146.88,71.59) -- (154.23,99.4) ;
\draw    (115.01,70.57) -- (115.01,102.93) ;
\draw  [fill={rgb, 255:red, 0; green, 0; blue, 0 }  ,fill opacity=1 ] (142.89,72.02) .. controls (142.65,69.76) and (144.25,67.73) .. (146.46,67.49) .. controls (148.66,67.25) and (150.64,68.89) .. (150.87,71.16) .. controls (151.1,73.42) and (149.51,75.45) .. (147.3,75.69) .. controls (145.1,75.93) and (143.12,74.29) .. (142.89,72.02) -- cycle ;
\draw    (146.88,41) -- (146.88,71.59) ;
\draw    (114.38,73.16) .. controls (118.67,62.73) and (120.54,82.61) .. (125.22,72.18) ;
\draw    (125.22,72.18) .. controls (129.51,61.76) and (131.38,81.64) .. (136.07,71.21) ;
\draw    (136.07,71.21) .. controls (140.36,60.78) and (142.23,80.66) .. (146.92,70.24) ;
\draw    (81.88,72.57) .. controls (86.17,62.14) and (88.04,82.02) .. (92.72,71.6) ;
\draw    (92.72,71.6) .. controls (97.01,61.17) and (98.88,81.05) .. (103.57,70.62) ;
\draw    (103.57,70.62) .. controls (107.86,60.2) and (109.73,80.08) .. (114.41,69.65) ;
\draw    (52.93,51.85) .. controls (61.93,45.56) and (53.18,63.3) .. (62.53,57.24) ;
\draw    (62.53,57.24) .. controls (71.53,50.95) and (62.78,68.69) .. (72.13,62.62) ;
\draw    (72.13,62.62) .. controls (81.13,56.34) and (72.38,74.08) .. (81.73,68.01) ;
\draw    (43.33,46.46) .. controls (52.33,40.17) and (43.58,57.91) .. (52.93,51.85) ;
\draw    (54.53,90.7) .. controls (52.72,79.5) and (64.6,95.11) .. (63.12,83.69) ;
\draw    (63.12,83.69) .. controls (61.32,72.49) and (73.19,88.1) .. (71.72,76.67) ;
\draw    (71.72,76.67) .. controls (69.91,65.47) and (81.79,81.08) .. (80.31,69.66) ;
\draw    (45.94,97.71) .. controls (44.13,86.52) and (56,102.13) .. (54.53,90.7) ;

\end{tikzpicture} & \raisebox{0.7cm}{$FCG$} & & & & \\
\hline
\multicolumn{6}{|c|}{

\tikzset{every picture/.style={line width=0.75pt}} 

\begin{tikzpicture}[x=0.75pt,y=0.75pt,yscale=-1,xscale=1]

\draw  [fill={rgb, 255:red, 0; green, 0; blue, 0 }  ,fill opacity=1 ] (211.68,38.25) .. controls (211.44,35.99) and (213.04,33.96) .. (215.25,33.72) .. controls (217.45,33.48) and (219.43,35.12) .. (219.66,37.38) .. controls (219.9,39.65) and (218.3,41.67) .. (216.09,41.91) .. controls (213.89,42.15) and (211.91,40.51) .. (211.68,38.25) -- cycle ;
\draw    (251.53,38.41) -- (215.67,37.82) ;
\draw    (186.59,18.91) .. controls (195.6,12.62) and (186.85,30.36) .. (196.19,24.3) ;
\draw    (196.19,24.3) .. controls (205.2,18.01) and (196.45,35.75) .. (205.79,29.69) ;
\draw    (205.79,29.69) .. controls (214.8,23.4) and (206.05,41.14) .. (215.39,35.08) ;
\draw    (188.2,57.77) .. controls (186.39,46.57) and (198.26,62.18) .. (196.79,50.75) ;
\draw    (196.79,50.75) .. controls (194.98,39.55) and (206.86,55.16) .. (205.39,43.74) ;
\draw    (205.39,43.74) .. controls (203.58,32.54) and (215.45,48.15) .. (213.98,36.72) ;
\draw  [fill={rgb, 255:red, 0; green, 0; blue, 0 }  ,fill opacity=1 ] (319.35,39.25) .. controls (319.11,36.99) and (320.71,34.96) .. (322.91,34.72) .. controls (325.12,34.48) and (327.09,36.12) .. (327.33,38.38) .. controls (327.56,40.65) and (325.97,42.67) .. (323.76,42.91) .. controls (321.56,43.15) and (319.58,41.51) .. (319.35,39.25) -- cycle ;
\draw    (359.2,39.41) -- (323.34,38.82) ;
\draw    (294.26,19.91) .. controls (303.27,13.62) and (294.52,31.36) .. (303.86,25.3) ;
\draw    (303.86,25.3) .. controls (312.87,19.01) and (304.12,36.75) .. (313.46,30.69) ;
\draw    (313.46,30.69) .. controls (322.47,24.4) and (313.72,42.14) .. (323.06,36.08) ;
\draw    (323.34,38.82) -- (295,55.15) ;
\draw  [fill={rgb, 255:red, 0; green, 0; blue, 0 }  ,fill opacity=1 ] (99.01,38.77) .. controls (98.78,36.51) and (100.38,34.48) .. (102.58,34.24) .. controls (104.78,34) and (106.76,35.64) .. (107,37.91) .. controls (107.23,40.17) and (105.63,42.2) .. (103.43,42.44) .. controls (101.22,42.68) and (99.25,41.04) .. (99.01,38.77) -- cycle ;
\draw    (73.93,19.43) .. controls (82.93,13.15) and (74.18,30.89) .. (83.53,24.82) ;
\draw    (83.53,24.82) .. controls (92.53,18.53) and (83.78,36.27) .. (93.13,30.21) ;
\draw    (93.13,30.21) .. controls (102.13,23.92) and (93.38,41.66) .. (102.73,35.6) ;
\draw    (75.53,58.29) .. controls (73.72,47.09) and (85.6,62.7) .. (84.12,51.27) ;
\draw    (84.12,51.27) .. controls (82.32,40.08) and (94.19,55.69) .. (92.72,44.26) ;
\draw    (92.72,44.26) .. controls (90.91,33.06) and (102.79,48.67) .. (101.31,37.24) ;
\draw  [fill={rgb, 255:red, 0; green, 0; blue, 0 }  ,fill opacity=1 ] (435.18,38.27) .. controls (434.94,36.01) and (436.54,33.98) .. (438.75,33.74) .. controls (440.95,33.5) and (442.93,35.14) .. (443.16,37.41) .. controls (443.4,39.67) and (441.8,41.7) .. (439.59,41.94) .. controls (437.39,42.18) and (435.41,40.54) .. (435.18,38.27) -- cycle ;
\draw    (475.03,38.43) -- (439.17,37.84) ;
\draw    (439.17,37.84) -- (411.5,54.65) ;
\draw    (113.87,38.92) .. controls (118.56,28.98) and (119.79,48.72) .. (124.88,38.82) ;
\draw    (124.88,38.82) .. controls (129.57,28.88) and (130.8,48.62) .. (135.89,38.72) ;
\draw    (102.86,39.01) .. controls (107.55,29.08) and (108.78,48.82) .. (113.87,38.92) ;
\draw    (411,23.15) -- (439.17,37.84) ;
\draw    (516.5,37.65) -- (544.99,37.74) ;
\draw  [fill={rgb, 255:red, 0; green, 0; blue, 0 }  ,fill opacity=1 ] (541,38.17) .. controls (540.76,35.91) and (542.36,33.88) .. (544.57,33.64) .. controls (546.77,33.4) and (548.75,35.04) .. (548.98,37.31) .. controls (549.22,39.57) and (547.62,41.6) .. (545.41,41.84) .. controls (543.21,42.08) and (541.23,40.44) .. (541,38.17) -- cycle ;
\draw    (544.99,37.74) -- (573,38.15) ;
\draw    (544.99,37.74) -- (545,59.65) ;
\draw    (545,15.15) -- (544.99,37.74) ;

\draw (161.17,29.33) node [anchor=north west][inner sep=0.75pt]    {$F:$};
\draw (271.5,30.83) node [anchor=north west][inner sep=0.75pt]    {$G:$};
\draw (45.17,30.52) node [anchor=north west][inner sep=0.75pt]    {$C:$};
\draw (386.67,31.35) node [anchor=north west][inner sep=0.75pt]    {$P:$};
\draw (493.33,30.85) node [anchor=north west][inner sep=0.75pt]    {$Q:$};

\end{tikzpicture}

}\\
\hline
    \end{tabular}
    \caption{Diagrams and associated coupling configurations for the $VV\phi\phi\phi$ process}
    \label{tab:Vphi5pt}
\end{table}

Although the gauge-invariance conditions for the couplings $\bar{M}^2$ and $\bar{P}$ in Eqs.~\eqref{eq:PbarConstraint}-\eqref{eq:MbarConstraint} are used to identify $Q$ involving vector indices, they are also useful in diagnosing tree unitarity of a system. For this reason, in the summary of tree unitarity conditions in Appendix~\ref{appendix:summary_conditions}, we also present the couplings $\bar{M}^2$ and $\bar{P}$ involving vector indices, together with the relations between the VEV-involving $\bar{P}$ and $Q$.

\section{Tree unitarity of dark photon scenarios}
\label{sec:dark_photon}

In Section~\ref{sec:lagrangian_construction}, we showed that the tree unitarity conditions for 4-pt amplitudes form the Lie algebra of a group, and that the conditions for higher-pt amplitudes are encoded in the gauge-invariance conditions. Collectively, these conditions provide a unified and compact formulation of tree unitarity. They allow one to directly check tree unitarity once the particle content is specified in the mass basis.

In this section, to demonstrate the utility of the explicit conditions summarized in Appendix~\ref{appendix:summary_conditions}, we examine the tree  unitarity of dark photon scenarios widely used in DM phenomenology. These scenarios introduce a massive spin-1 field that mediates interactions between the visible and dark sectors~\cite{Holdom:1985ag,Pospelov:2008zw,Fabbrichesi:2020wbt}.

A massive spin-1 field generically exhibits energy growth from its longitudinal mode, so tree unitarity is not guaranteed automatically. In SBGTs, this growth is canceled by Higgs exchange. Here, we analyze this issue systematically in three systems consisting of a massive dark photon $Z'$ and two DM species with spin 0 $(\phi_{\alpha},\phi_{\beta})$, spin 1/2 $(\psi_{\alpha},\psi_{\beta})$, or spin 1 $(V_{\alpha},V_{\beta})$. Since we consider dark photon scenarios, $Z'$ is assumed to couple to a pair of DM fields, i.e., the $VX_{\alpha}X_{\beta}$ couplings with $X=\phi,\psi,V$ are taken to be non-vanishing by definition.

Specifically, we examine whether tree unitarity can be maintained within a minimal field content in which both external states and internal propagating states are restricted to $Z'$ and the DM fields. We begin with a dark photon $Z'$ whose mass term is introduced by hand. In this case, we show that the theory leads to the elastic scenario, where the two DM species have equal masses for scalar and fermion cases. Next, we consider the dark photon $Z'$ acquiring its mass through SSB, which requires the introduction of an additional scalar field. This setup leads to the inelastic scenario with two distinct DM masses for scalar and fermion cases.

For the vector case, the situation is different. To generate the coupling $Z' V_{\alpha} V_{\beta}$, at least one scalar field responsible for SSB is required from the outset. We first show that introducing such a scalar results in equal masses for the $Z'$ and two vectors. However, unlike the scalar and fermion cases, introducing additional scalars responsible for SSB does not lead to an inelastic scenario. Instead, we show that introducing an additional massless vector allows one to differentiate the dark photon and the two vector masses.

\subsection{$Z'$--Scalar}

Let us first consider the $Z'$--scalar system including only a single massive vector boson $Z'$, which implies an Abelian gauge group. Specifically, the generator of the Abelian group can be expressed as
\begin{align}
T^{\capi{(Z')}}_{\capmi{IJ}}
=
\begin{pmatrix}
0&iG&\frac{i}{m_{Z'}}F_{\alpha}
\\
-iG&0&\frac{i}{m_{Z'}}F_{\beta}
\\
-\frac{i}{m_{Z'}}F_{\alpha}&-\frac{i}{m_{Z'}}F_{\beta}&0
\end{pmatrix}
\quad ({\cal I,J}=\alpha,\beta,Z'),
\label{eq:TZ'}
\end{align}
in terms of the $V\phi\phi$ and $VV\phi$ couplings, $G$ and $F$, where the column and row indices run over $\alpha,\beta,Z'$ in that order.

The 4-pt tree unitarity condition in Eq.~\eqref{eq:D_[T,T]} requires the generator $T^{\capi{(Z')}}$ to satisfy
\begin{align}
[T^{\capi{(Z')}}, T^{\capi{(Z')}}] = 0.
\end{align}
However, this relation holds automatically, implying that the $Z'$--scalar system satisfies the 4-pt tree unitarity condition without any adjustment of the couplings. We first consider the case in which the $Z'$ mass is introduced by hand, implying that there is no VEV in the system. Using Eq.~\eqref{eq:D_v=Tm}, we compute the VEV associated with the generator in Eq.~\eqref{eq:D_[T,T]}:
\begin{gather}
(T^{\capi{(Z')}}T^{\capi{(Z')}})(iT^{\capi{(Z')}}v )
=(F_{\alpha}^2+F_{\beta}^2+G^2)(iT^{\capi{(Z')}}v)
\nonumber
\\
\to \quad v_{i}=\frac{F_{i}}{(m_{Z'}^2G^2+F_{\alpha}^2+F_{\beta}^2)}
\quad (i=\alpha,\beta),
\label{eq:Z-scalar_vev}
\end{gather}
indicating that each component of the VEV is proportional to the coupling $F_{Z'Z'\phi_i}$ $(i=\alpha,\beta)$. Thus, by taking $F_i=0$ (and hence $v_i=0$), we obtain the reduced form of Eq.~\eqref{eq:TZ'}:
\begin{align}
T^{\capi{(Z')}}_{\capmi{IJ}}
=
\begin{pmatrix}
0&iG&0
\\
-iG&0&0
\\
0&0&0
\end{pmatrix}
\quad ({\cal I,J}=\alpha,\beta,Z').
\label{eq:TZ'2}
\end{align}

Using this generator, we examine the higher-pt tree unitarity conditions. The absence of a VEV $(v=0)$ implies that the quadratic and cubic couplings in the scalar potential in Eq.~\eqref{eq:D_bar_nobar} satisfy
\begin{align}
\bar{M}^2_{\capmi{IJ}}=M^2_{\capmi{IJ}},\quad \bar{P}_{\capmi{IJK}}
=P_{\capmi{IJK}}\quad 
\mbox{and}\quad M_{\capi{A}\capmi{I}}^2=P_{\capi{A}\capmi{IJ}}=0,
\end{align}
where couplings involving vector indices vanish (see Eqs.~\eqref{eq:D_M2_vector} and~\eqref{eq:D_P_vector}). Using the higher-pt tree unitarity condition for the quadratic coupling in Eq.~\eqref{eq:D_M2}, we obtain 
\begin{gather}
(iT^{\capi{(Z')}}_{i_1 j})
M^2_{j i_2}
+
(iT^{\capi{(Z')}}_{i_2 j})
M^2_{j i_1}
=[i T^{\capi{(Z')}},M^2]_{i_1i_2}=0 
\quad \to \quad m_{\alpha}=m_{\beta},
\end{gather}
which implies that $\phi_{\alpha}$ and $\phi_{\beta}$ have equal masses.

Since the generator in Eq.~\eqref{eq:TZ'2} acts only on the scalar sector, the condition for the cubic coupling $P$ in Eq.~\eqref{eq:D_P} implies
\begin{align}
&(iT^{\capi{(Z')}}_{i_1j})
P_{ji_2i_3}
+
\mbox{cyclic}(i_1,i_2,i_3)
=0,
\end{align}
which ensures that the quartic couplings involving vector indices vanish, i.e., $Q_{\capi{A}\capmi{IJK}}=0$. Thus, from the condition in Eq.~\eqref{eq:D_Q}, we obtain
\begin{align}
(iT^{\capi{(Z')}}_{i_1j})
Q_{j i_2i_3i_4}
+
\mbox{cyclic}(i_1,i_2,i_3,i_4)
=0.
\end{align}
Collecting all the conditions derived so far, we conclude that a $Z'$–scalar system with a manually introduced $Z'$ mass reduces to an elastic DM scenario. Furthermore, this system is equivalent to massive scalar QED.

To examine how to realize the inelastic scenario, we consider the case in which the $Z'$ mass arises from SSB. Using the relation in Eq.~\eqref{eq:D_lam=iTv}, we have
\begin{align}
\begin{pmatrix}
0&iG&\frac{i}{m_{Z'}}F_{\alpha}
\\
-iG&0&\frac{i}{m_{Z'}}F_{\beta}
\\
-\frac{i}{m_{Z'}}F_{\alpha}&-\frac{i}{m_{Z'}}F_{\beta}&0
\end{pmatrix}
\begin{pmatrix}
v_{\alpha} \\ v_{\beta}\\ 0
\end{pmatrix}
=
\begin{pmatrix}
0 \\ 0\\ m_{\capi{Z'}}
\end{pmatrix}
\;\; \; \to \;\; \;
G_{Z'\alpha\beta}=0,
\end{align}
forbidding the $Z'\phi_{\alpha}\phi_{\beta}$ interaction. This implies that
at least three scalars are required to realize the inelastic dark photon scenarios~\cite{Baek:2020owl}.

\subsection{$Z'$--Fermion}

In the system with the dark photon $Z'$ and two fermions $\psi_{\alpha}$ and $\psi_{\beta}$, we have the fermionic 4-pt tree unitarity condition in Eq.~\eqref{eq:D_HR-LH-TH}: 
\begin{gather}
H R
-L H=0,
\label{eq:mu_R^2+L^2_mu-2LmuR=0}
\end{gather}
where we have omitted the $Z'$ index for simplicity. For simplicity, we consider vector-like fermions with $R=L$. In the system with no scalar responsible for SSB, i.e., with no VEV, one finds $H=0$ by taking $v_i=0$ in Eq.~\eqref{eq:D_Ha}. This implies that the $Z'$--fermion system with no scalars responsible for SSB holds tree unitarity for 4-pt amplitudes.

To identify the intrinsic fermion mass matrix $\mu_0$ in Eq.~\eqref{eq:D_mu0}, we examine the relation in Eq.~\eqref{eq:D_Lmu0-mu0R}:
\begin{align}
[R,\mu_0]=0 \quad \mbox{with} \quad 
\mu_0=
\begin{pmatrix}
m_{\alpha}&0
\\
0&m_{\beta}
\end{pmatrix}.
\label{eq:elastic_Rmu}
\end{align}
Here, the identity component of $\mu_0$ does not contribute to Eq.~\eqref{eq:elastic_Rmu}, as it commutes with $R$. Thus, we use the parametrization:
\begin{gather}
R=
\begin{pmatrix}
r_1&r_0
\\
r_0^*&r_2
\end{pmatrix}
,\quad
\mu_{-}\equiv 
\begin{pmatrix}
0&0
\\
0&m_{\beta}-m_{\alpha}
\end{pmatrix},
\label{eq:R&mu}
\end{gather}
for the matrices $R$ and $\mu_{-}=\mu_0-m_{\alpha} I$. Note that the off-diagonal component of $R$ generates the $Z'\bar{\psi}_{\alpha}\psi_{\beta}$ interaction. Inserting Eq.~\eqref{eq:R&mu} into Eq.~\eqref{eq:elastic_Rmu}, a nonzero off-diagonal component of $R$ requires the mass difference to vanish, i.e., $m_{\alpha}=m_{\beta}$, leading to an elastic scenario. This system is equivalent to massive QED.

To realize the inelastic scenario, i.e. $m_{\alpha}\neq m_{\beta}$, we consider the $Z'$--fermion system with an additional scalar $h$ arising from SSB. Then, from Eq.~\eqref{eq:D_HR-LH-TH}, we have
\begin{align}
&HR
-LH
-T^{\capi{(Z')}}_{\capi{Z'} h}H_{h}=0,
\label{eq:inelastic_HR-LH}
\\
&H_{h}R
-LH_{h}
-T^{\capi{(Z')}}_{h \capi{Z'}}H=0,
\label{eq:inelastic_H_hR-LH_h}
\end{align}
where the generator $T^{\capi{(Z')}}$ is a $2\times 2$ matrix containing the $Z'Z'h$ coupling $F_h$:
\begin{gather}
T^{\capi{(Z')}}=
\begin{pmatrix}
0&\frac{i}{m_{Z'}}F_h
\\
-\frac{i}{m_{Z'}}F_h&0
\end{pmatrix},
\end{gather}
Here, following the same procedure as in Eq.~\eqref{eq:Z-scalar_vev}, one finds that the VEV is given by $v=F_h=m_{\capi{Z'}}$.

As in the $Z'$--fermion system without an additional scalar, we consider vector-like fermions $(R=L)$ for simplicity. Then, Eq.~\eqref{eq:inelastic_HR-LH} can be recast in the following form:
\begin{align}
[R,[R,H_h]]= H_h.
\label{eq:inelastic_[R[RH]]}
\end{align}
To examine Eq.~\eqref{eq:inelastic_[R[RH]]}, we take $H_h$ to be diagonal:\footnote{As shown in Eq.~\eqref{eq:D_mu0}, the matrix $H_h$ is not required to be diagonal, although $\mu_0+H_h v$ is diagonal.}
\begin{align}
H_h\equiv \mbox{diag}(h_1,h_2).
\label{eq:diag_H_h}
\end{align}
Then, inserting the explicit forms of $R$ and $H_h$ in Eqs.~\eqref{eq:R&mu} and~\eqref{eq:diag_H_h} into Eq.~\eqref{eq:inelastic_[R[RH]]}, we obtain
\begin{gather}
2|r_0|^2(h_1-h_2)
\begin{pmatrix}
1&0
\\
0&-1
\end{pmatrix}
=
\begin{pmatrix}
h_1&0
\\
0&h_2
\end{pmatrix}
\quad \to \quad h_1=-h_2\;\;\mbox{ and }\;\;|r_0|=\frac12.
\end{gather}
Recall that the intrinsic fermion mass matrix $\mu_0$ satisfying Eq.~\eqref{eq:elastic_Rmu} must be proportional to an identity matrix. Thus, from Eq.~\eqref{eq:D_mu0}, one observes that introducing a scalar responsible for SSB generates a mass splitting between the two fermions, leading to an inelastic scenario. A concrete example can be found in Ref.~\cite{Ko:2019wxq,Baek:2020owl}.

Collecting the results for the scalar and fermion cases, we observe the following. In the scalar case, the 4-pt tree unitarity condition in Eq.~\eqref{eq:D_[T,T]} is satisfied for both elastic and inelastic scenarios, whereas the fermionic condition in Eq.~\eqref{eq:D_HR-LH-TH} directly enforces the elastic scenario. This can cause confusion when examining only 4-pt unitarity, as the inelastic scalar case may suggest that the $Z'$--scalar system with two distinct scalar masses is UV safe without additional scalars, leading to incorrect conclusions. This provides a clear example of the necessity of examining full tree-level unitarity.

\subsection{$Z'$--Vector}

For the $Z'$--vector system, realizing the $Z'V_{\alpha}V_{\beta}$ coupling requires a non-Abelian gauge structure; introducing the $Z'$ mass by hand is incompatible with tree unitarity. For convenience, we label the three vectors as $V_1, V_2, V_3$, and later identify one of them as the dark photon $Z'$.

To obtain a massive three-vector system, we introduce an additional scalar responsible for SSB. Using Eq.~\eqref{eq:D_T}, the vector components of the generators are given by
\begin{align}
&T^{(a)}_{bc}=i g
\frac{(m_{a}^2-m_{b}^2-m_{c}^2)}{2m_{b}m_{c}}  \,\veps_{abc} 
\quad (a,b,c=V_1,V_2,V_3),
\end{align}
with the $V_1V_2V_3$ coupling $g$. For the generator $T^{\capi{(V_1)}}$, we can write
\begin{gather}
T^{\capi{(V_1)}}_{\capmi{IJ}}
=
\begin{pmatrix}
0&i\frac{m_1}{v_h}&0&0
\\
-i\frac{m_1}{v_h}& & &
\\
0& &T^{\capi{(V_1)}}_{ab}&
\\
0& & & 
\end{pmatrix}
,\quad 
({\cal I,J}=h,V_1,V_2,V_3)
\label{eq:TV1_1h}
\end{gather}
where the row and column indices are ordered as $h, V_1, V_2, V_3$, and $a, b$ denote vector indices. This parametrization is chosen to satisfy $(iT^{\capi{(a)}}v){\capmi{I}}=\delta_{\capmi{I}a}m_a$ with the VEV vector $v=(v_h,0,0,0)$. In this way, one can easily construct the other generators, $T^{\capi{(V_2)}}$ and $T^{\capi{(V_3)}}$. Inserting Eq.~\eqref{eq:TV1_1h} into the 4-pt tree unitarity condition in Eq.~\eqref{eq:D_[T,T]}, we find that all three vector masses must be identical:
\begin{gather}
[T^{(a)},T^{(b)}]-iC_{abc}T^{(c)}
= 0 \;\;\; \to \;\;\; m_1=m_2=m_3=\frac{g v}{2} \quad (g>0),
\end{gather}
implying the symmetry-breaking pattern of $SU(2) \to \varnothing$. This degeneracy is not a consequence of our parametrization, but follows directly from the underlying Lie algebra structure imposed by tree unitarity.

In analogy with the scalar and fermion cases, one may attempt to introduce two scalars responsible for SSB to realize a mass splitting such as $m_1 \neq m_2 = m_3$. However, this does not produce the desired structure, as the vector masses are tightly constrained by the gauge symmetry. Thus, we instead introduce an additional massless vector, enlarging the gauge symmetry. This allows nontrivial mixing that leads to mass splitting.

For the couplings among four vector particles, we take the structure constants to be
\begin{gather}
C_{V_1V_2V_3}=g_1,\quad C_{V_2V_3V_4}=g_2,
\end{gather}
with two independent couplings, $g_1$ and $g_2$. In this case, the generator $T^{\capi{(V_1)}}$ is
\begin{gather}
T^{\capi{(V_1)}}_{\capmi{IJ}}
=
\begin{pmatrix}
0&i \frac{m_i}{v_h} &0&0&0
\\
-i \frac{m_i}{v_h}& & & &
\\
0 & && &
\\
0& & &T^{\capi{(V_1)}}_{\capi{AB}} &
\\
0& & & &
\end{pmatrix}
,\quad 
({\cal I,J}=h,V_1,V_2,V_3,V_4)
\label{eq:TV1_1h(4V)}
\end{gather}
where we use the capital vector indices, $A$ and $B$, to denote the generators involving also the massless vector index. Here, we set $m_1=0$ for the vector $V_1$. Inserting Eq.~\eqref{eq:TV1_1h(4V)} into the 4-pt tree unitarity condition in Eq.~\eqref{eq:D_[T,T]}, we obtain the mass spectrum:
\begin{gather}
m_1=0,\quad  m_2=m_3=\frac{\sqrt{g_1^2+g_2^2}}{2}v ,\quad m_4=\frac{\sqrt{g_1^2+g_2^2}}{2g_2}v,
\end{gather}
which exhibits the mass splitting $m_2=m_3\neq m_4$.
Our setup with this mass splitting corresponds to the symmetry-breaking pattern $SU(2)\times U(1)\to U(1)$, as in the SM. In this case, the mass splitting arises from diagonalizing the vector mass matrix, rather than from introducing an additional scalar responsible for SSB. This highlights a key difference from the scalar and fermion cases, where SSB alone is sufficient to generate mass splittings.

This indicates that, in contrast to lower-spin cases, tree unitarity tightly constrains the vector sector; in other words, the vector mass spectrum is severely restricted by tree unitarity, which requires an underlying gauge structure. In particular, achieving a nondegenerate vector mass spectrum requires a nontrivial extension of the gauge sector, rather than a minimal SSB setup.

This nontrivial feature of the dark photon scenario with vector DM makes it challenging to construct a realistic model, since the additional vectors must be massive; otherwise, they would already have been observed experimentally. Therefore, constructing a consistent inelastic vector DM scenario requires going beyond simple extensions of the scalar sector and instead relies on a nontrivial gauge structure. Thus, we leave the realization of inelastic vector scenarios for future work.

So far, we have examined the tree unitarity of the massive vector dark matter (VDM) system mediated by a massive dark photon. Given the VDM linked to the SM sector, one can also consider the Higgs portal scenario, which has attracted significant interest~\cite{Hambye:2009fg,Baek:2012se,Lebedev:2011iq,Baek:2013dwa,Arcadi:2021mag}. In the EFT perspective, a simplified model can be cast as 
\begin{equation}
{\cal L}_{\rm VDM} = - \frac{1}{4} V_{\mu\nu} V^{\mu\nu}
+ \frac{1}{2} m_V^2 V_\mu V^\mu + \lambda_{HV} H^\dagger H V_\mu V^\mu ,
\label{eq:vdm}
\end{equation}
which involves only two parameters: the VDM mass $m_V$ and the Higgs portal coupling $\lambda_{HV}$. However, the tree-unitary Lagrangian [Eq.~\eqref{eq:sbgt_L}] implies that, to satisfy tree unitarity, the operator $H^\dagger H V_\mu V^\mu$ must originate from a covariant derivative structure. In this sense, Eq.~\eqref{eq:vdm} does not by itself define a tree-unitary theory. A tree-unitary completion can be achieved, for instance, by introducing a Higgs-like scalar whose covariant derivative contains the $V$ field~\cite{Hambye:2009fg,Baek:2012se,Lebedev:2011iq,Arcadi:2021mag}.
As discussed, the tree unitarity violation leads to amplitudes diverging as $m_V \rightarrow 0$. This causes a serious phenomenological problem in this model: the Higgs invisible decay width $(h \to VV)$ is divergent in the limit $m_V \rightarrow 0$, being unphysical. This disaster can be resolved if the VDM mass is generated through the dark Higgs mechanism~\cite{Baek:2014jga,Baek:2021hnl}.

Despite these problems, simplified VDM descriptions are still commonly employed as phenomenological benchmarks in collider and astroparticle studies. In collider analyses in particular, tree unitarity is not merely a formal requirement: once violated, it signals the breakdown of the EFT above its cutoff scale ($\Lambda = m_V$), beyond which the predictions become unreliable. Therefore, the interpretation of collider searches based on Eq.~\eqref{eq:vdm} is intrinsically cutoff-sensitive. Our results provide a direct criterion to diagnose this limitation. Thus, we leave a more systematic analysis of tree unitarity in VDM models, including fully consistent completions and their phenomenological implications, to future work.

To summarize our analysis, we present in Table~\ref{tab:Higgs_require} the requirements for an additional scalar responsible for SSB (a Higgs-like scalar) in elastic and inelastic DM scenarios, classified by the DM spin.

\begin{table}[tbp]
    \centering
    \begin{tabular}{|c|c|c|}
    \hline\hline
       \multicolumn{3}{c}{Requirement of Higgs-like scalar}
       \\
        \hline\hline
          spin & elastic DM  & inelastic DM  \\
         \hline
        0 & $\times$  & $\bigcirc$ \\
        \hline
        1/2 & $\times$  & $\bigcirc$ \\
        \hline
        1 & $\bigcirc$ & $\bigcirc$ \\
        \hline
    \end{tabular}
    \caption{Summary of the Higgs-like scalar requirement for elastic and inelastic dark photon scenarios for dark matter with spin up to 1.}
    \label{tab:Higgs_require}
\end{table}

\section{Tree unitarity without scalars}
\label{sec:higgsless}

In Section~\ref{sec:lagrangian_construction}, we showed that tree-unitary theories with an arbitrary but finite number of particles of spin up to 1 can be mapped to SBGTs supplemented by additional Abelian mass terms. In such setups, the cancelation of energy-growing terms in scattering amplitudes is typically achieved through scalar exchange, indicating that scalar degrees of freedom play a central role in maintaining tree unitarity. 

However, since the coupling conditions for tree unitarity derived in Section~\ref{sec:4-pt_tree_unitarity} were obtained in a general setup, they can also be used to infer the tree unitarity conditions even in theories without scalars. In this section, by examining several simple cases, we find that an infinite tower of particles is required.

Let us first consider the coupling condition in Eq.~\eqref{eq:relation_VVVV2} for the $VVVV$ process in the absence of scalar contributions, i.e., $F_{VV\phi}=0$. For simplicity, we set $a_1=a_3$ and $a_2=a_4$, and take all external vector masses to be equal, $m_i=m_a$ $(i=1,2,3,4)$. Then, we obtain
\begin{align}
    \sum_{b}C_{a_1a_2b}^2\left(3m_b^2-4m_a^2\right)=0\ .\label{eq:V4Higgsless}
\end{align}
However, this relation cannot be satisfied in general. In particular, if $m_a$ is the maximal vector mass in the spectrum, each term in the sum is non-positive, making the left-hand side strictly negative. This leads to a contradiction, indicating that the relation cannot be fulfilled within a finite set of vector states. Such a condition instead points toward the necessity of an infinite tower of vector states, as realized in extra-dimensional Higgsless constructions~\cite{Csaki:2003dt,Csaki:2003zu,Csaki:2003sh,Csaki:2004sz}.

Next, we consider the coupling condition in Eq.~\eqref{eq:relation_VpsiVpsi2} for the $V\bar{\psi}V\psi$ process with no scalar exchanges, i.e., $H_{\phi}=F_{VV\phi}=0$. Taking the external vectors and fermions to have equal masses within each sector, i.e., $m_1=m_3=m_a$ and $m_2=m_4=m_{\rm i}$, we obtain
\begin{align}
    \sum_{\rm j}4m_{\rm j}L^a_{\rm ij}R^a_{\rm ji}-2\left((L^a_{\rm ij})^2+(R^a_{\rm ij})^2 \right)m_{\rm i}=0\ .
\end{align}
One can confirm that the LHS above is strictly negative when $m_{\rm i}$ is the maximal fermion mass:
\begin{align}
    \sum_{\rm j}4m_{\rm j}L^a_{\rm ij}R^a_{\rm ji}-2\left((L^a_{\rm ij})^2+(R^a_{\rm ij})^2 \right)m_{\rm i}<-\sum_{\rm j}2m_{\rm j}(L^a_{\rm ij}-R^a_{\rm ij})^2\leq 0\ .
\end{align}
This again leads to a contradiction, indicating that the condition cannot be satisfied within a finite set of fermion states. As in the vector case, this suggests the necessity of an infinite tower of fermions, as realized in extra-dimensional Higgsless constructions~\cite{Csaki:2003dt,Csaki:2003zu,Csaki:2003sh,Csaki:2004sz}.

\section{Conclusion}
\label{sec:conclusion}

We have derived explicitly all tree unitarity conditions on couplings in theories with a finite number of massive and massless particles of spin up to 1. By presenting these conditions in a unified form, we have shown that they enable the diagnosis of the tree unitarity of a system using only its particle content in the mass basis. In particular, they allow tree unitarity to be assessed without reconstructing the full Lagrangian, while maintaining a direct connection to it.

To reach this result, we have shown that in tree-unitary theories, one can define appropriate smooth massless limits of amplitudes with respect to vector masses, and that any 4-pt amplitude whose high-energy growth is canceled by tree unitarity conditions is on-shell constructible. Thus, we have constructed the 4-pt amplitudes using the recently developed recursive framework, referred to as the ALT shift. We then collected all tree unitarity conditions for 4-pt amplitudes in the high-energy limit.

To systematically derive the tree unitarity conditions for higher-pt amplitudes, we adopted the St\"uckelberg formulation, which changes the field content. By showing the equivalence between tree-level amplitudes from the original and St\"uckelberg Lagrangians, we have identified that in the formulation, all tree-unitarity-violating contributions arise from higher-pt contact terms in the scalar potential. By eliminating these contact terms, we derive all higher-pt tree unitarity conditions, which are revealed as gauge-invariance conditions on the couplings in the scalar potential. Unlike in previous works, we derive all tree unitarity conditions in an explicit form without any transformation of the coupling basis. This allowed us to show that all tree-unitarity conditions are fully captured up to bosonic 5-pt amplitudes.

We have demonstrated the utility of our conditions by applying them to dark photon scenarios with two DM particles of spin up to 1. For scalar and fermion DM, we found that the inelastic scenario can be realized by introducing an additional scalar responsible for SSB. The need for such an introduction follows from tree unitarity of higher-pt amplitudes in the scalar case, whereas in the fermion case, it already follows from 4-pt amplitudes. This shows that imposing tree unitarity only at the level of 4-pt amplitudes can lead to incorrect conclusions.

For vector DM, we showed that, unlike the scalar and fermion cases, even a mass splitting between the dark photon and the two vector DM particles cannot be realized by merely introducing an additional scalar; instead, it requires an additional massless vector. 
This is because the mass spectrum is severely constrained by the gauge group structure. Next, by examining several simple processes in theories without scalars, we observed that satisfying tree unitarity requires the presence of an infinite tower of vectors and fermions.

Finally, we emphasize that our explicit results can be directly applied to phenomenological model building to assess tree unitarity in realistic scenarios. A natural next step is to develop a systematic approach to probing UV sensitivity beyond strict tree unitarity in theories with higher-dimensional operators, and further in those with higher-spin particles.

\acknowledgments
This work was supported by 
KIAS Individual Grants (QP090001) [J.J.] and (QP021402) [P.K.] through the 
Quantum Universe Center (QUC) at the Korea Institute for Advanced Study (KIAS), 
and (PG096402) [Y.-H.Z.] through the School of Physics at KIAS.
P.K was also supported by the National Research Foundation of Korea (NRF) funded by the Ministry of Education, Science and Technology (RS-2025-24803289) [PK].

\appendix

\section{Spinor-helicity formalism
}
\label{appendix:spinor_helicity_formalism}

In this appendix, we provide a concise review of the spinor-helicity formalism \cite{Elvang:2015rqa,Arkani-Hamed:2017jhn}, along with the conventions and relations employed in constructing the on-shell amplitudes and operators discussed in the main text.

\subsection{Spinor-helicity variables}

The $SL(2,\mathbb{C}$) group is the double cover of the $SO(1,3)$ Lorentz group, so we can express any on-shell amplitude as a function of two-component 
spinor-helicity variables, the building blocks in the spinor-helicity formalism. A complex momentum $p^\mu$ with energy $E=\sqrt{\vec p^{\,2}+m^2}$ can be expressed through the spinor variables (or helicity spinors) as:
\begin{align}
\begin{aligned}
p_{\alpha\dot{\alpha}}&=p^\mu \sigma_{\mu\alpha\dot{\alpha}}
=\left(
       \begin{array}{cc}
        E-p^3 & -p^1+ip^2 
        \\
        -p^1-ip^2 & E+p^3    
       \end{array}
 \right)_{\!\!\alpha\dot{\alpha}}
=\lambda^I_{\alpha}\tilde{\lambda}_{\dot{\alpha}I}
\\
\bar{p}^{\dot{\alpha}\alpha}&=p^\mu \bar{\sigma}_{\mu}^{\dot{\alpha} \alpha }
=\left(
       \begin{array}{cc}
        E+p^3 & p^1-ip^2 
        \\
        p^1+ip^2 & E-p^3 
       \end{array}
 \right)_{\!\!\dot{\alpha}\alpha}
=\tilde{\lambda}^{\dot{\alpha}}_{I}
\lambda^{\alpha I},
\end{aligned}
\label{eq:momentum}
\end{align}
where we adopt the signature $(1,-1,-1,-1)$, with the four-vector Pauli matrices defined as $\sigma^\mu_{\alpha\dot{\alpha}}=(\delta_{\alpha\dot{\alpha}},\vec\sigma_{\alpha\dot{\alpha}})$ and
$\bar{\sigma}^{\mu\dot{\alpha}\alpha}=(\delta^{\dot{\alpha}\alpha},-\vec\sigma^{\dot{\alpha}\alpha})$.
Here, $\alpha, \dot{\alpha} = 1,2$ denote the $SL(2,\mathbb{C})$ Lorentz indices, and $I=1,2$ labels the $SU(2)$ little-group index, thereby $p_{\alpha\dot{\alpha}}$ and $\bar{p}^{\dot{\alpha}\alpha}$ are invariant under the little group transformation,
\begin{align}
\lam^I_{\alpha}\rightarrow U^I_{\;\;\,J}    \lam^J_{\alpha}
,\quad
\tilde{\lam}_{\dot{\alpha}I}\rightarrow U_{I}^{\;\,\, J}\tilde{\lam}_{\dot{\alpha}J}
,\quad
\tilde{\lam}_{I}^{\dot{\alpha}}
\rightarrow U_{I}^{\;\,\, J}\tilde{\lam}_{J}^{\dot{\alpha}}
,\quad
\lam^{\alpha I}\rightarrow U^I_{\;\;\,J}    \lam^{\alpha J},
\end{align}
with $(U^{-1})^I_{\;\;\,J}=U_{J}^{\;\,\, I}$.
The raising and lowering of indices will be specified later. For a massless particle, the little group reduces to $U(1)$, removing the index $I$ as $\lam^I_{\alpha},\tilde{\lam}_{\dot{\alpha} I} \!\to\! \lam_{\alpha},\tilde{\lam}_{\dot{\alpha}}$ and $\tilde{\lam}^{\dot{\alpha}}_I,\lam^{\alpha I} \!\to\! \tilde{\lam}^{\dot{\alpha}},\lam^{\alpha}$, leading to
$p_{\alpha\dot{\alpha}} = \lam_{\alpha}\tilde{\lam}_{\dot{\alpha}}$ and $\bar{p}^{\dot{\alpha}\alpha} = \tilde{\lam}^{\dot{\alpha}}\lam^{\alpha}$.
Given this smooth reduction from massive to massless spinors, we mainly discuss the massive spinors, referring to the massless case whenever needed.

The determinants of expressions in Eq.~\eqref{eq:momentum} lead to the mass squired as
\begin{align}
\mbox{det}|p_{\alpha\dot\alpha}|
=
\mbox{det}|\bar{p}^{\alpha\dot\alpha}|
=\frac12 p_{\alpha\dot{\alpha}}
\bar{p}^{\dot{\alpha}\alpha}
=m^2,
\label{eq:p_square}
\end{align}
indicating $p_{\alpha\dot\alpha}$ and $\bar{p}^{\alpha\dot\alpha}$, the rank-2 matrix, while the massless case with $\mbox{det}|p_{\alpha\dot\alpha}|=0$ implies the rank-1 matrix. We take the conventions on the determinants of $\lam^{I}_{\alpha}$ and $\tilde{\lam}_{\dot{\alpha} I}$ as
\begin{align}
\mbox{det}|\lam^I_{\alpha}|
=m,\quad 
\mbox{det}|\tilde{\lam}_{\dot{\alpha} I}|=m,
\label{eq:spinor_determinants}
\end{align}
and on the momentum sign flip as
\begin{align}
\lam^I_{\alpha}(-p)
=-\lam^I_{\alpha}(p),
\quad
\tilde{\lam}_{\dot{\alpha}I}(-p)
=\tilde{\lam}_{\dot{\alpha}I}(p),
\label{eq:continutation}
\end{align}
with $\lam_{\alpha}(-p)
=-\lam_{\alpha}(p)$ and $\tilde{\lam}_{\dot{\alpha}}(-p)
=\tilde{\lam}_{\dot{\alpha}}(p)$ for the massless case, 
where we regarded $I$ as a column index and $\alpha$ and $\dot{\alpha}$ as row indices.

The super- and sub-indices of spinors can be raised and lowered using the two-component Levi-Civita tensor:
\begin{align}
\epsilon^{\alpha\beta}\lam^I_{\beta}
=\lam^{\alpha I},\;\;
\epsilon_{IJ}\lam^J_{\alpha}
=\lam_{\alpha I},\;\;
\epsilon_{\dot{\alpha}\dot{\beta}}
\tilde{\lam}^{\dot{\beta} I}
=\tilde{\lam}^I_{\dot{\alpha}},\;\;
\epsilon_{IJ}
\tilde{\lam}^{\dot{\alpha}J}
=\tilde{\lam}^{\dot{\alpha}}_I,
\end{align}
where the Levi-Civita tensor satisfies
\begin{align}
\epsilon^{\alpha\beta} \epsilon_{\beta\gamma}
=\delta^{\alpha}_{\gamma}
,\quad\;\;
\epsilon_{\alpha_1\alpha_2} \epsilon_{\beta_1\beta_2}
\epsilon^{\alpha_2\beta_2}
=-\epsilon_{\alpha_1\beta_1}.
\end{align}
Our convention~\cite{Elvang:2015rqa,Arkani-Hamed:2017jhn} of the Levi-Civita tensor is,
\begin{align}
\epsilon^{12}=-\epsilon^{21}
=\epsilon_{21}=-\epsilon_{12}
=1,
\end{align}
for $\epsilon^{\alpha\beta}$ and $\epsilon^{IJ}$ with the Lorentz and little group indices.

For compactness, we adopt the ``half-bracket" notation~\cite{Durieux:2019eor,Durieux:2020gip}:
\begin{align}
\lam_{\alpha}^I\equiv |p^I \rangle\equiv |\bs{p}\rangle\equiv |\bs{i}\rangle,\quad&
\lam^{\alpha I}\equiv \langle p^I |\equiv \langle \bs{p}|\equiv \langle \bs{i}|,
\nonumber
\\
\tilde{\lam}^{\dot{\alpha} I}\equiv |p^I]\equiv |\bs{p}]\equiv |\bs{i}],\quad &
\tilde{\lam}^{I}_{\dot{\alpha}}\equiv [p^I|\equiv [\bs{p}|\equiv [\bs{i}|,
\label{eq:massive_notation}
\end{align}
where the {\bf bold} symbol is introduced to omit the little-group index and to distinguish them from the massless ones, while $\bs{i}$ denotes the $i$-th particle label, mainly used in practical computations. For the massless spinors, we have
\begin{align}
\begin{aligned}
\lam_{\alpha}\equiv |p \rangle\equiv |i\rangle,\quad&
\lam^{\alpha }\equiv \langle p |\equiv \langle i|,
\\
\tilde{\lam}^{\dot{\alpha}}\equiv |p]\equiv |i],\quad &
\tilde{\lam}_{\dot{\alpha}}\equiv [p|\equiv [i|.
\end{aligned}
\label{eq:massless_notation}
\end{align}
Using Eqs.~\eqref{eq:massive_notation} and~\eqref{eq:massless_notation}, the momentum expressions in Eq.~\eqref{eq:momentum} can be expressed as
\begin{align}
\bs{p}=|p^I\rangle [p_I|,\quad 
\bar{\bs{p}}=|p_I] \langle p^I|
\quad \mbox{and} \quad 
p=|p\rangle [ p|,\quad 
\bar{p}=|p] \langle p|,
\end{align}
in the massive and massless cases, respectively. In addition, we can write the spinor products in a simple form:
\begin{align}
\begin{gathered}
\ab{p^Iq^J}
=\lam^{\alpha I}\chi^J_{\alpha}
=\epsilon_{\alpha\beta}
\lam^{\alpha I}\chi^{\beta J}
=-\epsilon_{\beta\alpha}
\chi^{\beta J} \lam^{\alpha I}
=-\chi^{\beta J} \lam^{I}_{\beta}
=-\ab{q^Jp^I},
\\
[p^{I}
q^J]
=\tilde{\lam}^{I}_{\dot{\alpha}}
\tilde{\chi}^{\dot{\alpha} J}
=\epsilon^{\dot{\alpha}\dot{\beta}}
\tilde{\lam}^{I}_{\dot{\alpha}}
\tilde{\chi}^{ J}_{\dot{\beta}}
=-\epsilon^{\dot{\beta}\dot{\alpha}}
\tilde{\chi}^{ J}_{\dot{\beta}}
\tilde{\lam}^{I}_{\dot{\alpha}}
=-
\tilde{\chi}^{ J}_{\dot{\beta}}
\tilde{\lam}^{\dot{\beta} I}
=-[q^Jp^{I}],
\end{gathered}
\label{eq:spinor_contraction}
\end{align}
and
\begin{align}
\langle p^I | \sigma^\mu |q^J]
=
[q^J | \bar{\sigma}^\mu |p^I\rangle,
\label{eq:asb_sigma_identity}
\end{align}
in terms of the spinors, $\lam(p)$ and $\chi(q)$. From Eq.~\eqref{eq:spinor_determinants}, we can extract the following identities:
\begin{align}
&\ab{p^I p^J}=-m\epsilon^{IJ},\qquad\quad
\sqb{p^I p^J}=m\epsilon^{IJ},
\label{eq:spinor_identities1}
\\
&\epsilon_{IJ}|p^I\rangle_{\alpha}{}^{\beta}\! \langle p^J|
=-m \delta^{\beta}_{\alpha},\quad 
\epsilon_{IJ}|p^I]^{\dot{\alpha}} {}_{\dot{\beta}}[p^J|
=m \delta^{\dot{\alpha}}_{\dot{\beta}},
\label{eq:spinor_identities2}
\end{align}
with $\ab{pp}=0$ and $\sqb{pp}=0$ for the massless case, ensured by Eq.~\eqref{eq:spinor_contraction}.

Introducing the basis in the little group space, we can expand the massive spinors, $\lam^I$ and $\tilde{\lam}^I$, respectively, in terms of two independent spinor variables as
\begin{align}
\lam^I_{\alpha}=\lam_{\alpha}\zeta^{-I}+\eta_{\alpha}\zeta^{+I},
\quad
\tilde{\lam}^I_{\dot{\alpha}}=
\tilde{\lam}_{\dot{\alpha}}\zeta^{+I}-\tilde{\eta}_{\dot{\alpha}}\zeta^{-I},
\label{eq:massive-two_massless}
\end{align}
where the two-component vectors $\zeta^{\pm}$ are
\begin{align}
\zeta^{-I}=
\begin{pmatrix}
0 \\ 1
\end{pmatrix}
,\quad
\zeta^{+I}=
\begin{pmatrix}
1 \\ 0
\end{pmatrix},
\end{align}
satisfying $\epsilon_{IJ}\zeta^{-I}\zeta^{+J}=1$. 
Here, $\zeta^{\mp J}$ are defined such that their components are directly related to the helicities. We also introduce the following half-bracket notation for the $i$-th particle:
\begin{align}
\lam_{\alpha}=|i\rangle=|p^2\rangle,\quad 
\tilde{\lam}_{\dot{\alpha}}=[i|=[p_2|
,\quad
\eta_{\alpha}=|\eta_i\rangle=|p^1\rangle,\quad \tilde{\eta}_{\dot{\alpha}}=[\eta_i|=-[p_1|,
\label{eq:two_bracket_notations}
\end{align}
which is convenient for analyzing the massless or high-energy limits of massive spinors.

Then, in terms of $p_{\alpha\dot{\alpha}}$, we have 
\begin{align}
p_{\alpha\dot{\alpha}}=|i\rangle [i|-|\eta_i\rangle [\eta_i|,
\label{eq:p_lam_eta}
\end{align}
and Eq.~\eqref{eq:spinor_identities1} leads to
\begin{align}
\ab{i\eta_i}=m_i,\quad \sqb{i\eta_i}=m_i.
\label{eq:lam_eta_m}
\end{align}

In practical computations in this work, we use the following explicit forms of spinor variables in Eq.~\eqref{eq:massive-two_massless}   
\begin{align}
\begin{aligned}
|i\rangle=|p^2\rangle=
\sqrt{E+p}\left(\begin{array}{c} -s^*_{\theta}
\\
c_{\theta}\end{array}\right),
&\quad
[i|=[p_2|
=\sqrt{E+p}\left(\begin{array}{c} -s_{\theta}
\\
c_{\theta}\end{array}\right),
\\
|\eta_i\rangle=|p^1\rangle=
\sqrt{E-p}\left(\begin{array}{c} c_{\theta}
\\
s_{\theta}\end{array}\right),
&\quad 
[\eta_i|=-[p_1|=
-\sqrt{E-p}\left(\begin{array}{c} c_{\theta}
\\
s^*_{\theta}\end{array}\right),
\end{aligned}
\label{eq:lam_eta_spinors_explicit}
\end{align}
in the case for the real momentum $p^\mu=(E,p\sin\theta\cos\phi,p\sin\theta\cos\phi,p\cos\theta)$, where $c_{\theta}=\cos\frac{\theta}{2}$ and $s_{\theta}=\sin\frac{\theta}{2}e^{i\phi}$, including the polar and azimuthal angles, $\theta$ and $\phi$. Using Eq.~\eqref{eq:lam_eta_spinors_explicit} and Eq.~\eqref{eq:continutation}, one finds that the spinors $\lam^I$ and $\tilde{\lam}^I$ with the real momentum are related by the complex conjugation as
\begin{align}
(\lam^I_{\alpha})^*=\pm \tilde{\lam}_{I\dot{\alpha}},\quad
(\tilde{\lam}^I_{\dot{\alpha}})^*=\mp \lam_{I\dot{\alpha}},
\label{eq:spinor_conjugate}
\end{align}
for the positive $(+ \mbox{ and } -)$ or negative $(- \mbox{ and } +)$ energy. This allows us to have
\begin{align}
\ab{i^I j^J}=\pm \sqb{j_J i_I},
\label{eq:spinor_conjugate2}
\end{align}
for the equal $(+)$ or for opposite $(-)$ energy signs.

The explicit forms of $\lambda^I$ and $\tilde{\lambda}^I$ for an arbitrary complex momentum can be obtained by extending Eq.~\eqref{eq:lam_eta_spinors_explicit} through the following substitutions:\footnote{For generality, one may also include a complex overall factor in each of $\lambda,\tilde{\lambda}$ and $\eta,\tilde{\eta}$, which mutually cancel in $p_{\alpha\dot{\alpha}}$.}
\begin{align}
\begin{gathered}
p\rightarrow \sqrt{(p^1)^2+(p^2)^2+(p^3)^2},\quad 
c_\theta \rightarrow \frac{p^3}{\sqrt{(p^1)^2+(p^2)^2+(p^3)^2}},
\\
s_\theta \rightarrow \frac{p^1+ip^2}{\sqrt{(p^1)^2+(p^2)^2+(p^3)^2}}
,\quad
s^*_\theta \rightarrow \frac{p^1-ip^2}{\sqrt{(p^1)^2+(p^2)^2+(p^3)^2}},
\end{gathered}
\label{eq:spinor_complexify}
\end{align}
showing that the energy power counting of $\lambda^I$ and $\tilde{\lambda}^I$ with Eq.~\eqref{eq:lam_eta_spinors_explicit} remains valid even for complex momenta in the massless limit.

\subsection{Dirac spinors and Polarization vectors}
\label{appendix:u_and_eps}

To translate on-shell amplitudes into the operators, we examine the Dirac spinors and polarization vectors in the spinor representation. Contracting $|p^I \rangle$ and $| p^I]$ to Eq.~\eqref{eq:spinor_identities2}, we obtain the Dirac equation for the helicity spinors:
\begin{align}
&\bar{p}|p^I\rangle=
m|p^I]
,\quad
p|p^I]=
m|p^I\rangle,
\quad
\langle p^I| p
=-m[p^I|
,\quad
[p^I| p
=-m\langle p^I|.
\label{eq:spinor_dirac_eq}
\end{align}
Using the above, we obtain
\begin{align}
\begin{array}{l}
(\slash\!\!\! p -m)u^I(p)=0,
\quad 
(\slash\!\!\! p +m)v_I(p)=0,
\\
\bar{u}_I(p)(\slash\!\!\! p -m)=0,
\quad 
\bar{v}^I(p)(\slash\!\!\! p +m)=0,
\end{array}
\quad \mbox{with}
\quad 
\slash\!\!\! p \equiv p^\mu \gamma_\mu,\quad 
\gamma^\mu=
\begin{pmatrix}
0&\sigma^\mu_{\alpha\dot{\beta}}
\\
\bar{\sigma}^{\mu \dot{\alpha}\beta}&0
\end{pmatrix},
\end{align}
for the Dirac spinors of which the solutions are given by
\begin{align}
\begin{aligned}
&u^I(p)
=
\begin{pmatrix}
\lam^I_{\alpha}
\\
\tilde{\lam}^{\dot{\alpha}I}
\end{pmatrix}
=
\begin{pmatrix}
|p^I\rangle
\\
|p^I]
\end{pmatrix}
,\qquad\qquad\quad\,
v_I(p)
=
\begin{pmatrix}
\lam_{I\alpha}
\\
-\tilde{\lam}^{\dot{\alpha}}_I
\end{pmatrix}
=
\begin{pmatrix}
|p_I\rangle
\\
-|p_I]
\end{pmatrix},
\\
&
\bar{u}_I(p)
=
\begin{pmatrix}
-\lam^{\alpha}_{I}
&
\tilde{\lam}_{\dot{\alpha} I}
\end{pmatrix}
=
\begin{pmatrix}
-\langle p_I|
&
[p_I|
\end{pmatrix}
,\quad
\bar{v}^I(p)
=-
\begin{pmatrix}
\lam^{I\alpha}
&
\tilde{\lam}^I_{\dot{\alpha}}
\end{pmatrix}
=-
\begin{pmatrix}
\langle p^I|
&
[ p^I|
\end{pmatrix},
\end{aligned}
\label{eq:uv_to_spinors}
\end{align}
with the helicity spinors. In the conventional representation~\cite{Jacob:1959at}, the charge conjugation is given by $i\gamma^2 u^{h*}= v^h$ with the helicity $h = \mp 1/2$. Based on this charge conjugation and Eq.~\eqref{eq:spinor_conjugate}, we determined the positions of the little-group indices in the $u^I$ and $v_I$ spinors, yielding the correspondence $I = 2,1 \leftrightarrow h = -1/2, +1/2$.

For ease of reference, we present the conventional expressions of $u$ and $v$ spinors:
\begin{align}
u^h(p)=
\begin{pmatrix}
\sqrt{E-2hp}\;\xi_h(\hat{p})
\\
\sqrt{E+2hp}\;\xi_h(\hat{p})
\end{pmatrix}
,\quad
v^h(p)=2h
\begin{pmatrix}
-\sqrt{E+2hp}\;\xi_{-h}(\hat{p})
\\
\sqrt{E-2hp}\;\xi_{-h}(\hat{p})
\end{pmatrix},
\label{eq:conventional_u_and_v}
\end{align}
for the real momentum where $\xi_h(\hat{p})$\; ($h=\mp 1/2$) are
\begin{align}
\xi_{-\frac12}(\hat{p})
=\left(\begin{array}{c} -s_{\theta}^*\\c_{\theta}\end{array}\right)
,\quad
\xi_{+\frac12}(\hat{p})
=\left(\begin{array}{c} c_{\theta}\\s_{\theta}\end{array}\right).
\end{align}

Unlike the Dirac spinors, the polarization vectors cannot be settled down smoothly to the massless one from the massive one due to the longitudinal one. We first present two useful identities in the discussion of the polarization vectors:
\begin{align}
\sigma^\mu_{\alpha\dot{\alpha}}
\bar{\sigma}^{\dot{\beta}\beta}_{\mu}
=2\delta^{\beta}_{\alpha}
\delta^{\dot{\beta}}_{\dot{\alpha}}
,\quad 
\sigma^\mu_{\alpha\dot{\alpha}}
\bar{\sigma}^{\nu \dot{\alpha}\alpha}
=2\eta^{\mu\nu}.
\label{eq:pauli_iden}
\end{align}
For the massless case, the polarization vectors with helicities $\pm 1$ can be defined by
\begin{align}
\varepsilon_{\mu}^{+}
=\frac{\langle r|\sigma_{\mu}|p]}{\sqrt{2}\langle pr\rangle},
,\quad
\varepsilon_{\mu}^{-}
=\frac{\langle p|\sigma_{\mu}|r]}{\sqrt{2}[ pr]},
\label{eq:massless_polarization}
\end{align}
in terms of a reference momentum $r$ satisfying $r\cdot p\neq 0$. Using the first one in Eq.~\eqref{eq:pauli_iden}, one finds $\epsilon^\pm \cdot \epsilon^{\mp}=1$, $\epsilon^\pm \cdot \epsilon^{\pm}=0$, and $\epsilon^{\pm}\cdot p=0$, the minimal condition on the polarization vectors. 
Contracting the Pauli matrices $\bar{\sigma}_{\mu}$ to Eq.~\eqref{eq:massless_polarization} using the second one in Eq.~\eqref{eq:pauli_iden}, we have the following alternative expressions:
\begin{align}
\varepsilon_{\alpha\dot{\alpha}}^{-}
=\frac{\sqrt{2}|p\rangle [ r|}{[ pr]}
,\quad
\varepsilon_{\alpha\dot{\alpha}}^{+}
=\frac{\sqrt{2}|r\rangle [p|}{\langle pr\rangle}.
\label{eq:eps_massless_spinor}
\end{align}
An arbitrary spinor $|q\rangle$ satisfies the Schouten identities for massless spinors,
\begin{align}
&|q\rangle \ab{pr}
+
|r\rangle \ab{qp}
+
|p\rangle \ab{rq}=0,
\\
&|q]\sqb{pr}
+
|r]\sqb{qp}
+
|p]\sqb{rq}=0.
\end{align}
Taking the shift $r\rightarrow r+p\equiv q$ and using the identities above, one finds
\begin{align}
\epsilon^+ \;\rightarrow \;
\epsilon^++\frac{\sqrt{2}}{\ab{pr}}p,
,\quad
\epsilon^- \;\rightarrow \;
\epsilon^-+\frac{\sqrt{2}}{\sqb{pr}}p
\label{eq:gauge_trans}
\end{align}
Any on-shell amplitude must be independent of the reference momentum. Therefore, the shift in Eq.~\eqref{eq:gauge_trans} induced by $r \to r+p$ leaves the amplitude invariant, implying that the momentum component of the polarization vector does not contribute to the on-shell amplitude — a direct consequence of gauge invariance. 
\\

For a massive vector boson, the polarization vectors are given with the massive spinor as
\begin{align}
\epsilon^{IJ}_{\mu}
=
\left\{
\begin{array}{ll}
\displaystyle
\frac{\langle p^I|\sigma_{\mu}|p^J]}{\sqrt{2}m}
&\mbox{for $I=J$}
\\[10pt]
\displaystyle
\frac{\langle p^I|\sigma_{\mu}|p^J]
+\langle p^J|\sigma_{\mu}|p^I]}{2\sqrt{2}m}
&\mbox{for $I\neq J$}
\end{array}
\right.,
\label{eq:massive_polarization}
\end{align}
symmetrized over the little-group indices $I,J$, satisfying $\epsilon^{IJ}_{\mu}\epsilon^{\mu I'J'}=(\veps^{II'}\veps^{JJ'}+\veps^{IJ'}\veps^{JI'})/2$ and $\epsilon^{IJ}_{\mu} p^\mu=0$.\footnote{We have consistently distinguished the polarization vectors and the Levi-Civita tensor by $\eps$ and $\veps$, respectively, but we explicitly write the four-vector index here to avoid any possible confusion.} This symmetrization leads the three independent polarization vectors, $\epsilon^{22}$, $\epsilon^{11}$, and $\epsilon^{12}=\epsilon^{21}$, with the helicities $-1, +1$, and $0$, respectively. Similarly to the case of a massless vector boson, contracting the Pauli matrices $\sigma_\mu$ to Eq.~\eqref{eq:massive_polarization} yields
\begin{align}
\epsilon^{IJ}_{\alpha\dot{\alpha}}
=
\left\{
\begin{array}{ll}
\displaystyle
\frac{\sqrt{2}| p^I\rangle [p^J|}{m}
&\mbox{for $I=J$}
\\[10pt]
\displaystyle
\frac{| p^I\rangle [p^J|+| p^J\rangle [p^I|}{\sqrt{2}m}
&\mbox{for $I\neq J$}
\end{array}
\right..
\label{eq:massive_polarization2}
\end{align}
Omitting the little group indices in the spinors, we simplifies Eq.~\eqref{eq:massive_polarization2},
\begin{align}
\epsilon^{IJ}_{\alpha\dot{\alpha}}
\equiv\frac{|\bs{p}\rangle [\bs{p}|}{\sqrt{2}m},
\label{eq:eps_massive_spinor}
\end{align}
according to the notational convention in Eq.~\eqref{eq:massive_notation}. Using Eq.~\eqref{eq:lam_eta_spinors_explicit}, we find the conventional expressions of polarization vectors for the real momentum:
\begin{align}
\eps^{\mp}_{\mu}=\frac{e^{\mp i\phi}}{\sqrt{2}}(0,\pm \hat{\theta}+i\hat{\phi}) 
\quad \;\;\mbox{and}\quad \;\;
\eps^{0}_{\mu}=\frac{1}{m}(p,-E \hat{p}),
\label{eq:conventional_polarization}
\end{align}
in terms of
\begin{align}
\hat{\theta}&=(\cos\theta \cos\phi,\cos\theta \cos\phi,\sin\theta),
\\
\hat{\phi}&=(-\sin\phi,\cos\phi ,0),
\end{align}
satisfying $\eps^\pm\cdot \eps^\mp=1$, $\eps^0\cdot \eps^0=-1$, and $\eps^{h}\cdot p=0$ $(h=\pm 1,0)$.

\subsection{3-pt on-shell amplitudes of massless particles}
\label{appendix:3-pt_massless_on-shell}

The 3-pt on-shell amplitudes of massless particles are uniquely determined by their helicities. Regarding all particles as incoming, from momentum conservation we obtain
\begin{align}
&|1\rangle [1|+|2\rangle [2|+|3\rangle [3|=0
\;\;\rightarrow \;\;
\left\{
\begin{array}{l}
\ab{12}\sqb{12}=0
\\
\ab{23}\sqb{23}=0
\\
\ab{31}\sqb{31}=0
\end{array}
\right.
\;\;\;\mbox{and}\;\;\;
\left\{
\begin{array}{l}
\ab{12}\sqb{32}=0
\\
\ab{23}\sqb{13}=0
\\
\ab{31}\sqb{21}=0
\end{array}
\right.,
\label{eq:momentum_cons_massless}
\end{align}
which enforce either $\ab{12} = \ab{23} = \ab{31} = 0$ or $\sqb{12} = \sqb{23} = \sqb{31} = 0$, implying that $|1\rangle \propto |2\rangle \propto |3\rangle$ or $|1] \propto |2] \propto |3]$. Therefore, the amplitudes can be constructed in a purely holomorphic (or anti-holomorphic) kinematic configuration without loss of generality.\footnote{The term “holomorphic” (“anti-holomorphic”) refers to the angle (square) bracket.} For the holomorphic case, we have
\begin{align}
\mathcal{M}_3(1^{h_1}2^{h_2}3^{h_3})=\ab{12}^{n_1}\ab{23}^{n_2}\ab{31}^{n_3},
\end{align}
in terms of an arbitrary integer powers $n_1,n_2,n_3$. The little-group scaling ensures to express the exponents in terms of the helicities $h_1,h_2,h_3$ as
\begin{align}
\begin{array}{l}
n_1+n_3=-2h_1
\\
n_2+n_1=-2h_2
\\
n_3+n_2=-2h_3
\end{array}
\quad\rightarrow\quad
\begin{array}{l}
n_1=h_1-h_2-h_3
\\
n_2=h_2-h_3-h_1
\\
n_3=h_3-h_1-h_2
\end{array}.
\end{align}
Since each of $h_i$ $(i = 1, 2, 3)$ is an integer or a half-integer and angular momentum conservation requires fermions to appear in pairs, $n_i$ $(i = 1, 2, 3)$ must be integers. Given that locality forbids 3-pt amplitudes with negative mass dimensions, \footnote{In nonlocal theories, inverse-derivative terms can generate amplitudes with negative mass dimensions.} the helicities are constrained as
\begin{align}
[\mathcal{M}_3(1^{h_1}2^{h_2}3^{h_3})]
=n_1+n_2+n_3=-(h_1+h_2+h_3)\equiv -h \geq 0,
\end{align}
yielding the conditions $h \leq 0$ ($h \geq 0$) for the holomorphic (anti-holomorphic) kinematics. Thus, the massless 3-pt on-shell amplitudes can be expressed in the following compact forms,
\begin{align}
\mathcal{M}_3
(1^{h_1}2^{h_2}3^{h_3})
=
\left\{
\begin{array}{ll}
\ab{12}^{h_3-h_1-h_2}
\ab{23}^{h_1-h_2-h_3}
\ab{31}^{h_2-h_3-h_1}
&\mbox{for} \; h\leq 0
\\[3pt]
\sqb{12}^{h_1+h_2-h_3}
\sqb{23}^{h_2+h_3-h_1}
\sqb{31}^{h_3+h_1-h_2}
&\mbox{for} \; h\geq 0
\end{array}
\right.,
\label{eq:3pt_massless_amps}
\end{align}
apart from the overall factors.

\subsection{Massless limits of spinor contractions}
\label{sec:massless_limits}

Using the explicit expressions in Eq.~\eqref{eq:lam_eta_spinors_explicit} with Eq.~\eqref{eq:spinor_complexify}, one observes that the spinor variables have the following leading terms in the massless limit,
\begin{gather}
|i\rangle \sim \sqrt{E}
,\quad |i] \sim \sqrt{E}
,\quad
|\eta_i\rangle \sim 
\frac{m}{\sqrt{E}}
,\quad |\eta_i] \sim \frac{m}{\sqrt{E}},
\label{eq:leading_term_massless_limit}
\end{gather}
for the $i$-th particle, where the the spinors $|i\rangle$ and $|i]$ reduce to the massless ones. In Section~\ref{appendix:3-pt_massless_on-shell}, it has been discussed that the momentum conservation enforces that the massless 3-pt on-shell amplitudes to be constructed only by holomolpic (anti-holomolpic) kinematic configurations. Specifically, using Eq.~\eqref{eq:momentum_cons_massless}, we have 
\begin{align}
|1]=\frac{\ab{23}}{\ab{31}}|2]=\frac{\ab{23}}{\ab{12}}|3]
\quad \mbox{or}\quad
|1\rangle=\frac{\sqb{23}}{\sqb{31}}|2\rangle=\frac{\sqb{23}}{\sqb{12}}|3\rangle,
\label{eq:holo_antiholo}
\end{align}
for holomorpic or antiholomolpic cases. Contracting $[ \eta_i|$ and $\langle \eta_i|$ respectively to the LHS and RHS in Eq.~\eqref{eq:holo_antiholo} for $i=1,2,3$ allows us to compute the massless limits of the following spinor products,
\begin{align}
\begin{split}
\frac{[1^12^2]}{m_2}=
\frac{[1\eta_2]}{\sqb{2\eta_2}} \to \frac{\ab{23}}{\ab{31}},
&\quad 
\frac{[2^13^2]}{m_3}=
\frac{[2\eta_3]}{\sqb{3\eta_3}}\to\frac{\ab{31}}{\ab{12}},
\quad 
\frac{[3^11^2]}{m_1}=
\frac{[3\eta_1]}{\sqb{1\eta_1}}\to\frac{\ab{12}}{\ab{23}},
\\
\frac{[1^22^1]}{m_1}=
\frac{[\eta_12]}{\sqb{1\eta_1}}\to-\frac{\ab{31}}{\ab{23}},
&\quad 
\frac{[2^23^1]}{m_2}=
\frac{[\eta_23]}{\sqb{2\eta_2}}\to-\frac{\ab{12}}{\ab{31}},
\quad 
\frac{[3^21^1]}{m_3}=\frac{[\eta_31]}{\sqb{3\eta_3}}\to-\frac{\ab{23}}{\ab{12}}.
\end{split}
\label{eq:massless_limits1}
\end{align}
and
\begin{gather}
\begin{gathered}
\frac{\ab{1^22^1}}{m_2}=
\frac{\ab{1\eta_2 }}{\ab{2\eta_2}}\to\frac{\sqb{23}}{\sqb{31}}
,\quad
\frac{\ab{2^23^1}}{m_3}=
\frac{\ab{2 \eta_3}}{\ab{3\eta_3}}\to\frac{\sqb{31}}{\sqb{12}}
,\quad
\frac{\ab{3^21^1}}{m_1}=
\frac{\ab{3\eta_1}}{\ab{1\eta_1}}\to\frac{\sqb{12}}{\sqb{23}},
\\
\frac{\ab{1^12^2}}{m_1}=
\frac{\ab{\eta_1 2}}{\ab{1\eta_1}}\to -\frac{\sqb{31}}{\sqb{23}}
,\quad
\frac{\ab{2^13^2}}{m_2}=
\frac{\ab{\eta_2 3}}{\ab{2\eta_2}}\to -\frac{\sqb{12}}{\sqb{31}}
,\quad
\frac{\ab{3^11^2}}{m_3}=
\frac{\ab{\eta_3 1}}{\ab{3\eta_3}}\to -\frac{\sqb{23}}{\sqb{12}},
\end{gathered}
\label{eq:massless_limits2}
\end{gather}
with $\sqb{i\eta_i}=m_i$ and $\ab{i\eta_i}=m_i$  in Eq.~\eqref{eq:lam_eta_m} (see Eq.~\eqref{eq:two_bracket_notations} for two half-bracket notations).

To verify the mass scaling behavior of the holomorphic (antiholomorphic) configuration in the massless limit, we begin by evaluating
\begin{align}
\ab{1^I 2^J}\sqb{2_{J} 1_I}=m_3^2-m_1^2-m_2^2
=\ab{12}\sqb{12}-\ab{\eta_1 2}\sqb{\eta_1 2}-
\ab{1 \eta_2}\sqb{1\eta_2}
+\ab{\eta_1 \eta_2}\sqb{\eta_1 \eta_2},
\end{align}
appearing in a 3-pt massive amplitude. Keeping only the leading terms in the massless limit, we obtain
\begin{align}
\sqb{12}\sim\frac{m_3^2-m_1^2-m_2^2}{\ab{12}}
,\quad
\ab{12}\sim \frac{m_3^2-m_1^2-m_2^2}{\sqb{12}},
\label{eq:[]_<>_suppress}
\end{align}
corresponding to the holomorphic and antiholomorphic configurations, respectively, and suppressed at the mass-squared order in the massless limit. In the same manner, one can find the mass scalings of other spinor products.

Using Eq.~\eqref{eq:lam_eta_spinors_explicit} with Eq.~\eqref{eq:spinor_complexify}, one finds that the transverse polarization vectors are convergent in the massless limit as 
\begin{gather}
\epsilon^{22}_{i}=\frac{\sqrt{2} |i^2\rangle[i^2|}{
m_i}
=
\frac{\sqrt{2} |i\rangle[ \eta_i|}{
\sqb{i\eta_i}}
,\quad
\epsilon^{11}_{i}=\frac{\sqrt{2} |i^1\rangle[i^1|}{
m_i}
=
\frac{\sqrt{2} |\eta_i\rangle[ i|}{
\ab{i\eta_i}}.
\end{gather}
On the other hand, the longitudinal one is divergent:
\begin{align}
\frac{| p^2\rangle [p^1|+| p^1\rangle [p^2|}{\sqrt{2}m}
=
\frac{| i\rangle [i|+| \eta_i\rangle [\eta_i|}{\sqrt{2}m}
\sim \frac{E}{m},
\end{align}
as expected.

\subsection{Useful identities}
\label{appendix:useful_identities}

We present several useful identities in the computations for deriving the results in the main text. We first cast the Schouten identities for massive spinors:
\begin{align}
&|\bs{1}\rangle\ab{\bs{23}}+
|\bs{2}\rangle\ab{\bs{31}}+
|\bs{3}\rangle\ab{\bs{12}}=0
,\quad |\bs{1}]\sqb{\bs{23}}+
|\bs{2}]\sqb{\bs{31}}+
|\bs{3}]\sqb{\bs{12}}=0.
\label{eq:massive_schouten}
\end{align}
For a 3-pt on-shell massive process with $\bs{p}_1+\bs{p}_2+\bs{p}_3=0$, using Eq.~\eqref{eq:massive_schouten}, we can compute the following,
\begin{align}
[\bs{232}\rangle
\ab{\bs{31}}\sqb{\bs{13}}
&=m_3 \ab{\bs{12}}\sqb{\bs{23}}\sqb{\bs{31}}
+m_1 \ab{\bs{23}}\sqb{\bs{31}}\sqb{\bs{12}}
-m_2 \ab{\bs{31}}\sqb{\bs{12}}\sqb{\bs{23}},
\\
\asb{\bs{232}}
\sqb{\bs{31}}\ab{\bs{13}}
&=m_3 \sqb{\bs{12}}\ab{\bs{23}}\ab{\bs{31}}
+m_1 \sqb{\bs{23}}\ab{\bs{31}}\ab{\bs{12}}
-m_2 \sqb{\bs{31}}\ab{\bs{12}}\ab{\bs{23}},
\end{align}
and also for the cyclically permutated cases. Eq.~\eqref{eq:asb_sigma_identity} ensures
\begin{align}
[\bs{232}\rangle
\ab{\bs{31}}\sqb{\bs{13}}+\mbox{c.p.t}
=
\asb{\bs{232}}
\sqb{\bs{31}}\ab{\bs{13}}+\mbox{c.p.t},
\end{align}
leading to
\begin{gather}
\begin{aligned}
m_3\ab{\bs{12}}\sqb{\bs{23}}\sqb{\bs{31}}
+\mbox{c.p.t.}
=m_3\sqb{\bs{12}}\ab{\bs{23}}\ab{\bs{31}}
+\mbox{c.p.t.},
\label{eq:m1m2m3_identity}
\end{aligned}
\end{gather}
where we used the Dirac equation in Eq.~\eqref{eq:spinor_dirac_eq}
This identity allows us to rewrite the 3-pt on-shell massive $VVV$ amplitude Eq.~\eqref{eq:VVV_amp} as
\begin{gather}
\frac{\ab{\bs{12}}\sqb{\bs{23}}\sqb{\bs{31}}}{m_1m_2}
+\mbox{c.p.t.}
=
\frac{[\bs{232}\rangle \ab{\bs{31}}\sqb{\bs{13}}}
{m_1m_2m_3}
+\mbox{c.p.t.},
\label{eq:3V-momentum_insert}
\end{gather}
where the RHS is the expression directly derived from the Feynman rule of $VVV$ operator.

\section{Tree level Hermicity}
\label{appendix:CPT}

The realness of the tree-level couplings follows from the unitarity of the $S$-matrix, $S = 1 + iT$. Using the relation $SS^\dagger = 1$, which ensures $T - T^\dagger = 0$ at tree level, we derive the realness. We begin by relating two equivalent expressions of an $n$-point amplitude.
\begin{align}
\ab{0|\mathcal{M}|p_1,h_1;\cdots;p_n,h_n}
=\mathcal{M}_n(\bs{1}^{\{I_1\}} \cdots \bs{a}^{\{I_a\}}\cdots \bar{\bs{b}}^{\{I_{b}\}}  \cdots \bs{n}^{\{I_n\}}),
\label{eq:hilbert_vs_spinor-helicity}
\end{align}
in the Hilbert-space and spinor-helicity representations, where the all-incoming convention corresponds to all initial states~\cite{Bachu:2023fjn}. Specifically, each particle in the two representations follows the correspondence: 
\begin{align}
|p_i,h_i\rangle \leftrightarrow \bs{i}^{\{I_i\}},
\end{align}
in terms of the helicity $h_i$ and the set of little-group indices $I_i$ of the $i$-th particle.

Using the expression of the LHS in Eq.~\eqref{eq:hilbert_vs_spinor-helicity} and CPT invariance, one finds that an $n$-pt tree-level amplitude follows\footnote{We intentionally don't use the expression $\mathcal{CPT}|p_1,h_1;\cdots;p_n,h_n\rangle$ to explicitly show the complex conjugation of the amplitude arising from the antiunitarity of the time-reversal.}  
\begin{align}
\ab{0|\mathcal{M}|p_1,h_1;\cdots;p_n,h_n}
=\Big[\ab{0|\mathcal{M}|p_1,h_1;\cdots;p_n,h_n}_{\mathcal{CPT}}\Big]^*
\label{eq:amp_CPT}
\end{align}
where the subscript ``$\mathcal{CPT}$’’ denotes the CPT-transformed state.  
To find the CPT transformation of each particle state, let us consider the CPT transformation of fields and their Fourier components.  
The antiunitarity of time reversal results in the complex conjugation of each coefficient of the state operator. Given this aspect, and using Eq.~\eqref{eq:conventional_u_and_v} and Eq.~\eqref{eq:conventional_polarization}, we find the following relations for the spin-$1/2$ and vector fields:
\begin{align}
\begin{array}{l}
u^{h*}(p)=-(-1)^{1/2-h}\gamma_5 \gamma^0\big[\bar{v}^{-h}(p)\big]^T,
\\
\bar{v}^{h*}(p)=(-1)^{1/2-h} \big[u^{-h}(p)\big]^T\gamma^0\gamma_5,
\\
\eps^{h*}_{\mu}(p)=-(-1)^{1-h}\eps^{-h}_{\mu}(p).
\end{array}
\end{align}
From the CPT transformations $\psi(x)\rightarrow -\gamma_5 \psi^*(-x)$ and $V_{\mu}(x) \rightarrow -V^*_{\mu}(-x)$, we find
\begin{align}
|p,h\rangle_{\mathcal{CPT}} \propto (-1)^{s-h}|p,-h\rangle_{\rm anti},
\end{align}
in terms of the spin $s$ and helicity $h$ of the particle,  
where the subscript “anti’’ denotes the antiparticle state.

The phase $(-1)^{s-h}$ determines how the Feynman rule of each particle is replaced when moving from the LHS to RHS of Eq.~\eqref{eq:amp_CPT}:  
\begin{align}
u^{I_i}(p_i) \;\;\rightarrow \;\; \bar{v}_{I_i}(p_i)
,\quad
\bar{v}^{I_i}(p_i) \;\;\rightarrow \;\; u_{I_i}(p_i)
,\quad 
\eps^{I_i J_i}(p_i) \;\;\rightarrow \;\; \eps_{I_i J_i}(p_i).
\end{align}
This allows us to rewrite Eq.~\eqref{eq:amp_CPT} directly in the spinor-helicity representation:
\begin{align}
&\mathcal{M}_n(\bs{1}^{\{I_1\}}\cdots \bs{a}^{\{I_a\}}  \cdots \bar{\bs{b}}^{\{I_{b}\}} \cdots \bs{n}^{\{I_n\}})
=\mathcal{M}_n^*(\bs{1}_{\{I_1\}} \cdots \bs{b}_{\{I_b\}} \cdots\bar{\bs{a}}_{\{I_{a}\}} \cdots  \bs{n}_{\{I_n\}}).
\end{align}
Consequently, using the spinor relations in Eq.~\eqref{eq:spinor_conjugate2}, one finds that all 3-pt amplitudes and 4-pt contact-interaction amplitudes require the bosonic (fermionic) couplings to be real (Hermitian) in the real-momentum limit.

\section{High-energy limits of spinor contractions}
\label{appendix:high-E_limits_spinor}

As discussed in Section~\ref{sec:tree_unitarity}, the High-energy limits of 4-pt on-shell amplitudes are investigated in the c.m. frame. The kinematic configurations of particles are set to be
\begin{align}
&(\theta_1,\phi_1)=(0,0),\quad 
(\theta_2,\phi_2)=(\pi,\pi),
\label{eq:kinematic1}
\\
&(\theta_3,\phi_3)=(\theta,\phi),\quad 
(\theta_4,\phi_4)=(\pi-\theta,\pi+\phi),\quad
\label{eq:kinematic2}
\end{align}
in terms of their polar and azimuthal angles. In this work, we take all the particles to be incoming so that $p_{1,2}=(E_{1,2},\vec p_{1,2})$ and $p_{3,4}=-(E_{3,4},\vec p_{3,4})$ with the positive $E_{1,2,3,4}$. The energy-momentum conservation compels the momenta and energies to be
\begin{align}
&|\vec p_1|=|\vec p_2|=\frac{E}{2}\sqrt{\bigg(1-\frac{m_1^2+m_2^2}{E^2}\bigg)^2-\frac{4m_1^2m_2^2}{E^2}},
\\
&|\vec p_3|=|\vec p_4|=\frac{E}{2}\sqrt{\bigg(1-\frac{m_3^2+m_4^2}{E^2}\bigg)^2-\frac{4m_3^2m_4^2}{E^2}},
\end{align}
and
\begin{align}
E_{1,2}=\frac{E+m_{1,2}^2-m_{2,1}^2}{2E}
,\quad
E_{3,4}=\frac{E+m_{3,4}^2-m_{4,3}^2}{2E}
\end{align}
in terms of the collision energy $E=\sqrt{(p_1+p_2)^2}=\sqrt{(p_3+p_4)^2}$.

In this work, our interest in the amplitudes is the leading order in energy. Thus, it is efficient to identify the leading terms of spinors and directly use them in the amplitude calculations. We first focus on the spinors diminishing in the high-energy limit. The kinematic configurations in Eqs.~\eqref{eq:kinematic1} and~\eqref{eq:kinematic2} enable us to approximate them as
\begin{align}
|\eta_1\rangle&\approx+\frac{m_1}{E}\bigg(1+\frac{m_1^2}{2E^2}\bigg)|2\rangle,
&
|\eta_1]&\approx-\frac{m_1}{E}\bigg(1+\frac{m_1^2}{2E^2}\bigg)|2],
\\
|\eta_2\rangle&\approx-\frac{m_2}{E}\bigg(1+\frac{m_2^2}{2E^2}\bigg)|1\rangle,
&
|\eta_2]&\approx+\frac{m_2}{E}\bigg(1+\frac{m_2^2}{2E^2}\bigg)|1],
\\
|\eta_3\rangle&\approx+\frac{m_3}{E}\bigg(1+\frac{m_3^2}{2E^2}\bigg)
e^{i\phi}|4\rangle ,
&
|\eta_3]&\approx-\frac{m_3}{E}\bigg(1+\frac{m_3^2}{2E^2}\bigg)
e^{-i\phi}|4],
\\
|\eta_4\rangle&\approx-\frac{m_4}{E}\bigg(1+\frac{m_4^2}{2E^2}\bigg)
e^{i\phi}|3\rangle,
&
|\eta_4]&\approx +\frac{m_4}{E}\bigg(1+\frac{m_4^2}{2E^2}\bigg)
e^{-i\phi}|3],
\end{align}
up to the second leading order in terms of the spinors growing in the high-energy limit. where we take the spinor notations, $\eta_i$ and $\lam_i \rightarrow i$ introduced in Eq.~\eqref{eq:massive-two_massless} except for the little group index for brevity. These replacements ensure that any amplitude includes only the contractions of spinors that grow with energy. For ease of reference, we list them as
\begingroup
\footnotesize
\begin{align}
&\langle 12\rangle\approx\left(E-\frac{m_1^2+m_2^2}{2E}-\frac{m_1^2m_2^2}{4E^3}\right),
& 
&[12]\approx-\left(E-\frac{m_1^2+m_2^2}{2E}-\frac{m_1^2m_2^2}{4E^3}\right),
\\
&\langle 34\rangle\approx\left(E-\frac{m_3^2+m_4^2}{2E}-\frac{m_3^2m_4^2}{4E^3}\right) e^{-i\phi},
&
&[34]\approx-\left(E-\frac{m_3^2+m_4^2}{2E}-\frac{m_3^2m_4^2}{4E^3}\right) e^{i\phi},
\\
&\langle 13\rangle\approx\left(E-\frac{m_2^2+m_4^2}{2E}-\frac{m_2^2m_4^2}{4E^3}\right)\sin\frac{\theta}{2}e^{-i\phi},
&
&[13]\approx\left(E-\frac{m_2^2+m_4^2}{2E}-\frac{m_2^2m_4^2}{4E^3}\right)\sin\frac{\theta}{2}e^{i\phi},
\\
&\langle 14\rangle\approx-\left(E-\frac{m_2^2+m_3^2}{2E}-\frac{m_2^2m_3^2}{4E^3}\right)
\cos\frac{\theta}{2}e^{-i\phi},
&
&[14]\approx-\left(E-\frac{m_2^2+m_3^2}{2E}-\frac{m_2^2m_3^2}{4E^3}\right)\cos\frac{\theta}{2}e^{i\phi},
\\
&\langle 23\rangle\approx\left(E-\frac{m_1^2+m_4^2}{2E}-\frac{m_1^2m_4^2}{4E^3}\right)\cos\frac{\theta}{2},
&
&[23]\approx\left(E-\frac{m_1^2+m_4^2}{2E}-\frac{m_1^2m_4^2}{4E^3}\right)\cos\frac{\theta}{2},
\\
&\langle 24\rangle\approx\left(E-\frac{m_1^2+m_3^2}{2E}-\frac{m_1^2m_3^2}{4E^3}\right)\sin\frac{\theta}{2},
&
&[24]\approx\left(E-\frac{m_1^2+m_3^2}{2E}-\frac{m_1^2m_3^2}{4E^3}\right)\sin\frac{\theta}{2},
\end{align}
\endgroup
up to the third leading order.

\section{4-pt amplitudes with no high-energy growth}
\label{appendix:amp_no_high-E}

In this section, we list the three 4-pt amplitudes with no high-$E$ growth. Although not used in the main text, they may be useful in practical analyses.

We first exhibit the $\phi\psi \phi \bar{\psi}$ and $\psi \bar{\psi}\psi \bar{\psi}$ amplitudes:
\begingroup
\small
\begin{align}
&\mathcal{M}_{4,s}
(\bs{1}_{i_1},\bs{2}_{\rm{i}_2},
\bs{3}_{i_3},\bar{\bs{4}}_{\rm{i}_4})
\nonumber
\\
&=\sum_{\rm{j}}
\Bigg(
\frac{1}{s_{12}-m^2_{\rm{j}}}
\Big(
(H_{i_3})_{\rm{i}_4 \rm{j}}
(H_{i_1}^\dagger)_{\rm{j} \rm{i}_2}
(m_{\rm{i}_2} \sqb{\bs{2}\bs{4}}-\asb{\bs{2}\bs{1}\bs{4}})
+
(H_{i_3}^{\dagger})_{\rm{i}_4 \rm{j}}
(H_{i_1})_{\rm{j} \rm{i}_2}
(m_{\rm{i}_4} \sqb{\bs{2}\bs{4}}-\asb{\bs{4}\bs{3}\bs{2}})
\nonumber
\\
&\qquad\qquad +(H_{i_3})_{\rm{i}_4 \rm{j}}
(H_{i_1})_{\rm{j} \rm{i}_2}
m_{\rm{j}} \sqb{\bs{2}\bs{4}}
+(H_{i_3}^\dagger)_{\rm{i}_4 \rm{j}}
(H_{i_1}^\dagger)_{\rm{j} \rm{i}_2}
m_{\rm{j}} \ab{\bs{2}\bs{4}}\Big)
+(\bs{1}_{i_1}\!\leftrightarrow \bs{3}_{i_3})\Bigg)
\nonumber
\\
&\quad -\sum_{b}\frac{iG_{bi_1i_3} }{m_b^2 (s_{13}-m^2_b)}
\Big\{ R^b_{\rm{i}_4\rm{i}_2}
\big[2m_b^2 \asb{\bs{412}}-(m_b^2+m^2_{i_1}-m^2_{i_3})
(m_{\rm{i}_2} \ab{\bs{24}}- m_{\rm{i}_4}\sqb{\bs{24}})\big]
\nonumber
\\
&\qquad \qquad\qquad\qquad\qquad
+L^b_{\rm{i}_4\rm{i}_2}
\big[2m_b^2 \asb{\bs{214}}-(m_b^2+m^2_{i_1}-m^2_{i_3})
(m_{\rm{i}_2} \sqb{\bs{24}}- m_{\rm{i}_4}\ab{\bs{24}})\big]
\Big\}
\nonumber
\\
&\quad -\sum_{\hat{b}}\frac{iG_{\hat{b}i_1i_3}}{ s_{13} }
\Big\{ R^{\hat{b}}_{\rm{i}_4\rm{i}_2}
\big[2\asb{\bs{412}}-(m_{\rm{i}_2} \ab{\bs{24}}- m_{\rm{i}_4}\sqb{\bs{24}})\big]
 +L^{\hat{b}}_{\rm{i}_4\rm{i}_2}
\big[
2 \asb{\bs{214}}-(m_{i_2}\sqb{\bs{24}}
-m_{i_4}\ab{\bs{24}})\big]\Big\}
\nonumber
\\
&\quad +\sum_{j} \frac{P_{j i_1 j_3}
\big(
(H_j)_{\rm{i}_4\rm{i}_2}
\sqb{\bs{24}}
+
(H_j^\dagger)_{\rm{i}_4\rm{i}_2}
\ab{\bs{24}}
\big)
}{s_{12}-m_j^2},
\end{align}
\endgroup
and
\begingroup
\small
\begin{align}
&\mathcal{M}_{4,s}
(\bs{1}_{\rm{i}_1},\bar{\bs{2}}_{\rm{i}_2},
\bs{3}_{\rm{i}_3},\bar{\bs{4}}_{\rm{i}_4})
\nonumber
\\
&=\Bigg(\sum_{b}
\frac{1}{m_b^2(s_{12}-m^2_{b})}
\Big\{\big[R^B_{\rm{i}_4\rm{i}_3}
(m_{\rm{i}_3} \sqb{\bs{3}\bs{4}}
- m_{\rm{i}_4} \ab{\bs{3}\bs{4}})
+
L^B_{\rm{i}_4\rm{i}_3}
(m_{\rm{i}_3} \ab{\bs{3}\bs{4}}
- m_{\rm{i}_4} \sqb{\bs{3}\bs{4}})\big]
\nonumber
\\
&\qquad\;\; \times \big[R^b_{\rm{i}_2\rm{i}_1}
(m_{\rm{i}_1} \sqb{\bs{1}\bs{2}}
- m_{\rm{i}_2} \ab{\bs{1}\bs{2}})
+
L^b_{\rm{i}_2\rm{i}_1}
(m_{\rm{i}_1} \ab{\bs{1}\bs{2}}
- m_{\rm{i}_2} \sqb{\bs{1}\bs{2}})\big]
\nonumber
\\[5pt]
&\qquad\;\; -2m_b^2
\big( R^b_{\rm{i}_4\rm{i}_3} L^b_{\rm{i}_2\rm{i}_1} 
\ab{\bs{2}\bs{3}} \sqb{\bs{1}\bs{4}}
+
R^b_{\rm{i}_2\rm{i}_1} L^b_{\rm{i}_4\rm{i}_3} 
\ab{\bs{1}\bs{4}} \sqb{\bs{2}\bs{3}}
+
R^b_{\rm{i}_2\rm{i}_1} R^b_{\rm{i}_4\rm{i}_3} 
\ab{\bs{1}\bs{3}} \sqb{\bs{2}\bs{4}}
+
L^b_{\rm{i}_2\rm{i}_1} L^b_{\rm{i}_4\rm{i}_3} 
\ab{\bs{2}\bs{4}} \sqb{\bs{1}\bs{3}}\big)\Big\}
\nonumber
\\[1pt]
&\quad -\sum_{\hat{b}}
\frac{2}{s_{12}}
\big( R^{\hat{b}}_{\rm{i}_4\rm{i}_3} L^{\hat{b}}_{\rm{i}_2\rm{i}_1} 
\ab{\bs{2}\bs{3}} \sqb{\bs{1}\bs{4}}
+
R^{\hat{b}}_{\rm{i}_2\rm{i}_1} L^{\hat{b}}_{\rm{i}_4\rm{i}_3} 
\ab{\bs{1}\bs{4}} \sqb{\bs{2}\bs{3}}
+
R^{\hat{b}}_{\rm{i}_2\rm{i}_1} R^{\hat{b}}_{\rm{i}_4\rm{i}_3} 
\ab{\bs{1}\bs{3}} \sqb{\bs{2}\bs{4}}
+
L^{\hat{b}}_{\rm{i}_2\rm{i}_1} L^{\hat{b}}_{\rm{i}_4\rm{i}_3} 
\ab{\bs{2}\bs{4}} \sqb{\bs{1}\bs{3}}\big)
\nonumber
\\
&\quad -\sum_{i}
\frac{\big(
(H_i)_{\rm{i}_2\rm{i}_1}
\sqb{\bs{1}\bs{2}}
+
(H_i^\dagger)_{\rm{i}_2\rm{i}_1}
\ab{\bs{1}\bs{2}}
\big)
\big(
(H_i)_{\rm{i}_2\rm{i}_1}
\sqb{\bs{3}\bs{4}}
+
(H_i^\dagger)_{\rm{i}_2\rm{i}_1}
\ab{\bs{3}\bs{4}}
\big)
}{s_{12}-m_i^2}
\Bigg) - (\bs{1}_{\rm{i}_1}\!\leftrightarrow \bs{3}_{\rm{i}_3}).
\end{align}
\endgroup
For the $\phi\phi\phi\phi$ amplitude, we have
\begin{align}
&\mathcal{M}_{4,s}
(\bs{1}_{i_1} \bs{2}_{i_2} \bs{3}_{i_3} \bs{4}_{i_4})
\nonumber
\\
&=-\sum_{B}\frac{G_{\capi{B} i_1 i_2}G_{\capi{B} i_3 i_4}
\big[(m_{i_1}^2-m_{i_2}^2)(m_{i_3}^2-m_{i_4}^2)+m_{\capi{B}}^2(s_{13}-s_{14})\big]
}{m_{\capi{B}}^2(s_{12}-m_{\capi{B}}^2)}-\sum_{j}\frac{P_{j i_1i_2}P_{j i_3i_4}}{
(s_{12}-m_j^2)},
\end{align}
in the $s$-channel. Deriving the $t$ and $u$-channel amplitudes through the exchanges $\bs{2}_{i_2}\leftrightarrow \bs{3}_{i_3}$ and $\bs{2}_{i_2}\leftrightarrow \bs{4}_{i_4}$, we obtain the $\phi\phi\phi\phi$ amplitude as
\begin{align}
&\mathcal{M}_{4}
(\bs{1}_{i_1} \bs{2}_{i_2} \bs{3}_{i_3} \bs{4}_{i_4})
\nonumber
\\
&=
\mathcal{M}_{4,s}
(\bs{1}_{i_1} \bs{2}_{i_2} \bs{3}_{i_3} \bs{4}_{i_4})
+
\mathcal{M}_{4,s}
(\bs{1}_{i_1} \bs{3}_{i_3} \bs{2}_{i_2} \bs{4}_{i_4})
+
\mathcal{M}_{4,s}
(\bs{1}_{i_1} \bs{4}_{i_4} \bs{2}_{i_2} \bs{3}_{i_3})
-Q_{i_1i_2i_3i_4},
\label{eq:SSSS_full_amp}
\end{align}
where $Q_{i_1i_2i_3i_4}$ is the dimensionless coupling of the contact $\phi\phi\phi\phi$ operator.

\section{Summary of Tree Unitarity Conditions}
\label{appendix:summary_conditions}

We summarize the tree unitarity conditions, only with which one can diagnose tree unitarity of a given particle content. Note that for any coupling expression, if no expression is specified for a given set of indices, it is understood to vanish.
\vspace{0.2cm}

\noindent
{\bf Notations of indices}
\begin{gather}
\{a,b,c\}\mbox{ : massive vector}
,\quad 
\{\hat{a},\hat{b},\hat{c}\}\mbox{ : massless vector}
\nonumber
\\
\{i,j,k\}\mbox{ : scalar}
,\quad 
\{{\rm i},{\rm j},{\rm k}\}\mbox{ : spin-1/2 fermion}
\nonumber
\\
\{A,B,C\}\mbox{ : collection of massive and massless vectors}
\nonumber
\\
\{{\cal I},{\cal J},{\cal K}\}\mbox{ : collection of vectors 
and scalar}
\nonumber
\end{gather}

\noindent
{\bf Tree unitary Lagrangian in the mass basis}
\begingroup
\small
\begin{align}
\mathcal{L}&=-\frac{1}{4}
(\partial_{\mu} V_{\capi{A}\nu}-\partial_{\nu} V_{\capi{B}\mu}
-C_{\capi{ABC}} V_{\capi{B}\mu}V_{\capi{C}\mu})^2
+
\bar{\psi}_R i(\slash \!\!\! \partial +i \slash \!\!\! V_{\capi{A}}
R^{\capi{A}})\psi_R
+
\bar{\psi}_L i(\slash \!\!\! \partial +i \slash \!\!\! V_{\capi{A}}
L^{\capi{A}})\psi_L
\nonumber
\\
&\quad
+\frac{1}{2}V_{\capi{A}\mu}V_{\capi{B}}^\mu
\bigg[ m_{\capi{A}}^2\delta_{\capi{AB}}
+2F_{\capi{AB}i}\phi_i+
\bigg(\frac{1}{m_c^2}F_{\capi{A}ci}F_{\capi{B}cj}
+G_{\capi{A}ki}G_{\capi{B}kj}\bigg)
\phi_i \phi_j\bigg]
\nonumber
\\
&\quad +\frac{1}{2}(\partial_{\mu}\phi_i)^2
-G_{\capi{A}ij}V_{\capi{A}\mu}\partial^\mu\phi_i \phi_j
-\mathcal{V}(\phi)
-\big[\bar{\psi}_L  \mathcal{Y}(\phi) \psi_R +\mbox{h.c.}\big],  \qquad (\mu_{\mathrm{ij}}=m_{\mathrm{i}}\delta_{\mathrm{ij}})
\end{align}
\endgroup

\begin{align}
\mathcal{V}(\phi)&=
\frac12 M_{ij}^2\phi_{i}\phi_{j}+\frac{1}{6}P_{ijk}\phi_{i}\phi_{\capmi{J}}\phi_{k}
+\frac{1}{24}Q_{ijkl}\phi_{i}\phi_{J}\phi_{k}\phi_{l},
\label{eq:D_V}
\\
\mathcal{Y}(\phi)&=\mu+H_{i}\phi_{i},
\label{eq:D_Y}
\end{align}

\noindent
{\bf Unification and optimization of couplings}
\vspace{0.2cm}

$a_1$) Bosonic couplings
\begin{gather}
T^{\capi{(A)}}_{bc}=i C_{\capi{A}bc}
\frac{m_{\capi{A}}^2-m_{b}^2-m_{c}^2}{2m_{b} m_{c}},
\quad T^{(\hat{a})}_{\hat{b}\hat{c}}=-i C_{\hat{a}\hat{b}\hat{c}},
\nonumber
\\
T^{(a)}_{ib}=
-T^{(a)}_{bi}=
\frac{i}{m_b}
F_{ab i}
,\quad
T^{\capi{(A)}}_{ij}
=iG_{\capi{A}ij},
\label{eq:D_T}
\end{gather}

$a_2$) Supplementary formulas for bosonic couplings
\begin{gather}
T^{\capi{(A)}} \lambda^{\capi{(B)}}
-
T^{\capi{(B)}} \lambda^{\capi{(A)}}
=iC_{\capi{ABC}}\lambda^{\capi{(C)}},
\label{eq:D_[T,lam]}
\\
\lam^{\capi{(A)}}_{\capmi{I}}=
\delta_{\capmi{I}\capi{A}}m_{\capi{A}}=(iT^{\capi{(A)}}v)_{\capmi{I}}\qquad (v_{\capi{A}}=0),
\label{eq:D_lam=iTv}
\\
v_{\capmi{I}}
=-\Big(\sum_{b}T^{(b)}T^{(b)}\Big)^{-1}_{\capi{IJ}}
\sum_{a}(iT^{a}_{\capmi{J}a})m_{a},
\label{eq:D_v=Tm}
\end{gather}

$b$) Fermionic couplings
\begin{gather}
\mu=\mu_{0}+H_iv_i,
\label{eq:D_mu0}
\\
L^a \mu_0-\mu_0 R^a=0
\label{eq:D_Lmu0-mu0R}
\\
H_a
=\frac{i}{m_a}(L^a H_i -H_i R^a) v_i,
\label{eq:D_Ha}
\end{gather}

$c_1$) Couplings in the scalar potential
\begin{align}
\bar{Q}_{\capmi{IJKL}}&=Q_{\capmi{IJKL}},
\nonumber
\\
\bar{P}_{\capmi{IJK}}
&=P_{\capmi{IJK}}-Q_{\capmi{IJKL}}v_{\capmi{L}},
\nonumber
\\
\bar{M}^2_{\capmi{IJ}}&=M_{\capmi{IJ}}^2-
P_{\capmi{IJK}}
v_{\capmi{K}}+\frac12 Q_{\capmi{IJKL}}
v_{\capmi{K}}v_{\capmi{L}}.
\label{eq:D_bar_nobar}
\end{align}

$c_2$) Supplementary formulas for the couplings in the scalar potential
\begin{gather}
\bar{M}^2_{\capmi{IJ}}v_{\capmi{J}}=-M^2_{\capmi{IJ}}v_{\capmi{J}},
\\
M^2_{\capmi{I}\capi{A}}=0,
\\
P_{\capi{A}\capmi{IJ}}v_{\capmi{J}}=Q_{\capi{A}\capmi{IJK}}v_{\capmi{J}}v_{\capmi{K}},
\\
P_{\capi{A}\capmi{IJ}}v_{\capmi{I}}v_{\capmi{J}}=Q_{\capi{A}\capmi{IJK}}v_{\capmi{I}}v_{\capmi{J}}v_{\capmi{K}}
=0,
\end{gather}

\noindent
{\bf Tree unitraity conditions of 4-pt amplitudes}
\begin{gather}
[T^{\capi{(A)}},T^{\capi{(B)}}]_{\capi{\mathcal{I}\mathcal{J}}}=
iC_{\capi{ABC}}T^{\capi{(C)}}_{\capi{\mathcal{I}\mathcal{J}}},
\label{eq:D_[T,T]}
\\
[R^{\capi{A}},R^{\capi{B}}]=iC_{\capi{ABC}}R^{\capi{C}}
,\quad
[L^{\capi{A}},L^{\capi{B}}]=iC_{\capi{ABC}}L^{\capi{C}},
\label{eq:D_[R,R]}
\\
H_{\capi{\mathcal{I}}}R^{\capi{A}}
-L^{\capi{A}}H_{\capi{\mathcal{I}}}
-T^{\capi{A}}_{\capi{\mathcal{I}\mathcal{J}}}H_{\capi{\mathcal{J}}}=0,
\label{eq:D_HR-LH-TH}
\end{gather}

\noindent
{\bf Tree unitraity conditions of higher-pt amplitudes}
\begin{align}
&(iT^{\capi{(A_1)}}_{\capmi{I}_2\capmi{J}})
\bar{M}^2_{\capmi{J}\capmi{I}_3}
+
\mbox{cyclic}(\mathcal{I}_2,\mathcal{I}_3)
=0,
\label{eq:D_M2}
\\
&(iT^{\capi{(A_1)}}_{\capmi{I}_2\capmi{J}})
\bar{P}_{\capmi{J}\capmi{I}_3\capmi{I}_4}
+
\mbox{cyclic}(\mathcal{I}_2,\mathcal{I}_3,\mathcal{I}_4)
=0,
\label{eq:D_P}
\\
&
(iT^{\capi{(A_1)}}_{\capmi{I}_2\capmi{J}})
Q_{\capmi{J}\capmi{I}_3\capmi{I}_4\capmi{I}_5}
+
\mbox{cyclic}(\mathcal{I}_2,\mathcal{I}_3,\mathcal{I}_4,\mathcal{I}_5)
=0,
\label{eq:D_Q}
\end{align}

\noindent
{\bf Couplings $\bar{M}^2$, $\bar{P}$, and $Q$ with vector indices}

\begin{gather}
\bar{M}_{\capi{A_1A_2}}^2
=
-\frac{(iT^{\capi{(A_1)}}M^2 v)_{\capi{A_2}}}{m_{\capi{A_1}}}
,\quad
\bar{M}_{\capi{A_1}i_2}^2
=
-\frac{(iT^{\capi{(A_1)}}M^2 v)_{i_2}}{m_{\capi{A_1}}}, 
\label{eq:D_M2_vector}
\\
\bar{P}_{\capi{A_1A_2}i_3}
=\frac{(iT^{\capi{(A_1)}}_{\capi{A_2}j_1})}{m_{\capi{A_1}}}
\bar{P}_{i_3j_1 j_2}v_{j_2}
,\quad
\bar{P}_{\capi{A_1}i_2i_3}
=\frac{(iT^{\capi{(A_1)}}_{i_2 j_1})}{m_{\capi{A_1}}}
\bar{P}_{i_3 j_1 j_2} v_{j_2}
+
(i_2\leftrightarrow i_3),
\label{eq:D_P_vector}
\end{gather}
\begin{align}
Q_{\capi{A_1A_2A_3A_4}}
&=\frac{[iT^{\capi{(A_1)}},[iT^{\capi{(A_4)}},M^2]]_{\capi{A_2A_3}}}{2m_{\capi{A_1}}m_{\capi{A_4}}}+\mbox{cyclic}(A_2,A_3,A_4),
\nonumber
\\
Q_{\capi{A_1A_2A_3}i_4}
&=\frac{[iT^{\capi{(A_1)}},[iT^{\capi{(A_3)}},M^2]]_{\capi{A_2}i_4}}{m_{\capi{A_1}}m_{\capi{A_3}}}+(A_2\leftrightarrow A_3)
-
\frac{(iT^{\capi{(A_1)}}[iT^{\capi{(A_3)}},M^2])_{i_4\capi{A_2}}}{m_{\capi{A_1}}m_{\capi{A_3}}},
\nonumber
\\
Q_{\capi{A_1A_2}i_3i_4}
&=\frac{[iT^{\capi{(A_1)}},[iT^{\capi{(A_2)}},M^2]]_{i_3i_4}}{m_{\capi{A_1}}m_{\capi{A_2}}}+\frac{(iT^{\capi{(A_1)}}_{\capi{A_2B}})[iT^{\capi{(B)}},M^2]_{i_3i_4}}{m_{\capi{A_1}}m_{\capi{B}}}
+\frac{(iT^{\capi{(A_1)}}_{\capi{A_2}j})}{m_{\capi{A_1}}}P_{ji_3i_4},
\nonumber
\\
Q_{\capi{A_1}i_2i_3i_4}
&=\frac{(iT^{\capi{(A_1)}}_{i_2\capi{B}})[iT^{\capi{(B)}},M^2]_{i_3i_4}}{m_{\capi{A_1}}m_{\capi{B}}}
+\frac{(iT^{\capi{(A_1)}}_{i_2j})}{m_{\capi{A_1}}}P_{ji_3i_4}
+\mbox{cyclic}(i_2,i_3,i_4),
\label{eq:D_Q_vector}
\end{align}
%
%
%



 \bibliographystyle{JHEP}
 \bibliography{biblio.bib}


\end{document}